\documentclass[12pt,twoside,beltcrest]{ociamthesis}

\usepackage{amsmath}
\usepackage{amssymb}
\usepackage{cancel}
\usepackage{cite}
\usepackage{dsfont}

\let\origdoublepage\cleardoublepage\newcommand
{\clearemptydoublepage}{
  \clearpage
  {\pagestyle{empty}\origdoublepage}
}

\newcommand{\nn}{\nonumber}
\newcommand{\ph}{\phantom}

\newcommand{\be}{\begin{equation}}
\newcommand{\ee}{\end{equation}}
\newcommand{\bea}{\begin{eqnarray}}
\newcommand{\eea}{\end{eqnarray}}
\newcommand{\bs}{\begin{subequations}}
\newcommand{\es}{\end{subequations}}

\newcommand{\calk}{{\mathcal K}}
\newcommand{\cM}{{\cal M}}
\newcommand{\cL}{{\cal L}}
\newcommand{\cO}{{\cal O}}
\newcommand{\cV}{{\cal V}}
\newcommand{\cK}{{\cal K}}
\newcommand{\cW}{{\cal W}}
\newcommand{\cX}{{\cal X}}
\newcommand{\cZ}{{\cal Z}}
\newcommand{\cG}{{\cal G}}

\newcommand{\im}{\mathrm{Im}\,}
\newcommand{\re}{\mathrm{Re}\,}
\newcommand{\tr}{\mathrm{tr}}

\title{Generalized Compactification in \\[1ex]
       Heterotic String Theory}

\author{Cyril Matti}
\college{Lincoln College}

\degree{Doctor of Philosophy}
\degreedate{Trinity 2011}

\begin{document}

\baselineskip=21pt plus1pt

\setcounter{secnumdepth}{3}
\setcounter{tocdepth}{3}

\maketitle
\clearemptydoublepage
\begin{dedication}
\emph{\`A ma famille.}
\end{dedication}
\clearemptydoublepage
\begin{dedication}
{\emph{``Tout change autour de nous. Nous changeons nous-m\^emes et nul ne peut s'assurer qu'il aimera demain ce qu'il aime aujourd'hui."}} \\
\begin{flushright}
--- Jean-Jacques Rousseau
\end{flushright}
\end{dedication}
\clearemptydoublepage
\begin{abstractseparate}

\vspace{1.5cm}
In this thesis, we consider heterotic string vacua based on a warped product of a four-dimensional domain wall and a six-dimensional internal manifold preserving only two supercharges. Thus, they correspond to half-BPS states of heterotic supergravity. The constraints on the internal manifolds with $SU(3)$ structure are derived. They are found to be a generalization of half-flat manifolds with a particular pattern of torsion classes and they include half-flat manifolds and Strominger's complex non-Kahler manifolds as special cases. We also verify that heterotic compactifications on half-flat mirror manifolds are based on this class of solutions.

Furthermore, within this context, we construct specific examples based on six-dimensional nearly-Kahler homogeneous manifolds and non-trivial vector bundles thereon. Our solutions are based on three specific group coset spaces satisfying the half-flat torsion class conditions. It is shown how to construct line bundles over these manifolds, compute their properties and build up vector bundles consistent with supersymmetry and the heterotic anomaly cancellation. It turns out that the most interesting solutions are obtained from $SU(3)/U(1)^2$. This space supports a large number of vector bundles leading to consistent heterotic vacua with GUT group and, for some of them, with three chiral families.

\end{abstractseparate}
\clearemptydoublepage
\begin{acknowledgements}

It is an immense pleasure to express my deepest gratitude for my supervisor Professor Andr\'e Lukas, without whom I would have never written this thesis. I would also like to thank my collaborator Michael Klaput, who participated in the elaboration of part of this work.

Many thanks as well to my colleagues and friends Hwasung Lee, Seung-Joo Lee and Maxime Gabella for invaluable support and discussion throughout the whole period of my D.Phil. research.

My thanks also goes to Yang Hui He, James Gray, Lara Anderson and Eran Palti for helping me with their knowledge of string theory, along with everybody who stopped by Office $6.4$ for interesting discussion and for providing a friendly work environment: Maxime Gabella, Chris McCabe, Andreas Athenodorou, Joao Rosa, Andrei Constantin and Georgios Giasemidis. Furthermore, I am grateful to everybody from the HA group for sharing their enthusiasm of mathematics and for introducing me into the realm of algebraic geometry (even though not being use in this thesis).

Last but not least, thanks to Katie, Ruth and especially Liz for moral support and the good times spent together.

Nothing would have taken place without the generous scholarship of the Berrow Foundation in association with Lincoln College. I also acknowledge the Swiss National Science Foundation for providing me precious funding for the last year of my research.

\end{acknowledgements}
\clearemptydoublepage

\begin{romanpages}
\tableofcontents
\clearemptydoublepage
\listoftables
\clearemptydoublepage
\end{romanpages}

\chapter{Introduction}
\section{String vacuum}

String theory is based on the assumption that the fundamental constituents of the universe are made out of strings rather than point particles. While this starting point may seem simple, its consequences are far from undemanding. For instance, world-sheet supersymmetry needs to be included in order to have a tachyon-free theory and, in addition, quantum excitations require a specific number of space-time dimensions, namely ten for the case of a critical theory of superstrings. The motivation for such a setting lies in the quest of a consistent theory of quantum gravity. As a matter of fact, the well-established framework of gauge fields theories cannot be extended using standard methods to include the gravitational force, as it leads to a non-renormalizable theory. On the other hand, the quantum particle mediating gravity, the graviton, appears naturally within the string spectrum in a consistent manner.

String theory has been very successful towards numerous avenues. For example, it encompasses Einstein's General Relativity equations for the space time metric and presents features compatible with modern developments of theoretical physics such as: extra dimensions~\cite{Matti:2006fx, Shifman:2009df}, supersymmetry~\cite{Sohnius:1985qm, Wess:1992cp} or Grand Unified Theory (GUT) groups~\cite{Georgi:1982jb}. However, it is still lacking of explicit examples leading to the complete Standard Model of particles which is necessary to embody a physical theory. It has also other drawbacks such as the existence of several different theories that, even though they are related through dualities, still represent a vast degeneracy of vacua. Hence, it should merely be considered as a framework rather than a single theory. In this thesis, we will focus on the case of heterotic string theory~\cite{Gross:1985fr} which is characterized by a ten-dimensional space-time and a vector bundle over it with gauge group $SO(32)$ or $E_8\times E_8$, the latter being our primary source of attention.\\

The goal being to describe a theory of particle physics, a mechanism must take place in order to reduce the number of dimensions down to the four observable ones. The standard paradigm considers the extra six dimensions to be a compact internal manifold with characteristic length $l_c$ small enough not to be observed. The energy scale necessary to probe such a space is of the order of $\sim1/l_c$. Therefore, at a regime below this energy such as, say, at energies reached by the LHC\footnote{The Large Hadron Collider (LHC) at CERN currently reaches the highest energy scale obtained by man-made physics experiment.}, the internal dimensions remain unobserved. The corresponding dimensional reduction leading to the four-dimensional effective theory goes under the name of Kaluza-Klein mechanism~\cite{Duff:1986hr}. A legitimate question arises then about which manifold is appropriate for such a purpose. The simple answer is to consider backgrounds which are vacuum solutions of the theory. Nonetheless, a lot of freedom still remains about the specific choice of internal geometry.

The precise nature of this six-dimensional manifold plays a crucial role because, although invisible, its geometry is responsible for various properties of the effective four-dimensional theory. For instance, an adequate choice can spontaneously break some of the ten-dimensional supersymmetry at very high energy (as close as the Planck scale). In that case, it is customary to keep $N=1$ supersymmetry in four dimensions on phenomenological grounds. In addition, its topology determines the net number of particle generations, their field content and their corresponding Yukawa couplings~\cite{Strominger:1985ks}. Furthermore, for the particular case of heterotic string where the gauge fields are provided by the vector bundle, the choice of background geometry allows one to break the gauge groups $SO(32)$ or $E_8\times E_8$ down to a more phenomenological GUT group. Indeed, the compactness of the internal space enables a non-trivial fibration of the bundle fibers, thus, turning on vacuum expectation values for the gauge fields. The four-dimensional effective (physical) gauge group is then given by the commutant of the vector bundle structure group with the full $SO(32)$ or $E_8\times E_8$ group as these are the modes in which the gauge fields can still fluctuate effortlessly. A careful choice leads to an appropriate GUT group.

Heterotic string theory displays itself as a very promising candidate leading to a four-dimensional $N=1$ supergravity with GUT group, thanks to its gauge bundle. In turns, this effective theory has phenomenologically attractive characteristics in that it incorporates a natural supersymmetric extension of the Standard Model --- namely the Minimal Supersymmetric Standard Model --- and allows for a unification of the gauge couplings. However, the existence of a huge degeneracy of the theory's vacuum presents difficulties. The obstacle is two-fold: first, the selection of a six-dimensional manifold offers unfortunately far too many possibilities and then, the choice of a vector bundle over it multiply the options. Identifying the appropriate vacuum reproducing the desired phenomenology remains one of the main challenges of string theory and is under remarkable investigations~\cite{Lee, Braun:2005ux, Braun:2005bw, Bouchard:2005ag, Anderson:2009mh, He:2009wi, Anderson:2011cza, Anderson:2011ns, He:2011rs}.

The subject of this thesis resides within the context of exploring heterotic vacua on general grounds. Its goal is to present new directions of research, including new compactification options interesting to pursue model building. For this purpose, we look at backgrounds breaking more supersymmety than the standard lore of four-dimensional $N=1$ supergravity. We study solutions of heterotic supergravity comprising of a domain wall and an internal compact six-dimensional space.

\section{Motivations and outline}

Part of any program aiming at realistic string models must be the stabilization of moduli and this remains difficult for heterotic compactifications. In type II models a combination of NS and RR flux allows one, at least in principle, to stabilize all complex structure moduli and the dilaton while, thanks to the no-scale structure, keeping the theory in a Minkowski vacuum~\cite{Giddings:2001yu, Grana:2005jc}. On the other hand, only NS-NS flux is available in heterotic compactifications. This stabilizes the complex structure moduli only and we need to use topological properties of the internal space to stabilize the remaining fields~\cite{Dasgupta:1999ss}. Moreover, the heterotic flux superpotential, unlike its IIB counterpart, does not allow itself to be tuned to small values by a careful choice of the flux integers. These features mean that it will be difficult at best to achieve a scale separation between the string and the flux scale in heterotic Calabi-Yau models with flux.

The previous discussion suggests that heterotic models with stable moduli may require compactifications on more general manifolds with $SU(3)$ structure where some of the ``missing" RR flux is replaced by the intrinsic torsion of the manifold. Studying such more general backgrounds for heterotic compactifications is the main purpose of this thesis. One such class has been identified early on by Strominger~\cite{Strominger:1986uh}. It is obtained by assuming a maximally symmetric four-dimensional space and four preserved supercharges. In this case, it turns out that the associated internal six-dimensional manifolds have $SU(3)$ structure and are complex but, in general, no longer Kahler. In this thesis, we will generalize this discussion by relaxing both initial assumptions. We will allow the four-dimensional space to deviate from maximal symmetry. More specifically, we will allow it to be a domain wall and we will only require two preserved supercharges for the $10$-dimensional solution.

Why are we interested in backgrounds which violate the conventional requirement of a four-dimensional maximally symmetric space? The simple answer is that there exists conditions where the lowest order in $\alpha'$ flux superpotential in four dimensions leads to a runaway direction for some of the moduli and the simplest solution consistent with this feature is a four-dimensional domain wall. This happens for heterotic compactifications on half-flat manifolds as studied in Refs.~\cite{Gurrieri:2004dt, Gurrieri:2007jg, deCarlos:2005kh, Gurrieri:2005af} as well as for the crude heterotic Calabi-Yau compactification with flux.

In general, world sheet effects --- such as instantons --- need to be considered in order to stabilize all moduli. This is typically the case for the dilaton mode. Therefore, we should phenomenologically require a four-dimensional maximally symmetric space only after all relevant effects have been included, including non-perturbative ones. When studying $10$-dimensional perturbative string solutions, we could then allow for more general four-dimensional spaces, keeping in mind the possibility of a non-perturbative ``lift" to a maximally symmetric four-dimensional space. This allows us to study some more general backgrounds as generally considered as potentially interesting solutions.\\

With this motivations in mind, we will study $10$-dimensional solutions of the heterotic string which consist of a warped product of a six-dimensional internal space and a four-dimensional domain wall that preserves only two supercharges --- thus, they are half-BPS from a four-dimensional $N=1$ point of view. There are two main questions we would like to answer in this context. First, what are the allowed internal six-dimensional spaces in such a setting? This question will be answered using the $G$-structure formalism~\cite{Hitchin:2001rw, Chiossi:2002tw, LopesCardoso:2002hd} applied to the heterotic case for the groups $G_2$ and $SU(3)$. This leads to a significant generalization of the class of manifolds found by Strominger. Secondly, we would like to show the consistency of certain heterotic compactifications on half-flat mirror manifolds \cite{Gurrieri:2004dt, Gurrieri:2005af, Gurrieri:2007jg, deCarlos:2005kh} which has been carried out in the absence of a full $10$-dimensional solution. This will be done by verifying that such half-flat mirror manifolds are allowed internal geometries within our generalized setting and that the domain wall solutions in the associated four-dimensional $N=1$ supergravity do indeed lift up to the correct $10$-dimensional solutions.

Furthermore, having established the existence of such backgrounds, we will turn to finding explicit examples, mainly to allow for the study of the gauge sector. The presence of gauge bundles is one of the distinctive features of heterotic string compactifications and is responsible for many of the physically interesting properties as well as technical complications of heterotic models. For Calabi-Yau compactifications, the internal metric is not known explicitly (which makes it difficult to find gauge connections), however, this problem is largely circumnavigated by using techniques from algebraic geometry. On the other hand, for the case of half-flat mirror manifolds, the lack of an integrable complex structure adds a complication in that the powerful tools of algebraic geometry cannot be directly applied. Notwithstanding, we will focus on a small class of half-flat coset manifolds which are suitable for heterotic compactifications and have the advantage of allowing for an explicit computation of most of the relevant gauge field quantities. We will consider the three coset spaces $SU(3)/U(1)^2$, $Sp(2)/SU(2)\times U(1)$ and $G_2/SU(3)$. In Refs.~\cite{Nolle:2010nn, Lechtenfeld:2010dr, Chatzistavrakidis:2008ii, Chatzistavrakidis:2009mh}, heterotic compactifications on these manifolds have been studied focusing on the gravitational sector of the theory. In this thesis, we want to study gauge bundles over these spaces in order to construct consistent heterotic compactifications. Their group origin facilitates computations as metrics and gauge connections can be explicitly constructed and the relevant equations of the $10$-dimensional $N=1$ supergravity can be checked directly.\\

In chapter~\ref{chapterreview}, we start with a review of heterotic supergravity, mainly to set our conventions. We present in section~\ref{sectionaction} the general action of the bosonic heterotic background fields and their corresponding equations of motion. We subsequently describe the two kinds of compact internal six-dimensional manifolds preserving four supercharges and leading to maximally symmetric solutions in four dimensions. They are the Calabi-Yau manifolds, which are manifolds of $SU(3)$ holomony, and the more general solutions having $SU(3)$ structure that are characterized by the Strominger system of differential equations. Thereafter, we turn to the four-dimensional theory resulting from a Kaluza-Klein dimensional reduction and describe in section~\ref{sectionCY} the corresponding theory for the Calabi-Yau case. Finally, due to the lack of explicit examples of manifolds satisfying the Strominger system, we only present general aspects of the related four-dimensional theory and motivates, in section~\ref{sectionflux}, the Kahler potentials and superpotential coming from an adequate choice of field truncation. This is then illustrated with the case of half-flat mirror manifolds.

In chapter~\ref{chapterDW}, we present our solutions corresponding to the half-BPS compactifications that preserve only two supercharges from the ten-dimensional supersymmetry and show the consistency of mirror half-flat compactification. We start in section~\ref{sectionansatz} with the presentation of the corresponding metric ansatz. We then review the four-dimensional domain wall solutions of $N=1$ supergravity in section~\ref{section4dDW}. This is subsequently used to demonstrate that the effective four-dimensional domain walls do lift up to the correct ten-dimensional solutions. In the last sections, we illustrate this for the three cases of mirror half-flat manifolds with vanishing NS-NS flux, mirror half-flat manifolds with non-trivial flux and, finally, Calabi-Yau compactification with non-vanishing flux.

In chapter~\ref{chaptercoset}, we go further by picking up specific examples of mirror half-flat manifolds and study the gauge sector. We start with a short description of the formalism being used, that is coset space geometry. We then present the three explicit examples which are the six-dimensional nearly-Kahler homogeneous spaces $SU(3)/U(1)^2$, $Sp(2)/SU(2)\times U(1)$ and $G_2/SU(3)$. These manifolds are shown to be solutions of the heterotic string with vanishing flux and constant dilaton. In the last two sections of the chapter, we build vector bundles over these coset spaces and derive examples satisfying the equations of motion and the heterotic anomaly cancellation condition. We thus provide complete examples of new heterotic non-Calabi-Yau vacua that can be used for model building. The work of chapter~\ref{chapterDW} and chapter~\ref{chaptercoset} is entirely based on the papers~\cite{Lukas:2010mf} and~\cite{Klaput:2011mz}, respectively:
\begin{itemize}
\item A.~Lukas, C.~Matti,
  {\em {G-structures and Domain Walls in Heterotic Theories}},
  \newblock JHEP {\bf 1101 } (2011)  151,
  [arXiv:1005.5302 [hep-th]].
\item M.~Klaput, A.~Lukas, C.~Matti,
  {\em {Bundles over Nearly-Kahler Homogeneous Spaces in Heterotic String Theory}},
  \newblock JHEP {\bf 1109 } (2011)  100,
  [arXiv:1107.3573 [hep-th]].
\end{itemize}

Finally, in chapter~\ref{chapterconclusion}, we present our conclusion and draw possible future directions of research. Technical appendices are also included for the convenience of the reader. Appendix~\ref{appendixconventions} summarizes conventions and notations used throughout the thesis. Appendix~\ref{appendixtorsion} consists of a short review about manifolds with $G$-structure focusing on the case of $G_2$ and $SU(3)$ in $6$ and $7$ dimensions. In appendix~\ref{appendixCY}, we collect the relevant formulas for the moduli space of Calabi-Yau manifolds which apply to mirror half-flat manifolds as well. Last, appendix~\ref{appendixcoset} describes the geometry of six-dimensional nearly-Kahler homogeneous spaces by presenting the general formalism and listing the relevant data being used.
\clearemptydoublepage
\chapter{Review of heterotic supergravity}\label{chapterreview}

We start with a short review of heterotic supergravity. First, we study the ten-dimensional theory and, then, the effective four-dimensional supergravity resulting from compactification of the vacuum geometry. It is well established that, in the low-energy limit $\alpha'\rightarrow 0$, the massless modes of heterotic string theory are correctly approximated by a ten-dimensional $N=1$ supergravity~\cite{Becker:2007zj, Polchinski:1998rr}. Our main focus will be the nature of the vacuum of such a theory. Its geometry is responsible for breaking some of the original ten-dimensional supercharges and, in the standard analysis, leads to a well-defined $N=1$ supergravity in four dimensions. Furthermore, the background geometry determines most of the phenomenological characteristics of the effective supergravity obtained after dimensional reduction such as, for instance, the field content, the Yukawa couplings and the number of generations.

In this chapter, we will present the bosonic sector of the heterotic string supergravity. These are the fields which specify the background geometry. Indeed, the fermionic fields must vanish on a classical vacuum. We begin with the action and its equations of motion and describe the corresponding constraints necessary for having a supersymmetric vacuum. Then, we describe the standard compactifications of the Calabi-Yau solution and the Strominger system. We also present the general structure of $N=1$ four-dimensional supergravity together with the specific theories resulting from dimensional reduction on Calabi-Yau manifolds. Finally, we end up with a discussion of the effective four-dimensional theory resulting from compactifications with non-trivial fluxes.

\section{Ten-dimensional effective supergravity}\label{sectionaction}

\subsection{Bosonic sector lagrangian and equations of motion}

As previously stated, from a ten-dimensional perspective, the vacuum geometry is determined by the bosonic fields only. For the heterotic string, the background field spectrum consists of the ten-dimensional metric $g_{MN}$, the dilaton $\hat\phi$ and the NS-NS rank two anti-symmetric tensor field $\hat{B}_{MN}$. We also have the Yang-Mills gauge fields $A_M$ with corresponding gauge group $SO(32)$ or $E_8\times E_8$ and field strength
\begin{equation}
  \hat F=dA+A\wedge A \;.
\end{equation}
(Conventions about forms and their corresponding tensors are summarized in appendix~\ref{appendixconventions}). In addition, we can associate a three-form field strenght $\hat H$, the NS-NS flux, to the two-form field $\hat B$ as follows
\begin{equation}\label{H}
  \hat H=d\hat B+\frac{\alpha'}{4}\left(\omega_L-\omega_{YM}\right) \;,
\end{equation}
where the Lorentz and the Yang-Mills Chern-Simons terms are, respectively,
\begin{align}
  \omega_L&=\tr\left(\omega\wedge d\omega+\frac{2}{3}\omega\wedge\omega\wedge\omega\right) \;,\\
  \omega_{YM}&=\tr\left(A\wedge dA+\frac{2}{3}A\wedge A\wedge A\right) \;,
\end{align}
with the spin connection $\omega$ and the Yang-Mills connection $A$. For later purposes, it is also useful to introduce the connection
\begin{equation}\label{Bismut}
  \nabla^{(H)}_MV^N=\nabla_MV^N+\frac{1}{2}{\hat H}^N_{\phantom{N}MP}V^P \;,
\end{equation}
where $V^N$ is any vector field and $\nabla_M$ is the covariant derivative associated to the Levi-Civita connection of the metric $g$. This corresponds to a connection with torsion given by the NS-NS field $\hat H$. To the first order in $\alpha'$, the bosonic part of the string frame action is given by
\be\label{action}
\begin{aligned}
  S=-\frac{1}{2\kappa^2_{10}}\int_{M_{10}}e^{-2\hat\phi}\bigg[&\hat{R}*\textbf{1}-4d\hat\phi\wedge *d\hat\phi+\frac{1}{2}\hat H\wedge *\hat H \\
  &+\frac{\alpha'}{4}e^{\hat\phi}\left({\rm tr}\hat F\wedge *\hat F-{\rm tr}\hat{R}^-\wedge*\hat R^-\right)\bigg] \;,
\end{aligned}
\ee
where $\kappa_{10}$ is the ten-dimensional Planck constant and $\hat R^-$ is the curvature corresponding to the connection $\nabla^{-}_MV^N=\nabla_MV^N-\frac{1}{2}{\hat H}^N_{\phantom{N}MP}V^P$. The resulting equations of motion governing the dynamic of these fields are
\bs\label{Einstein}
\begin{align}
  &\hat{R}_{MN}-\frac{1}{4}\hat H_{PQM}\hat H^{PQ}_{\phantom{PQ}N}+2\nabla_M\partial_N\hat\phi\nn\\&\hspace{2cm}+\frac{\alpha'}{4}\left(\hat R^-_{MPQR}\hat R^{-\phantom{N}PQR}_N-\tr\hat F_{MP} \hat F_N^{\phantom{N}P}\right)+\cO\left(\alpha'^2\right)=0 \;, \\
  &\nabla^2\hat\phi-2g^{MN}\partial_M\hat\phi\partial_N\hat\phi+\frac{1}{12}\hat H_{MNP}\hat H^{MNP}\nn\\&\hspace{2cm}+\frac{\alpha'}{16}\left(\tr\hat F_{MN}\hat F^{MN}-\tr\hat R^-_{MN}\hat R^{-MN}\right)+\cO\left(\alpha'^2\right)=0 \;, \\
  &\nabla_M\left(e^{-2\hat\phi}\hat H^M_{\phantom{M}PQ}\right)+\cO\left(\alpha'^2\right)=0 \;, \\
  &\nabla^{(H)}_M\left(e^{-2\hat\phi}\hat F^M_{\phantom{M}N}\right)+\cO\left(\alpha'^2\right)=0 \;.
\end{align}
\es
These equations correspond to Einstein equations with field strength given by the dilaton and the NS-NS flux.

\subsection{Killing spinor equations and Bianchi identity}

The fermionic partners of the above fields are the gravitino $\psi_M$, the dilatino $\lambda$ and the gauginos $\chi$, all of which being ten-dimensional Majorana-Weyl spinors. Their supersymmetry transformations are obtained by
\begin{subequations}
\begin{align}\label{dpsi} 
  \delta\psi_M&=\left(\nabla_M+\frac{1}{8}{\cal\hat H}_M\right)\epsilon \;,\\\label{dlambda} 
  \delta\lambda&=\left(\cancel{\nabla}\hat\phi+\frac{1}{12}{\cal\hat H}\right)\epsilon \;,\\\label{dchi}
  \delta\chi&=\hat F_{MN}\Gamma^{MN}\epsilon \;,
\end{align}
\end{subequations}
where $\epsilon$ is a ten-dimensional Majorana-Weyl spinor parameterizing the transformations. Here, and in the following, we use the short-hand notation ${\cal\hat H}_M=\hat{H}_{MNP}\Gamma^{NP}$ and ${\cal\hat H}=\hat{H}_{MNP}\Gamma^{MNP}$ for the contraction of the field strength $\hat{H}$ with products of $10$-dimensional gamma matrices $\Gamma^M$. Supersymmetric vacuum is characterized by vanishing supersymmetry transformations $\delta\psi_M=0$, $\delta\lambda=0$ and $\delta\chi=0$. In particular, it implies that $\epsilon$ is covariantly constant with respect to the connection $\nabla^{(H)}_M$. It is well known~\cite{Ivanov:2009rh} that such a requirement is sufficient for having solutions to the bosonic equations of motion~\eqref{Einstein}, including the first order terms $\cO(\alpha')$, provided that the following Bianchi identity is satisfied as well,
\begin{equation}\label{bianchi}
  d\hat H=\frac{\alpha'}{4}\left(\tr\hat F\wedge \hat F-{\rm tr}\hat R^-\wedge \hat R^-\right) \;.
\end{equation}
This identity comes from taking the exterior derivative of the definition~\eqref{H} and is necessary for anomaly cancellations. A further investigation about these conditions reveals that the right-hand side term appears at the first order in $\alpha'$ and contributes to an exact form. Hence, to find solutions at zeroth order in the string tension, we can satisfy this equation in cohomology only. In our study, we concentrate on the case of the $E_8\times E_8$ heterotic string and, therefore, the vector bundle splits into a visible and a hidden sector. Consequently, we can write the Bianchi identity as
\begin{equation}\label{biancoho}
  \left[{\rm tr}\hat R^-\wedge\hat R^-\right]=\left[{\rm tr} F\wedge F+{\rm tr}\tilde F\wedge\tilde F\right] \;,
\end{equation}
where $F$ and $\tilde F$ denotes the visible and the hidden sector contributions respectively and the square bracket indicates cohomology classes in $H^4$. Of course, proceeding this way implies that our solutions get $\cO(\alpha')$ corrections and further analysis is required to provide a full solution to the field equations~\eqref{Einstein}. However, the goal of this thesis is to discuss solutions to the above Killing spinor equations and the Bianchi identity at zeroth order in $\alpha'$. This would be the basis for finding in subsequent work a complete solution perturbatively at all orders, see Refs.~\cite{Witten:1986kg,Gillard:2003jh,Martelli:2010jx} for a general discussion of the $\alpha'$ expansion.

\section{$N=1$ compactification}

Our world being four-dimensional, we need a procedure to reduce the $10$-dimensional theory down to four dimensions. The standard paradigm is to consider six out of the ten dimensions being compactified. By this way, for a sufficiently small scale of the internal radii, the universe will appear four-dimensional at low-energy. In this section, we present the two solutions of heterotic supergravity with compact internal space leading to maximally symmetric $N=1$ four-dimensional vacua. As this is a well established subject, we will be brief and skip the calculation details to focus on the relevant aspects for our later purposes.

\subsection{Calabi-Yau compactification}

What we look for in solutions of the vacuum  geometry is some realistic setting that will lead to an attractive phenomenological four-dimensional theory. The conventional way to proceed is to consider backgrounds that break some of the $16$ supercharges of $\epsilon$ and conserve only $4$ in order to lead to $N=1$ supergravity in four dimensions. The easiest solution is to consider vanishing NS-NS flux $\hat H=0$ and a constant dilaton $d\hat\phi=0$. This implies no field strength in Einstein equations~\eqref{Einstein} and imposes the ten-dimensional space-time to be Ricci flat. For this reason, the metric ansatz is chosen to be the product of a four-dimensional Minkowski space $M_{1,3}$ with some six-dimensional internal compact manifold
\be\label{10dCY}
  M_{10}=M_{1,3}\times M_{CY} \;,
\ee
where $M_{CY}$ is Ricci flat. The requirement to preserve four supercharges implies for the spinor $\epsilon$ to decompose under the ansatz~\eqref{10dCY} in such a way that only a singlet survives on the six-dimensional geometry. The Killing spinor equation~\eqref{dpsi} implies for this singlet to be covariantly constant and, as a consequence, the Levi-Civita connection $\nabla$ takes value into its stability subgroup. Hence, $M_{CY}$ must be a manifold of $SU(3)$ holonomy. Alternatively, we can build tensors $J$ and $\Omega$ from the supersymmetry spinor which means for the internal space to have a reduced structure group. (A short review of structure groups and torsion classes is given in appendix~\ref{appendixtorsion}.) Equation~\eqref{dpsi} with $\hat H=0$ is then equivalent to
\be
  dJ=0 \;, \quad d\Omega=0 \;.
\ee
This results in vanishing torsion classes and means for the internal six-dimensional space to be complex, Kahler and have $SU(3)$ holonomy. Such solutions are the famous Calabi-Yau manifolds~\cite{Candelas:1985en}.

Furthermore, including the gauge sector, the full solution is a vector bundle whose base space is the above Calabi-Yau manifold. For the gauge fields to preserve the four supercharges of the $N=1$ effective theory, the vanishing gauginos variation~\eqref{dchi} implies,
\begin{equation}
  \Omega\,\neg\,\hat F=0 \;, \quad J\,\neg\,\hat F=0 \;,
\end{equation}
where $\hat F$ is the curvature of the gauge bundle over $M_{CY}$ and $\neg$ if the contraction over two indices. The first equation implies that $A$ is a connection on some holomorphic vector bundle, while the second is equivalent to saying that this vector bundle is slope-stable with slope zero as a result from the Donaldson-Uhlenbeck-Yau theorem~\cite{Green:1987mn}. These statements provide practical ways of finding solutions. Moreover, the Bianchi identity~\eqref{bianchi} must also be satisfied by solutions of the above system. Classification of such vacua is an active field of research and an increasing database of solutions leading to attractive phenomenological properties is under construction~\cite{Anderson:2011ns}.

\subsection{Strominger system}

One might argue that the ansatz of vanishing $\hat H$-flux and constant dilaton $\hat\phi$ is too strong and one wants to look for more general solutions leading to $N=1$ four-dimensional supergravity. This has been carried on in the celebrated paper by Strominger~\cite{Strominger:1986uh}. It has been found that, in this case, the space-time must be a warped product
\be\label{10dStro}
  M_{10}=M_{1,3}\times_W M_{\cal S} \;.
\ee
The non-vanishing flux and the dilaton will now act as a field strength tensor for the metric in Einstein equations~\eqref{Einstein} and the internal manifold does not remain Ricci flat. Nonetheless, preserving four supercharges as for the Calabi-Yau case implies again the existence of tensors $J$ and $\Omega$ on the internal six-dimensional space $M_{\cal S}$. This time, it is the full connection $\nabla^{\left(H\right)}$ that takes value in the stability subgroup of the supersymmetry spinor and, thus, the manifold has $SU(3)$ structure. The Killing spinor equations can be re-written as,
\be\label{strominger}
\begin{aligned}
  &d\Omega=2d\hat\phi\wedge\Omega \;,& &dJ=2d\hat\phi\wedge J -*\hat H \;,\\
  &0=\hat H\wedge\Omega \;,& &*2d\hat\phi=-\hat H\wedge J \;,
\end{aligned}
\ee
and, together with the equations of the gauge sector, correspond to the Strominger system~\cite{Held:2010az}. The internal manifold can be characterized by the following conditions on its torsion classes,
\begin{equation}\label{stroclass}
  \cW_1=\cW_2=0 \;, \quad 2\cW_4=\cW_5=2d\hat\phi \;,
\end{equation}
where $\cW_3$ remains arbitrary. This means that the space $M_{\cal S}$ is still complex whereas not Kahler any more. It is also conformally balanced from the condition on $\cW_4$ and $\cW_5$. The equations for the gauge curvature remain the same,
\begin{equation}\label{HYM}
  \Omega\,\neg\,\hat F=0 \;, \quad J\,\neg\,\hat F=0 \;,
\end{equation}
and must also be supplemented by the anomaly cancellation condition~\eqref{bianchi}. The absence of the Kahler property makes the search for solutions of such system difficult due to the lack of techniques analogous to the powerful theorems of the Calabi-Yau context. While there exists a vast database of Calabi-Yau manifolds, building solutions of the Strominger system still remains a laborious challenge.

\section{Dimensional reduction and 4d supergravity}\label{sectionCY}

Having established the ten-dimensional vacuum geometry, we can now look at the corresponding effective four-dimensional theory. The Kaluza-Klein mechanism provides a way to reduce the ten dimensions required for consistency of string theory down to the four physical ones. The main idea is to expand the ten-dimensional fields into a set of eigenfunctions of the internal operator and then integrate~\eqref{action} over the compact internal manifold to obtain a four-dimensional action. We can thus integrate out the heavy modes (the one having non-zero eigenvalues under the internal operator) to obtain a finite truncated set of fields. In other terms, the proposition of the Kaluza Klein dimensional reduction procedure is that some excitations are not visible up to some effective energy scale set by the size of the internal space.

\subsection{$N=1$ supergravity in four dimensions}

As previously explained, the choice of background solutions has been made such that part of the ten-dimensional supersymmetry is broken by the geometry. Its purpose is to break it down to an $N=1$ supergravity in four dimensions. For this reason, it is natural to start by looking at the generic features of four-dimensional $N=1$ supergravity theories~\cite{Wess:1992cp}. In the following, we concentrate on the bosonic part of chiral multiplets which are the relevant fields for the gravity sector of heterotic string. In this thesis, we will not discuss the vector multiplets of the gauge sector from the four-dimensional perspective.

The bosonic part of a general supergravity action can always be divided up into kinetic terms (with derivatives) and a potential term. It takes the following form,
\be\label{action4dsugra}
  S=-\frac{1}{\kappa_4}\int \frac{1}{2}R+K_{IJ^*}g^{\mu\nu}\partial_\mu A^I\partial_\nu \bar A^{J^*}+V \;,
\ee
where $\kappa_4$ is the four-dimensional Planck constant, $R$ is the Ricci scalar of the four-dimensional metric, $A^I$ are the chiral superfields of the theory and $V$ is the scalar potential. Now, supersymmetry implies a very stringent structure for the respective terms. For instance, the kinetic terms imply contraction with a metric $K_{IJ^*}$ which must be Kahler, that is, it can be written in terms of a Kahler potential $K$ as follows,
\be\label{KIJ}
  K_{IJ^*}=\frac{\partial}{\partial A^I}\frac{\partial}{\partial \bar A^{J^*}}K\left(A^I,\bar A^{J^*}\right) \;.
\ee
The derivatives are taken with respect to the chiral superfields $A^I$ and~\eqref{KIJ} means that the space of fields itself (where the fields are being seen as coordinates) is a Kahler manifold. Furthermore, the scalar potential can be written in terms of a superpotential $W$ being a holomorphic function of the superfields $A^I$,
\be
  V=e^{K}\left(K^{IJ^*}D_IWD_{J^*}W^*-3|W|^2\right) \;,
\ee
where $K^{IJ^*}$ is the inverse of the Kahler metric and the Kahler covariant derivative is given by
\be
  D_IW=\partial_IW+K_IW \;.
\ee
Here $\partial_I\equiv\partial/\partial A^I$ and, in general, a capital subscript means a derivative with respect to the corresponding field $K_I\equiv\partial K/\partial A^I$, a notation we will adopt from hereon. Any four-dimensional $N=1$ supergravity is uniquely determined by the data of the Kahler potential $K$ and the superpotential $W$. The lagrangian of the theory is then given by the expression~\eqref{action4dsugra}.

Now, consider the above four-dimensional $N=1$ supergravity theory with chiral superfields $(A^I,\chi^I)$, Kahler potential $K$ and superpotential $W$. We also write the corresponding gravitino $\psi_\mu$. Then, the full supersymmetric action corresponding to~\eqref{action4dsugra} is invariant with respect to the following supersymmetry transformations,
\bs\label{4dtrans}
\begin{align}
  \delta\chi^I&=i \sqrt{2} \sigma^\mu \bar{\zeta} \partial_\mu A^I-\sqrt{2}e^{K/2}K^{IJ^*}D_{J^*}W^*\zeta \;, \\\label{KSE4d2}
  \delta\psi_\mu&=2 \mathcal{D}_\mu\zeta+ie^{K/2}W\sigma_\mu\bar{\zeta} \;,
\end{align}
\es
where the Weyl spinor $\zeta$ parameterizes supersymmetry and $(\sigma^{\mu})=(\mathds{1}_2,\sigma^{\alpha})$ with the Pauli matrices $\sigma^{\alpha}$. Moreover, the covariant derivative $\mathcal{D}_\mu$ is defined by
\begin{equation}
  \mathcal{D}_\mu=\partial_\mu+\omega_\mu+\frac{1}{4}\left(K_I\partial_\mu A^I-K_{I^*}\partial_\mu \bar A^{I^*}\right) \;,
\end{equation}
with $\omega_\mu$ the spin connection. These transformations must vanish for a supersymmetric vacuum. For a maximally symmetric solution, this implies,
\be\label{maxsusy}
  D_IW=0 \;, \quad W=0 \;.
\ee
Later on, we will come back to the Killing spinor equations~\eqref{4dtrans} and generalize the conditions~\eqref{maxsusy} to half-BPS states that correspond to domain wall solutions which are non-maximally symmetric and preserve only two supercharges of the spinor $\zeta$.

\subsection{Calabi-Yau dimensional reduction}\label{CYreduction}

Let us now specify the effective theory resulting from compactifications of heterotic string theory on Calabi-Yau manifolds~\cite{Witten:1985xb}. It comes from considering the full ten-dimensional action and integrating out the internal six-dimensional compact space. The spectrum is truncated keeping only the massless modes. For this purpose, we introduce the four-dimensional dilaton $\phi$ as a new variable and rescale the four-dimensional part of the metric with it
\be\label{rescaling}
  g_{\mu\nu}^{(4)}=e^{-2\phi}g_{\mu\nu}\;,\quad \phi=\hat{\phi}-\frac{1}{2}{\rm ln}\,\mathcal{V} \;,
\ee
where $\mathcal{V}$ is the volume of the internal manifold $\mathcal{V}=\int d^6x \sqrt{g_6}$ with $g_6$ the internal six-dimensional metric. This rescaling is taken in order to obtain a properly normalized Ricci scalar in four dimensions and, thus, have an effective action in the Einstein frame.

It has been shown that the moduli space of a Calabi-Yau manifold $X$ is in one-to-one correspondence with harmonic forms on itself~\cite{Candelas:1990pi}. The metric deformations $\delta g$ can be split according to the nature of their indices as the mixed components $\delta g_{a\bar b}$ decouple from the pure components $\delta g_{\bar a\bar b}$. Here, and in the following, we use a bar to differentiate anti-holomorphic indices from holomorphic ones. By definition of the Kahler moduli $v^i$ (corresponding to the deformations of the Kahler form $J$) and the complex structure moduli $Z^a$ (corresponding to the deformations of the holomorphic three-form $\Omega$), they are given by
\begin{align}
  \delta g_{a\bar b}&=-i\omega_{ia\bar b}\delta v^i \;,\\
  \delta g_{\bar a\bar b}&=-\frac{1}{||\Omega||^2}\bar\Omega_{\bar a}^{\ph{\bar a}cd}\left(\chi_e\right)_{cd\bar b}\delta Z^e \;,
\end{align}
where $\omega_i$ are $h^{\left(1,1\right)}$ basis forms of the second cohomology group $H^2\left(X\right)$ and $\chi_a$ are a basis of harmonic $\left(2,1\right)$-forms on $X$. (More details about the moduli space of Calabi-Yau manifolds is given in appendix~\ref{appendixCY}.) Therefore, the real scalar fields $v^i$ and the complex scalar fields $Z^a$ will be constituents of the massless modes surviving the Kaluza-Klein reduction. Plugging the dilaton definition~\eqref{rescaling} into the action~\eqref{action}, we find after integrating out the internal manifold and keeping only massless modes,
\be
\begin{aligned}
  S_4=-\frac{1}{2\kappa_4^2}\int\bigg(&R*1+2d\phi\wedge*d\phi+\frac{1}{2}e^{4\phi}da\wedge*da \\
  &+2K_{ij}dT^i\wedge*d\bar T^{j}+2K_{ab}dZ^a\wedge*d\bar Z^{b}\bigg) \;.
\end{aligned}
\ee
We also defined the axion $a$ to be the Hodge dual of the four-dimensional part of the $\hat H$ fields and introduced the complex superfields,
\begin{equation}
  T^i=b^i+iv^i \;, \quad Z^a=c^a+iw^a \;,
\end{equation}
where $b^i$ are the $\hat B$ field moduli and the complexified Kahler moduli come from the combination $\hat B+iJ$. The complex structure moduli are simply given in terms of their real $c^a$ and imaginary $w^a$ parts by definition. The associated Kahler metrics are,
\be
  K^{(1)}_ {ij}=\frac{1}{4\mathcal{V}}\int\omega_i\wedge*\omega_j \;,\quad K^{(2)}_{a b}=-\frac{\int\chi_a\wedge\bar\chi_b}{\int\Omega\wedge\bar\Omega} \;,
\ee
and they come from the Kahler potentials,
\be
  K^{(1)}=-\ln\left(\frac{4}{3}\int J\wedge J\wedge J\right) \;,\quad K^{(2)}=-\ln\left(i\int\Omega\wedge\bar{\Omega}\right) \;.
\ee
We can see that these potentials are both given by the logarithm of the volume of the internal Calabi-Yau manifold. In order to unveil the full supersymmetric aspects of the above action, we need to combine the dilaton $\phi$ and the axion $a$ in a proper way to get a complex chiral superfield. The correct choice with the corresponding Kahler potential is given by
\be
  S=a+ie^{-2\phi} \;, \quad K^{(S)}=-\ln\left(i\left(\bar S-S\right)\right)=-\ln\left(2e^{-2\phi}\right) \;.
\ee
Thus, we obtain a four-dimensional supergravity theory with the chiral superfields $\left(S,T^i,Z^a\right)$ where the Kahler potential $K$ and the superpotential $W$ are given by,
\begin{equation}
  K=K^{(S)}+K^{(1)}+K^{(2)} \;, \quad W=0 \;.
\end{equation}
This implies that all the fields are massless (as expected). This feature has drawbacks and is known as the moduli stabilization problem. Such flat directions of the superpotential are not desired as, first, the corresponding massless particles are not observed and, then, nothing fixes the scale of the internal manifold. This motivates looking at compactification with fluxes which we discuss in the next section.

\section{Dimensional reduction with fluxes}\label{sectionflux}

\subsection{Field truncation and superpotential}

The inclusion of non-vanishing flux implies a non-trivial stress-energy tensor in Einstein equations~\eqref{Einstein} and we can anticipate mass terms in the four-dimensional theory. Now the shortage of explicit solutions of the Strominger system is a major limitation to study string phenomenology. One way to circumnavigate this problem has been to consider Calabi-Yau manifolds with non-vanishing $\hat H$-flux nonetheless. The argument is that the flux only back-reacts on the metric as a small disturbance away from the Calabi-Yau properties. For flux parameters small enough compared to the volume of the internal space, we can indeed argue in favor of a scale separation between the string and the flux scale. However, while this works well in type II theories, the nature of the heterotic flux superpotential does not allow itself to be tuned to small values. (Nevertheless, we will see in this thesis that such solutions exist provided that more supersymmetry is broken.) In this section, we would like to discuss generic features of non-Calabi-Yau compactifications leading to $N=1$ supergravity in four dimensions.

The non-vanishing superpotential leads to a legitimate question as to what field truncation is appropriate. For the Calabi-Yau case of section~\ref{CYreduction}, we retain only massless fields and there is a clear cut off scale determined by the masses of the remaining fields. For the case of $G$-structure manifolds, we can expect, from the superpotential, a truncation where some of the massive modes need to be included in order to have a supersymmetric theory. It can be shown~\cite{Falcitelli:1994fu} that for an $SU(3)$ structure manifold, the Ricci scalar decomposes according to
\begin{equation}
  R=R_{\rm CY}+R_{\perp} \;,
\end{equation}
where $R_{\rm CY}$ has the property of a Calabi-Yau Ricci scalar (no intrinsic torsion) and $R_{\perp}$ is a perpendicular component. We want to introduce a hierarchy of scale between the Kaluza-Klein scale and the light modes (including massive ones) that stay in the effective supergravity. For this, we can assume the $R_{\perp}$ component of the metric to be a small perturbation and, thus, the intrinsic torsion parameters to be small with respect to a large volume.

In practice, the truncation implies the identification of an adequate set of forms which may be not harmonic in general. Ref.~\cite{Grana:2005ny} has shown that such a set is given by the forms of table~\ref{truncationtable} and we will refer to such basis as the truncation basis.
\begin{table}[t]
\begin{center}
\begin{tabular}{|c|c|}
  \hline
  Vector space & Basis forms \\
  \hline\hline
  $\Lambda^1_{\rm eff}$ & $1$ \\
  $\Lambda^2_{\rm eff}$ & $\omega_i$ \\
  $\Lambda^3_{\rm eff}$ & $\alpha_A,\;\beta^A$ \\
  $\Lambda^4_{\rm eff}$ & $\tilde\omega^i$ \\
  $\Lambda^6_{\rm eff}$ & $*1$ \\
  \hline
\end{tabular}
\parbox{6in}{\caption{\it\small Finite-dimensional subspace $\Lambda^p_{\rm eff}$ of the space of differential $p$-forms $\Lambda^p$ with respect to their basis forms constituting the field truncation of metric deformations.}\label{truncationtable}}
\end{center}
\end{table}
In this table, each subsets $\Lambda^p_{\rm eff}$ of $p$-forms must be closed under the Hodge $*$ operator and must be subject to the following intersections,
\bs\label{intersections}
\begin{align}\label{omtilintersec}
  \int\omega_i\wedge\tilde\omega^j=\delta_i^{\ph{i}j} \;&,\quad \omega_i\wedge\alpha_A=\omega_i\wedge\beta^B=0 \;,\\
  \int\alpha_A\wedge\beta^B=\delta_A^{\ph{A}B} \;&, \quad \int\alpha_A\wedge\alpha_B=\int\beta^A\wedge\beta^B=0 \;.
\end{align}
\es
The $SU(3)$ structure forms can be expanded on the above basis as follows,
\be
  J=v^i\omega_i \;, \quad \Omega=\cZ^A\alpha_A+\cG_A\beta^A \;,
\ee
where $v^i$ are the analogue of the Kahler moduli, $\cZ^A$ are complex homogeneous coordinates and $\cG_A$ the pre-potential in complete analogy with the Calabi-Yau case. We should emphasize that the form $J$ is not necessarily Kahler any more but we still refer to the ``Kahler'' moduli by abuse of language. It is a remarkable fact that the geometry of the space of deformation corresponding to $v^i$ and $\cZ^A$ holds a similar structure to the one of Calabi-Yau manifolds. Indeed, Hitchin functionals provide a natural Kahler potential given by,
\begin{align}\label{Ktor1}
  K^{\left(1\right)}&=-\ln\left(i\left(\bar T^I\mathcal{F}_I-T^I\bar{\mathcal{F}}_I\right)\right) \;, \\\label{Ktor2}
  K^{\left(2\right)}&=-\ln\left(i\left(\bar \cZ^A\mathcal{G}_A-\cZ^A\bar{\mathcal{G}}_A\right)\right) \;,
\end{align}
for the ``Kahler'' and ``complex structure'' moduli respectively. Here, $I=\{0,i\}$ and  $T^i$ are again the complexified combinations of the form $J$ with the $\hat B$-field while $T^0$ is a scaling parameter that can be set to one. More details on the corresponding fields for the Calabi-Yau case are given in appendix~\ref{appendixCY}. The special nature of such potentials implies for the geometry to be special Kahler with pre-potential $\mathcal{F}_I$ and $\mathcal{G}_A$. Moreover, the dilaton mode is defined in the same way as for the Calabi-Yau case and corresponds to,
\be\label{Ktor3}
  S=a+ie^{-2\phi} \;, \quad K^{\left(S\right)}=-\ln\left(i\left(\bar S-S\right)\right) \;,
\ee
where the four-dimensional dilaton $\phi$ is again defined as in~\eqref{rescaling}. Let us draw attention to the fact that examples of such manifolds are scarce. However, the geometry of deformations is well defined on general grounds despite the absence of explicit metrics. On the other hand, the truncation is only expected for consistency of supersymmetric four-dimensional theories and suffers from the lack of explicit cases to study.

Having written down the field content and the nature of the Kahler potentials, let us determine what is the adequate superpotential. It can be computed by looking at the fermionic sector of the ten-dimensional action and reducing the gravitino mass term. This leads to an explicit formula of what the superpotential must be. Indeed, the generic corresponding term of a supergravity lagrangian in four dimensions is
\be\label{gravitinomass}
  S_{\varPsi}=-\frac{1}{2}\int e^{\kappa^2_4K/2}W\varPsi_\mu^\dagger\sigma^{\mu\nu}\varPsi_\nu+{\rm h.c.} \;,
\ee
for a gravitino $\varPsi$. In ten dimensions, the gravitino $\psi$ is given by a Majorana-Weyl spinor and can be split according to the metric decomposition ansatz~\eqref{10dStro},
\be
  \psi_\mu=e^{\phi/2}\left(\varPsi_\mu\otimes\eta_++\bar\varPsi_\mu\otimes\eta_-\right) \;,
\ee
where $\varPsi_\mu$ is the related four-dimensional part of the full gravitino and $\eta_\pm$ are the globally defined Weyl spinors defining the $SU(3)$ structure. Also, the Majorana-Weyl condition implies $\eta^*_+=\eta_-$. The rescaling factor $e^{\phi/2}$ is chosen to have canonical kinetic terms in four dimensions so that the mass term is correctly normalized. We can now perform a dimensional reduction of the ten-dimensional supergravity action~\cite{Gurrieri:2004dt, Benmachiche:2008ma}, again using the definitions~\eqref{rescaling}. The gravitino appears in a kinetic and an interaction term, we obtain
\begin{align}\nn
  S&=-\frac{1}{24}\int_{10}e^{-\hat\phi/2}\hat H_{MNP}\,\psi_L\Gamma^{LMNPQ}\psi_Q-\int_{10}\psi_M\Gamma^{MNP}\nabla_N\psi_P \\\label{gravitinomass4d}
  &=-\int_4e^{\kappa^2_4K/2}\varPsi_\mu^\dagger\sigma^{\mu\nu}\varPsi_\nu\left(\frac{1}{12}\int_6\hat H_{abc}\,\eta_-^\dagger\gamma^{abc}\eta_++\frac{1}{2}\int_6\eta_-^\dagger\gamma^a\nabla_a\eta_++{\rm h.c.}\right) \;,
\end{align}
where the Kahler potential $K=K^{\left(1\right)}+K^{\left(2\right)}+K^{\left(S\right)}$ is the one defined previously in Eqs.~\eqref{Ktor1}, ~\eqref{Ktor2} and~\eqref{Ktor3}. The subscripts of the integrals indicate on which space they are taken. The two terms in bracket can be further expressed in terms of the structure forms $\left(J,\Omega\right)$ according to their definitions (see Eqs.~\eqref{jomdefspin} in appendix~\ref{appendixtorsion} for details):
\bs
\begin{align}
  \frac{1}{12}\int_6\hat H_{abc}\,\eta_-^\dagger\gamma^{abc}\eta_+=&\frac{1}{2}\int_X\Omega\wedge \hat H \;,\\
  \int_6\eta_-^\dagger\gamma^a\nabla_a\eta_+=&i\int_X \Omega\wedge dJ \;.
\end{align}
\es
The first term is just the definition of $\Omega$ while the second term comes from the covariance condition of the spinor $\nabla_a\eta=\tau_{abc}\gamma^{bc}\eta$ together with keeping only the non-vanishing terms. A comparison of~\eqref{gravitinomass} with~\eqref{gravitinomass4d} then leads to the superpotential
\begin{equation}\label{GukovW}
  W=\int_{X}\Omega\wedge\left(\hat{H}+idJ\right) \;.
\end{equation}
This corresponds to the expected type from the Gukov-Vafa formula~\cite{Gukov:1999ya, LopesCardoso:2003af}. We clearly see that it vanishes for the Calabi-Yau case where $\hat H$ and $dJ$ are equal to zero.

\subsection{Mirror half-flat manifolds}\label{halflatcomp}

In this sub-section, we want to apply the formalism developed above to some specific examples, namely mirror half-flat manifolds, and work out the effective four-dimensional theory explicitly. Essentially, these manifolds arise as mirrors of type II Calabi-Yau compactifications with electric NS-NS flux~\cite{Gurrieri:2002wz}. Specifically, consider a mirror pair $\hat X$, $\tilde{X}$ of Calabi-Yau manifolds and compactification of type IIB string theory on $\tilde{X}$ with electric NS-NS flux $\tilde{H}=e_i\tilde{\beta}^i$, where $i=1,\ldots, h^{2,1}(\tilde{X})$, the $\tilde{\beta}^i$ are part of the standard symplectic three-form basis on $\tilde{X}$ and $e_i$ are integer flux parameters. Then, mirror symmetry suggests the existence of a manifold $X$, closely related to the mirror Calabi-Yau manifolds $\hat X$, so that compactification of IIA on $X$ (without flux) is mirror to the IIB compactification on $\tilde{X}$ with flux $\tilde{H}$. In other terms, we expect the corresponding effective Lagrangians to lead to the same theory $\cL^{\left(\rm{IIA}\right)}(X)\equiv\cL^{\left(\rm{IIB}\right)}(\tilde{X})$. It has been shown that manifolds of this type must be half-flat, that is they are $SU(3)$ structure manifolds with structure forms satisfying,
\be\label{halflat}
  d\Omega_-=0 \;, \quad J\wedge\ dJ=0 \;,
\ee
and we will refer to the corresponding manifold $X$ as half-flat mirror manifold.

Although the explicit mirror map is unknown, mirror symmetry allows one to conjecture a number of properties for half-flat mirror manifolds which, in turn, facilitate explicit dimensional reduction on such spaces. Usually, these properties can be formulated in terms of related properties of the associated Calabi-Yau manifold $\hat X$. In particular, $X$ carries a set of two-forms $\{\omega_i\}$, where $i,j,\ldots = 1,\ldots , h^{1,1}(\hat X)$ and a symplectic basis of three-forms $\{\alpha_A,\beta^A\}$, where $A,B,
\ldots = 0,\ldots , h^{2,1}(\hat X)$, so that the $SU(3)$ structure forms $(J,\Omega)$ can be expanded as
\begin{equation}\label{Jexpansion}
  J=v^i\omega_i \;, \quad \Omega={\cal Z}^A\alpha_A-\mathcal{G}_A\beta^A \;.
\end{equation}
Of course these forms correspond to the ones of table~\ref{truncationtable} and, by abuse of terminology, we will also refer to the $v^i$ and ${\cal Z}^A$ as Kahler and complex structure moduli, respectively. We also introduce the affine complex structure moduli $Z^a={\cal Z}^a/{\cal Z}^0$, where $a,b,\ldots =1,\ldots ,h^{2,1}(\hat X)$. Many of the standard Calabi-Yau moduli space results therefore still apply and the ones relevant in the present context are summarized in appendix~\ref{appendixCY}. For a non-Calabi-Yau manifold $J$ and $\Omega$ are no longer closed and, given the above expansion, the same must be true for at least some of the forms $\{\omega_i\}$ and $\{\alpha_A,\beta^A\}$. It turns out from mirror symmetry that the only non-closed forms are
\begin{equation}\label{hfmdef}
  d\omega_i=e_i\beta^0 \;, \quad d\alpha_0=e_i\tilde{\omega}^i \;,
\end{equation}
where the intrinsic torsion parameters $e_i$ correspond to the NS-NS flux parameters of the mirror symmetry set up discussed above. Moreover, $\{\tilde{\omega}^i\}$ corresponds to the set of four-forms dual to $\{\omega_i\}$ defined in the previous sub-section and satisfy the intersections~\eqref{omtilintersec}. It is straightforward to verify that
\begin{equation}\label{dJ}
  dJ=v^ie_i\beta^0 \;, \quad d\Omega={\cal Z}^0e_i\tilde\omega^i \;,
\end{equation}
and that $J$ and $\Omega$ indeed satisfy the defining half-flat conditions~\eqref{halflat}.

Heterotic compactifications on half-flat mirror manifolds have been studied in Refs.~\cite{Gurrieri:2004dt, Gurrieri:2007jg} and, here, we briefly review the main results. Even though some steps have already been explained previously, we repeat the whole procedure for the sake of clarity. We begin with the reduction ansatz and the relation between the ten- and four-dimensional fields. The six-dimensional internal space is taken to be the half-flat mirror space $X$ with metric $g_{ab}$ associated to the $SU(3)$ structure $(J,\Omega)$. In terms of the internal volume ${\mathcal{V}=\int d^6x\sqrt{g}}$, the four-dimensional dilaton $\phi$ is given by
\begin{equation}\label{dilaton}
  \phi=\hat{\phi}-\frac{1}{2}{\rm ln}\mathcal{V} \;,
\end{equation}
where $\hat{\phi}=\hat{\phi}(x^\mu)$ is the zero mode of the $10$-dimensional dilaton. The ansatz for the $10$-dimensional metric then reads
\begin{equation}\label{ansatzHF}
  ds_{10}^2=e^{2\phi}g^{(4)}_{\mu\nu}\,dx^\mu dx^\nu+g_{ab}\,dx^adx^b \;,
\end{equation}
where the dilaton factor in front of the four-dimensional part has been included so that $g^{\left(4\right)}_{\mu\nu}$ is the four-dimensional Einstein-frame metric. Moreover, we introduce the NS-NS flux expansion on the truncation basis
\begin{equation}\label{Bzero}
  \hat{B}=B+b^i\omega_i \;, \quad \hat{H}=H+db^i\wedge\omega_i+b^id\omega_i \;,
\end{equation}
where $b^i$ are axionic scalars and $B$ is a four-dimensional two-form with field strength $H=dB$ which can be dualized to the universal axion $a$. Note that, even thought we are considering the case without ``explicit" flux, a non-zero flux is induced from the last term in Eq.~\eqref{Bzero} as a consequence of the differential relations~\eqref{hfmdef} for half-flat mirror manifolds. These various scalar fields form the lowest components of four-dimensional chiral supermultiplets in the usual way:
\begin{equation}\label{superfields}
  S=a+ie^{-2\phi} \;, \quad T^i=b^i+iv^i \;, \quad Z^a=c^a+iw^a \;.
\end{equation} 
Their Kahler potentials are given by the same expressions as for Calabi-Yau compactifications coming from the logarithm of the internal volume,
\bs\label{hfk}
\begin{align}
  K^{(1)}&=-\ln\left(\frac{4}{3}\int J\wedge J\wedge J\right) \;,\\ K^{(2)}&=-\ln\left(i\int\Omega\wedge\bar{\Omega}\right) \;,\\
  K^{(S)}&=-\ln\left(i\left(\bar S-S\right)\right) \;.
\end{align}
\es
In addition, the superpotential is obtained from the Gukov-Vafa type formula~\eqref{GukovW},
\begin{equation}
  W=\int_{X}\Omega\wedge\left(\hat{H}+idJ\right) \;.
\end{equation} 
Even though we are not considering explicit flux, the $\hat{H}$ term has to be included in this formula to correctly incorporate the flux induced by the structure of the half-flat mirror manifolds (see Eq.~\eqref{Bzero}). For half-flat mirror manifolds and vanishing flux, this superpotential takes the form,
\begin{equation}\label{hfw}
  W=\,e_iT^i \;,
\end{equation}
where the relations~\eqref{dJ} and \eqref{Bzero} have been used.

We can easily see that the related four-dimensional vacuum cannot be maximally symmetric as the $F$-term equations~\eqref{maxsusy} has no solutions. The above dimensional reduction has been performed despite the absence of a full ten-dimensional solution of the heterotic equations of motion. In the next chapter, we will remedy this boldness by showing that background solutions can be found provided that we look at half-BPS states of the supergravity. The corresponding geometry is given by a warp product of a domain wall and an internal half-flat manifold. Furthermore, we will show that domain wall solutions of the four-dimensional supergravity given above can be lift up to this ten-dimensional solution and, thus, put heterotic half-flat compactification on solid footings.
\clearemptydoublepage
\chapter{Domain wall background geometries}\label{chapterDW}

In the previous chapter, we reviewed the standard background geometries leading to $N=1$ supergravity compactification. However, we also realized the difficulties of flux compactifications due to the lack of known explicit manifolds that fulfill the required constraints. In addition, with mirror half-flat manifolds, we also encountered vacuum solutions that lead to perfectly well defined four-dimensional supergravities without solving the $10$-dimensional equations of motion. In this chapter, we would like to understand the origin of these mirror half-flat manifolds and show that they satisfy the heterotic equations of motion as long as they are properly fibered over the external four-dimensional space-time.

To achieve this goal, we need to bypass the constraints of $N=1$ supercharges preserved to exhibit more background solutions than the usual Calabi-Yau and Strominger system. Indeed, the metric decomposition ansatz~\eqref{10dCY} and~\eqref{10dStro} are the most general ones with four preserved supercharges and maximally symmetric four-dimensional space. We must therefore relax this condition to find new classes of solutions. The next to simplest kind of backgrounds are arguably half-BPS supersymmetric domain walls. Such geometries preserve only two supercharges and are BPS states of an $N=1$ supergravity. We will see that mirror half-flat manifolds fall into this category as well as Calabi-Yau manifolds with non trivial NS-NS flux. We will also derive the general conditions for such a class of solutions.\clearpage

Why are such solutions which violate four-dimensional Lorentz symmetry interesting? The simple answer is motivated by the existence of compactifications leading to $N=1$ supergravity where the lowest order in $\alpha'$ flux superpotential in four dimensions leads to a runaway potential for some of the moduli and, therefore, the simplest solution consistent with this feature is a four-dimensional domain wall. This happens for the two previously discussed cases of the crude heterotic Calabi-Yau compactification with fluxes and mirror half-flat manifolds. In general, world-sheet effects need to be considered in order to stabilize all moduli. This is typically the case for the dilaton mode. Therefore, one should phenomenologically requires a four-dimensional maximally symmetric space only after all relevant effects, including non-perturbative ones, have been taken into account. When studying $10$-dimensional perturbative string solutions, we could allow for more general four-dimensional spaces, keeping in mind the possibility of a non-perturbative ``lift" to a maximally symmetric four-dimensional space. This allows us to study more general backgrounds than usually considered as potentially interesting solutions.

Having this in mind, we look for backgrounds that only preserve two supercharges out of the original sixteen from the ten-dimensional Majorana-Weyl spinor. A classification of heterotic vacua according to the number of unbroken supercharges has been achieved in Ref.~\cite{Gran:2005wf}. The equations of motion impose that the spinor parameterizing the supersymmetry transformations is covariantly constant with respect to the connection~\eqref{Bismut}. The aforementioned classification is then based on the stabilizer of such covariantly constant spinor. Two options arise for the case of two unbroken supercharges: one with stability subgroup $SU(4)\ltimes R^8$ and the other one with $G_2$. The former has not been found appropriate to a domain wall decomposition of space-time and we will therefore concentrate on the latter. We must thus consider heterotic compactification on seven-dimensional spin manifolds having $G_2$ structure group. However, we eventually want to make contact with phenomenology and, consequently, impose that only six dimensions are compact.

In the next section, we will explain the metric ansatz corresponding to this idea. Thereafter, we will review four-dimensional domain wall solutions of $N=1$ supergravity on general grounds. This is the basis used to demonstrate the consistency of mirror half flat manifolds as solutions of heterotic string theory. This will be shown in section~\ref{hfcomp} by verifying that the four-dimensional domain walls lift up to the correct associated ten-dimensional solutions. The last two sections apply the same principles to more general solutions with fluxes, first with mirror half-flat manifolds and, then, with Calabi-Yau manifolds.

\section{Ten-dimensional half-BPS background and domain wall ansatz}\label{sectionansatz}

We start with a ten-dimensional metric ansatz being a warped product of a seven dimensional manifold $Y$ with a $2+1$-dimensional Minkowski space. This is motivated by the fact that we want a domain wall with a maximally symmetric world volume. Furthermore, we want only six compact dimensions in order not to depart too much from phenomenology. This imposes for the seven-dimensional space to decompose as $Y=I\times X$, where $X$ is the six-dimensional compact internal manifold and $I$ is some interval parametrized by $y$. This leads to the general ansatz
\begin{equation}\label{ansatz10dgeneral}
  ds_{10}^2=e^{2\tilde\Delta\left(x^m\right)}\left(\eta_{\alpha\beta}\,dx^\alpha dx^\beta+e^{2\Delta\left(x^m\right)}dx^3dx^3+g_{ab}\left(x^m\right)dx^adx^b\right) \;,
\end{equation}
where $\Delta$ and $\tilde\Delta$ are two warp factors, $x^\alpha$ are the domain wall world volume coordinates with the Minkowski metric $\eta_{\alpha\beta}$, the transverse direction is $x^3$ and the internal space coordinates $x^a$. The dependence on $x^m$ comes from the consideration of $Y$ as a seven-dimensional manifold and $m$ runs over all the seven-dimensional indices. Our index conventions are summarized in appendix~\ref{appendixconventions}. There is thus two pictures to have in mind. The first one is a metric that decomposes as $10=3+7$ being a warped product of a Minkowski space in three dimensions and an internal seven-dimensional manifold $Y$. The second one is to consider a warped product of a four-dimensional domain wall with an internal six-dimensional space $X$ corresponding to the decomposition $10=4+6$. Both viewpoints will be useful. 

In addition, we want to preserve $2+1$-dimensional Lorentz invariance of the domain wall world volume and demand that the flux parameters are subject to the conditions
\begin{equation}\label{fluxansatz}
  \hat{H}_{\alpha MN}=0 \;, \quad \partial_\alpha\hat{\phi}=0 \;.
\end{equation}
This requirement can still allow a space-filling three-form $\hat H_{\alpha\beta\gamma}$ on the domain wall world volume. However we discard this option to make contact with the flux compactifications we want to consider. For the same reason we will set $\hat H_{3MN}=0$.

We should also provide the ansatz for the spinor $\epsilon$ which parameterizes the $10$-dimensional supersymmetry transformations. Since we are interested in solutions with two preserved supercharges, we should assume the existence of a globally defined seven-dimensional Majorana spinor $\eta$ on $Y$. In analogy with the decomposition of the metric~\eqref{ansatz10dgeneral}, we write
\begin{equation}\label{spindec}
  \epsilon\left(x^m\right)=\rho\otimes\eta\left(x^m\right)\otimes\theta \;,
\end{equation}
where $\theta$ is an eigenvector of the third Pauli matrix $\sigma^{3}$ appearing from dimensional considerations and $\rho$ is a (constant) Majorana spinor in $2+1$ dimensions whose components represent the two preserved supercharges of the supersymmetry. Details about spinor conventions can be found in appendix~\ref{appendixconventions}. In what follows it will be useful to write $\eta$ in terms of two chiral six-dimensional spinors $\eta_\pm$ as
\begin{equation}\label{spinoransatz}
  \eta\left(x^m\right)=\frac{1}{\sqrt{2}}\left(\eta_+\left(x^m\right)+\eta_-\left(x^m\right)\right) \;.
\end{equation}
This decomposition corresponds to the seven-dimensional space being $Y=I\times X$. It implies the existence of spinors $\eta_\pm$ on the internal compact six-dimensional manifold.\\

Before embarking on a detailed analysis of the above ansatz, we would like to draw a simple conclusion. From the gravitino Killing spinor equation, $\delta\psi_m=0$, together with Eqs.~\eqref{dpsi}, \eqref{Bismut} and \eqref{spindec} we have
\begin{equation}\label{KSE7d}
  \nabla_m^{\left(H\right)}\eta=0 \;.
\end{equation} 
Hence, the internal spinor $\eta$ is covariantly constant with respect to the connection with torsion $\nabla^{(H)}$. Further, for two metrics related by a conformal re-scaling $\hat{g}=e^{2\tilde\Delta}\tilde{g}$, we have the relation $\hat{\nabla}_M=\tilde{\nabla}+\frac{1}{2}{\Gamma_M}^N\partial_N\tilde\Delta$ between the two respective Levi-Civita connections. After a short calculation, the external part of the gravitino Killing spinor equation, together with Eq.~\eqref{ansatz10dgeneral} leads to
\begin{equation}
  \delta\psi_\alpha=\frac{1}{2}\,{\Gamma_\alpha}^m\partial_m\tilde\Delta\epsilon= 0 \;.
\end{equation}
Therefore, the warp factor $\tilde\Delta$ is constant. For convenience, we can set it to zero which simplifies our metric ansatz~\eqref{ansatz10dgeneral} as follows,
\begin{equation}\label{ansatz10d}
  ds_{10}^2=\eta_{\alpha\beta}\,dx^\alpha dx^\beta+e^{2\Delta\left(x^m\right)}dy^2+g_{ab}\left(x^m\right)dx^adx^b \;.
\end{equation}
This concludes our set-up. We will now analyze the resulting solutions using the formalism of $SU(3)$ (and $G_2$) structures, beginning with the simplest case of vanishing flux and constant dilaton and, subsequently, considering more general cases. But before doing so, let us review the general features of four-dimensional supergravity domain walls which will be necessary for our analysis.

\section{Four-dimensional domain walls}\label{section4dDW}

As a preparation, we would like to discuss the general context of our four-dimensional solutions, that is the formalism of half-BPS $N=1$ supergravity domain walls, mainly following~\cite{Eto:2003bn} (but see also~\cite{Wess:1992cp, Cvetic:1996vr}). Consider again a four-dimensional $N=1$ supergravity theory with chiral superfields $(A^I,\chi^I)$, Kahler potential $K$, superpotential $W$ and gravitino $\psi_\mu$. The Killing spinor equations are given by the supersymmetry transformations of the fermionic fields being set to zero. From Eqs.~\eqref{4dtrans}, we obtain,
\bs\label{KSE4d}
\begin{align}
  &i\sqrt{2} \sigma^\mu \bar{\zeta} \partial_\mu A^I-\sqrt{2}e^{K/2}K^{IJ^*}D_{J^*}W^*\zeta=0 \;,\\
  &2\mathcal{D}_\mu\zeta+ie^{K/2}W\sigma_\mu\bar{\zeta}=0 \;,
\end{align}
\es
for the gauginos and gravitino transformations respectively. Supersymmetric vacua are given by solutions of this system of equations. We can easily see that a maximally symmetric solution must satisfy $D_IW=0$ and $W=0$. However, these can be relaxed for a domain wall as the fields $A^I$ are allowed to have a dependence on the world volume transverse direction and, thus, allow for more general solutions of~\eqref{KSE4d}.

Consequently, we should split the coordinates as $(x^\mu)=(x^\alpha ,y)$ where $\alpha,\beta,\ldots = 0,1,2$ label the directions longitudinal to the domain wall and $y$ is the transverse coordinate\footnote{This choice of labeling implies that the Pauli matrix $\sigma^2$ is assigned to $x^3$ whereas $\sigma^2$ corresponds to $x^3$. This somewhat confusing designation is made to be consistent with the notations of Ref.~\cite{Eto:2003bn} and still be concordant with our ten-dimensional decomposition.}. We should start accordingly with a metric ansatz
\begin{equation}\label{4dmetric}
  ds_4^2=e^{-2B}\left(\eta_{\alpha\beta}\,dx^\alpha dx^\beta +dy^2\right) \;,
\end{equation}
where $B=B(y)$ is a warp factor. In addition, all scalar fields together with the spinor $\zeta$ are functions of $y$ only. The spin connection of this metric is
\begin{equation}
  \omega_0=-\frac{1}{2}B '\sigma^{2}\;,\quad \omega_1=-i\frac{1}{2}B ' \sigma^{3} \;, \quad \omega_2=i\frac{1}{2}B '\sigma^{1} \;, \quad\omega_3=0 \;.
\end{equation}
Here, and in the following, we use a prime to denote the derivative with respect to the transverse direction $y$ to avoid over-clustering the equations. We should also make an ansatz for the spinor $\zeta$ parameterizing the supersymmetry transformations. We choose it to satisfy the constraint
\begin{equation}\label{1/2BPS}
  \zeta\left(y\right)=\sigma^{2}\bar{\zeta}\left(y\right) \;,
\end{equation}
which reduces the number of independent spinor components to two, corresponding to half-BPS solutions.

We can now plug all the above ansatz into the general Killing spinor equations~\eqref{KSE4d}. After a small calculation, and some re-arrangement, we find that they specialize to
\bs \label{DW}
 \begin{align}\label{DW1}
  \partial_yA^I&=-ie^{-B}e^{K/2}K^{IJ^*}D_{J^*}W^* \;,\\\label{DW2}
  B'&=ie^{-B}e^{K/2}W \;,\\\label{DW3}
  {\rm Im}\left(K_I\partial_yA^I\right)&=0 \;,\\\label{DW4}
  2\zeta'&=-B'\zeta \;.
 \end{align}
\es
These impose the dependence of the bosonic fields $A^I$ and the warp factor $B$ on the direction transverse to the wall. The last equation~\eqref{DW4} will not be discussed as it gives the normalization of the spinor parameter $\zeta$ with respect to $y$ which we are not concerned about. Solutions of Eqs.~\eqref{DW} give us the vacuum geometry of the four-dimensional effective theories we want to consider.

\section{Vanishing flux and half-flat compactifications}\label{hfcomp}

After this preparatory set up, we would like to present solutions of the heterotic supergravity equations of motion that enter the class of half-BPS ansatz described by the metric~\eqref{ansatz10d}. We start by focusing on the specific case of vanishing flux and constant dilaton, that is
\begin{equation}
  \hat{H}=0 \;, \quad \hat{\phi}=\mbox{constant} \;.
\end{equation}
As a first step, we will look at the structure of the $10$-dimensional solution. We find that the six-dimensional internal space $X$ is restricted to be half-flat while the four-dimensional domain wall is described by Hitchin's flow equations. These results are then related to the four-dimensional $N=1$ supergravity obtained from compactifications on half-flat mirror manifolds. In particular, within these four-dimensional effective supergavity theories, we find an explicit half-BPS domain wall solution which precisely matches the domain wall present in the $10$-dimensional geometry. This shows that heterotic compactifications on half-flat mirror manifolds are indeed consistent in the sense of there being an associated solution of the full $10$-dimensional theory.

\subsection{The ten-dimensional solution}

In the absence of flux, the internal gravitino Killing spinor equation~\eqref{KSE7d} reads
\begin{equation}\label{g2kse}
  \nabla_m\eta=0 \;,
\end{equation}
where we recall that $\nabla$ is the ordinary Levi-Civita connection. Hence, $\eta$ is a covariantly constant spinor on the seven-dimensional space $Y$. This implies that the Levi-Civita connection of $Y$ has holonomy $G_2$ (or smaller) and that its metric must be Ricci-flat. Of course, it is immediately clear that, in the absence of stress energy, a product of a $2+1$-dimensional Minkowski space and a seven-dimensional manifold with $G_2$ holonomy solves the $10$-dimensional Einstein equations. 

We can also describe this situation in terms of $G_2$ structures on $Y$. (See appendix~\ref{appendixtorsion} for a brief review on $G$-structures and torsion classes.) We can think of such a $G_2$ structure as being defined by a three-form $\varphi$ and its (seven-dimensional) Hodge dual $\varPhi=*_{7}\,\varphi$ on $Y$. In terms of the spinor $\eta$, the components of these forms can be written as
\begin{equation}\label{3form}
  \varphi_{mnp}=-i \eta^\dagger\gamma_{mnp}\eta \;, \quad \varPhi_{mnpq}=\eta^\dagger\gamma_{mnpq}\eta \;.
\end{equation}
(Conventions about forms are summarized in appendix~\ref{appendixconventions}.) The space $Y$ has holonomy $G_2$ (or smaller) if and only if the $G_2$ structure is torsion-free, that is, if it satisfies
\begin{equation}\label{notorsion}
  d_7\,\varphi=d_7\,\varPhi=0 \;,
\end{equation}
where $d_7$ is the seven-dimensional exterior derivative. These two equations are indeed equivalent to the above Killing spinor equations~\eqref{g2kse}.

When imposing the constraint of only six compact internal dimensions motivated by phenomenology, these equations further decompose into the $6+1$ split of our metric ansatz~\eqref{ansatz10d}. For this purpose, we introduce a one-form in the direction of the special coordinate $y$,
\begin{equation}
  v=e^\Delta dy \;.
\end{equation}
Its exterior derivative satisfy,
\begin{equation}
  dv=\varUpsilon\wedge v \;, \quad \varUpsilon=d\Delta \;.
\end{equation}
In terms of the six-dimensions chiral spinors $\eta_\pm$ defined in~\eqref{spinoransatz}, we can introduce the following contractions with gamma matrices,
\begin{equation}
  J_{ab}=\mp i\eta_\pm^\dagger\gamma_{ab}\eta_\pm \;, \quad \Omega_{abc}=\eta_+^\dagger\gamma_{abc}\eta_- \;,
\end{equation} 
where the indices are only taken with respect to the internal coordinates. The corresponding forms define an $SU(3)$ structure on the six-dimensional space $X$ for every fixed value of $y$. The definition of the $G_2$ structure~\eqref{3form} and the spinor decomposition~\eqref{spinoransatz} then immediately lead to the well-known relations
\begin{equation}\label{phireduction}
  \varphi=v\wedge J+\Omega_- \;, \quad \varPhi=v\wedge\Omega_++\frac{1}{2}J\wedge J \;,
\end{equation}
where $\Omega_\pm$ are the real and imaginary parts of $\Omega=\Omega_++i\Omega_-$. These relations express the $G_2$ structure on $Y$ in terms of the $SU(3)$ structure on the six-dimensional space $X$ and the one-form $v$ in the $y$-direction. Overall, it defines an $SU(3)$ structure on the seven-dimensional space $Y$.

The vanishing torsion conditions~\eqref{notorsion} for the $G_2$ structure can be re-written according to this decomposition,
\bs
\begin{align}\label{HF1}
  d\Omega_-&=0 \;,\\\label{HF2}
  J\wedge\ dJ&=0 \;,\\\label{Hitchin1Theta}
  d\Omega_+&=e^{-\Delta} J\wedge\partial_{y} J-\varUpsilon\wedge \Omega_+ \;,\\\label{Hitchin2Theta}
  dJ&=e^{-\Delta}\partial_{y}\Omega_--\varUpsilon\wedge J \;.
\end{align}
\es
The first two of these equations imply that the $SU(3)$ structure on the six-dimensional space $X$ is restricted to be half-flat. In terms of torsion classes, a half-flat $SU(3)$ structure can also be characterized by the following conditions
\begin{equation}\label{Jtorsion}
  \cW_{1-}=\cW_{2-}=\cW_4=\cW_5=0 \;.
\end{equation}
This can be seen by comparison with the general expressions for $dJ$ and $d\Omega$ in terms of torsion classes given in Eqs.~\eqref{su3torsion}. Here, and in the following, we use subscripts $\pm$ to denote the real and imaginary parts of torsion classes in the same way as we do for the form $\Omega$. Note that, unlike for Strominger's class of solutions~\eqref{stroclass}, $\cW_1$ and $\cW_2$ are non-zero in general and, as a consequence, the manifold $X$ does not necessarily admit an integrable complex structure. A further comparison between Eqs.~\eqref{Hitchin2Theta} and \eqref{Jtorsion} reveals that
\begin{equation}\label{warpcon}
  \varUpsilon=0 \;.
\end{equation}
Hence, the warp factor $\Delta$ is constant and can therefore be conveniently set to zero. 

Thus, the background geometry solving the equations of motion~\eqref{Einstein} can be summarized by the following. First, the $10$-dimensional string-frame metric is
\begin{equation}\label{hfmetric}
  ds^2_{10}=\eta_{\alpha\beta}\,dx^\alpha dx^\beta+dy^2+g_{ab}\left(y,x^d\right)dx^adx^b \;.
\end{equation}
This comes from the metric ansatz~\eqref{ansatz10d} and the condition~\eqref{warpcon}. Then, $g_{ab}$ must be a metric associated to a half-flat $SU(3)$ structure given by forms $J$ and $\Omega$. Finally, from Eqs.~\eqref{Hitchin1Theta} and \eqref{Hitchin2Theta}, the $y$-dependence of this $SU(3)$ structure is described by Hitchin's flow equations~\cite{Chiossi:2002tw},
\begin{equation}\label{hfe}
  d\Omega_+=J\wedge\partial_yJ\; ,\quad dJ=\partial_y\Omega_- \;.
\end{equation}
We should note that these flow equations do not guarantee for the volume of the internal manifold to be bounded nor to remain large everywhere in $y$. However, we ignore such issues in our present analysis as we don't consider the domain wall to be the ``final'' solution. From a physics point of view, the metric~\eqref{hfmetric} should be interpreted as a product of a six-dimensional half-flat space $X$ with metric $g_{ab}$ and a four-dimensional domain wall with world-volume coordinates $x^\alpha$ and transverse direction $y$. This shows that half-flat spaces can indeed be considered as solutions of the heterotic string provided that they are ``paired up" with an external domain wall solution rather than a maximally symmetric four-dimensional space-time. 

The existence of these solutions justifies heterotic compactifications on half-flat manifolds, as carried out in Refs.~\cite{Gurrieri:2004dt, Gurrieri:2007jg, deCarlos:2005kh, Gurrieri:2005af, Ali:2006gd, Ali:2007ra}, and suggests the existence of half-BPS domain wall solutions in the associated four-dimensional $N=1$ supergravity which should match the domain wall part of the metric~\eqref{hfmetric}. We will now verify this picture explicitly. The effective four-dimensional $N=1$ theories which originate from such compactifications has been reviewed in chapter~\ref{chapterreview}. It is left to find explicit half-BPS domain wall solutions of these supergravity theories and show that they match the $10$-dimensional solutions just obtained.

\subsection{The four-dimensional solution}

Having established the vacuum geometry from a ten-dimensional point of view, we need to check that the corresponding four-dimensional theory has a compatible vacuum solution. Thus, we would like to solve the domain wall Killing spinor equations \eqref{DW} for the specific supergravity theories obtained from compactification on half-flat mirror manifolds. We recall that the (chiral) superfields of these theories consist of $(A^I)=(S,T^i,Z^a)$. They are split up according to their real and imaginary parts
\begin{equation}\label{STZreal}
  S=a+ie^{-2\phi} \;, \quad T^i=b^i+iv^i \;, \quad Z^a=c^a+iw^a \;.
\end{equation} 
In addition, the corresponding Kahler potentials and the superpotential are given by Eqs.~\eqref{hfk} and \eqref{hfw}, respectively. In the following, we will make use of the properties of the moduli space geometry summarized in appendix~\ref{appendixCY}. 

It is not difficult to see by inspection that the superpotential needs to be purely imaginary as a consequence of the second equation~\eqref{DW2}. This implies,
\be\label{biei}
  b^ie_i=0 \;.
\ee
It then follows that the right-hand side of~\eqref{DW1} is purely imaginary as well. This implies that the real parts of all superfields must be constant,
\begin{equation}\label{imsol}
  a\sim b^i\sim c^a\sim {\rm constant} \;.
\end{equation}
Eq.~\eqref{DW3} becomes thus trivial. By comparing the first equation~\eqref{DW1} for the dilaton $S$ with the second equation~\eqref{DW2}, we find that $B'=\phi'$. So, without loss of generality, we can set
\begin{equation}
  B=\phi \;.
\end{equation} 
Having established this, we can work out the flow equations for the remaining imaginary parts. First, we can re-write Eq.~\eqref{DW1} for the $S$, $T^i$ and $Z^a$ fields respectively,
\vspace{-5mm}
\bs\label{DWstep}
\begin{align}\label{DWstep1}
  &\phi'=-\frac{3}{2}\frac{1}{\sqrt{\cK\tilde\cK}}\,\im W \;,\\\label{DWstep2}
  &\cK_i\left(\frac{\cK'}{\cK}+2\phi'\right)-\cK'_i=-2\sqrt{\frac{\calk}{\tilde\calk}}e_i \;,\\\label{DWstep3}
  &\tilde\cK_a\left(\frac{\tilde\cK'}{\tilde\cK}+2\phi'\right)-\tilde\cK'_a=0 \;.
\end{align}
\es
Here, $\calk=\calk_{ijk}v^iv^jv^k$ is the Kahler moduli pre-potential where $\calk_{ijk}$ are the intersection numbers of $X$ and $\calk_{i}=\calk_{ijk}v^jv^k$. Analogously, for the complex structure module, the pre-potential is given by $\tilde{\calk}=\tilde{\calk}_{abc}w^aw^bw^c$ with the intersection numbers $\tilde{\calk}_{abc}$ of the mirror Calabi-Yau $\tilde{X}$ (see appendix~\ref{appendixCY} for details). The last two equations have been obtained by multiplying~\eqref{DW1} on the left by the adequate Kahler metric and writing everything explicitly in terms of the pre-potentials. Eqs.~\eqref{DWstep} can still be simplified further. We can contract~\eqref{DWstep2} with $v^i$ and compare it to~\eqref{DWstep1}. We can also contract~\eqref{DWstep3} with $w^a$ to find an expression for $\phi'$ in terms of $\tilde\cK$ and $\tilde\cK'$ which we then plug back into~\eqref{DWstep3}. This gives the following set of equations completely equivalent to the system~\eqref{DWstep},
\begin{equation} \label{dweq}
  \phi '=-\frac{1}{2}\frac{\calk '}{\calk} \;,\quad \calk_{ij}\partial_yv^j=\sqrt{\frac{\calk}{\tilde\calk}}e_i \;, \quad \partial_yw^a=-2\phi' w^a \;.
\end{equation} 

They can easily be integrated. For this purpose, let us define a new coordinate corresponding to a rescaling by the dilaton,
\be
  d\tilde{y}=e^{2\phi}dy \;.
\ee
We can thus find solutions of the set of equations~\eqref{dweq}, we obtain,
\begin{equation}\label{sol}
  \calk=\calk_0e^{-2\phi} \;, \quad \calk_i=2\sqrt{\frac{\calk_0}{\tilde{\calk}\left({\bf k}\right)}}e_i\tilde{y}+\calk_{0i} \;, \quad w^a=k^a e^{-2\phi} \;,
\end{equation} 
where $\calk_0$, $\calk_{0i}$ and $k^a$ are integration constants and $\tilde{\calk}({\bf k})$ denotes the complex structure pre-potential as a function of the $k^a$. To find the explicit solution in terms of the Kahler moduli $v^i$, one has to invert the relations
\begin{equation}
  \calk_i=\calk_{ijk}v^jv^k \;,
\end{equation}
which can only be done on a case by case basis. This concludes our analysis of the four-dimensional background geometry and we now turn on to show the correspondence between the four- and the ten-dimensional solutions.

\subsection{Comparison between ten and four dimensions}

We would like to show that the four-dimensional domain wall~\eqref{dweq} indeed matches our $10$-dimensional solution~\eqref{hfmetric} in a way similar to what happens in the context of type IIA~\cite{Mayer:2004sd, Smyth:2009fu}. We start by re-writing the four-dimensional domain wall Killing spinor equations~\eqref{dweq} in term of $10$-dimensional language by introducing the fields $\cZ^0$ and $\hat\phi$. To this end, we insert the following definitions,
\begin{equation}
  \calk=\left({\cal Z}^0\right)^2\tilde\calk \;, \quad \hat{\phi}'=\phi'+\frac{1\calk'}{2\calk} \;.
\end{equation}
From a four-dimensional point of view it can be understood as a field re-definition whereas, from a $10$-dimensional point of view, the first of these arises from the compatibility relation~\eqref{su3def} while the latter is simply the definition of the dilaton~\eqref{dilaton}.

It is straightforward to see that the four-dimensional domain wall equations~\eqref{dweq} are equivalent to the following,
\begin{equation}\label{DWmatch}
  \hat\phi'=0 \;, \quad \calk_{ij}\partial_yv^j={\cal Z}^0 e_i \;, \quad \left({\cal Z}^0\omega^a\right)'=0 \;.
\end{equation}
It is also useful to recall from Eqs.~\eqref{biei} and~\eqref{imsol} the constraints these equations imply on the real parts of the superfields, namely
\begin{equation}\label{Realpartmatch}
  a\sim b^i\sim c^a\sim {\rm constant} \;, \quad b^ie_i=0 \;.
\end{equation}
We can now turn to the $10$-dimensional solution~\eqref{hfmetric} and show that it corresponds to the above system. To do this, we insert the defining relations of mirror half-flat manifolds~\eqref{dJ} into Hitchin's flow equations~\eqref{hfe}. We can easily see that the first flow equation for $J$ gives,
\begin{equation}\label{viflow}
  \calk_{ij}\partial_yv^j={\cal Z}^0 e_i \;,
\end{equation}
which is equivalent to the corresponding domain wall equation in~\eqref{DWmatch}. From the second Hitchin flow equation, we obtain three equations,
\begin{equation}\label{matchflow}
  {\cal Z}^0\omega^a={\rm constant} \;, \quad c^a={\rm constant} \;, \quad v^ie_i=\frac{1}{6}\left({\cal Z}^0\tilde\calk\right)' \;,
\end{equation}
which correspond to the components of the basis three-forms $\alpha_a$, $\beta^a$ and $\beta^0$, respectively. The first two equations are identical equations in~\eqref{DWmatch} and~\eqref{Realpartmatch}. The third one does not provide any more information as it is a contracted version of~\eqref{viflow} together with the condition ${\cal Z}^0\omega^a={\rm constant}$ and the compatibility relation $\calk=\left({\cal Z}^0\right)^2\tilde\calk$. It simply tells us that the flow equations are compatible with the relation~\eqref{su3def} between the $SU(3)$ structure forms $J$ and $\Omega$. Finally, we need to realize that the last conditions in~\eqref{DWmatch} and~\eqref{Realpartmatch} ensure the vanishing of $\hat H$ and a constant dilaton $\hat \phi$ from a $10$-dimensional point of view.

The set of four-dimensional equations~\eqref{dweq} and~\eqref{Realpartmatch} are, thus, completely equivalent to the set of ten-dimensional equations~\eqref{viflow} and~\eqref{matchflow}. This means that the four-dimensional domain wall solution of the effective supergravity can be lifted up to the corresponding $10$-dimensional flow equations.\clearpage

\section{Non-vanishing flux and half-flat compactifications}\label{halflatHcomp}

We will now extend the discussion of the previous section by including non-vanishing NS-NS flux as well as a non-constant dilaton. First, we derive the corresponding generalization of Hitchin's flow equations from the $10$-dimensional perspective. Then, we discuss the relation to domain wall solutions in the four-dimensional effective supergravity compactified on mirror half-flat manifolds.

\subsection{The ten-dimensional solution}

As before, we begin by working out the constraints on the $G_2$ structure of the seven-dimensional space $Y$. The starting point is the seven-dimensional part of the gravitino Killing spinor equation and the dilatino Killing spinor equation. From \eqref{dpsi} and \eqref{dlambda} they read
\bs
\begin{align}
  \nabla_m\eta=-\frac{1}{8}\hat{\cal H}_m\eta \;, \\ 
  \cancel{\nabla}\hat\phi\,\eta=-\frac{1}{12}{\cal\hat H}\,\eta \;.
\end{align}
\es
We proceed in the usual way by multiplying the above equations and their hermitian conjugates with anti-symmetrised products of gamma matrices times $\eta$ or $\eta^\dagger$ in order to obtain equations for tensors as described at the end of appendix~\ref{appendixconventions}. With the definitions~\eqref{3form} of the $G_2$ structure forms $\varphi$ and $\varPhi$, this leads to the following set of equations
\bs
\begin{align}
  4\nabla_{[m}\hat\phi\,\varphi_{npq]}&=-3\hat H_{v[mn}{\varphi_{pq]}}^v+\frac{1}{12}\epsilon_{mnpqrst}\hat H^{rst} \\
  \nabla_{m}\varphi_{npq}&=\frac{3}{2}\hat H_{ms[n}{\varphi_{pq]}}^s \\
  \nabla_{[m}\hat\phi\,\varPhi_{npqr]}&=\hat H_{s[mn}{\varPhi_{pqr]}}^s \\
  \nabla_{m}\varPhi_{npqr}&=-2\hat H_{ms[n}{\varPhi_{pqr]}}^s \\
  \epsilon_{mnpqrst}\nabla^t\hat\phi&=-10\hat H_{[mnp}\varphi_{qrs]} \\
  \hat H_{[mnp}\varPhi_{qrst]}&=0 \;.
\end{align}
\es
We can further contract them with the basis of the co-tangent space to obtain differential forms~\cite{Friedrich:2001yp, Gauntlett:2002sc, Gran:2005wf}. A combination of the first two equations, the third with the fourth and the last two equations on their own, they can then be written in the following manner
\bs\label{g2gen}
\begin{align}
  d_7\varphi&=2 d_7\hat\phi\wedge\varphi-*_7\,\hat H \\
  d_7\varPhi&=2d_7\hat\phi\wedge\varPhi \\
  *_7\,d_7\hat\phi&=-\frac{1}{2}\hat H\wedge\varphi \\
  0&=\hat H \wedge\varPhi \;,
\end{align}
\es
where $*_7$ and $d_7$ are the seven-dimensional Hodge star and exterior derivative, respectively. The first two of these equations characterize the $G_2$ structure on $Y$ and are the generalization of the torsion-free conditions~\eqref{notorsion} which appeared in the case without flux. The last two equations are constraints for the flux and the dilaton.

From these results, we can deduce the structure of $G_2$ torsion classes ${\cal X}_1,\ldots ,{\cal X}_4$. By comparing with the general relations \eqref{g2torsion}, it follows that ${\cal X}_1={\cal X}_2=0$. The class ${\cal X}_3$ is determined by the corresponding component of the flux $\hat{H}$ and the class ${\cal X}_4$ by the derivative $d_7\hat{\phi}$ of the dilaton and the corresponding component of $\hat{H}$. In other terms, this means that the $G_2$ structure is integrable and conformally balanced~\cite{Friedrich:2001yp}.

We can now split up these equations into $6+1$ dimensions and express them in terms of the $SU(3)$ structure on $X$. Since we are motivated by compactifications to four dimensions and to simplify matters, we set all remaining components of the flux breaking four-dimensional Lorentz symmetry to zero, that is,
\begin{equation}\label{Hy0}
  \hat{H}_{3mn}=0 \;.
\end{equation}
We recall that the relation between $G_2$ and $SU(3)$ structures is given by Eq.~\eqref{phireduction}. Using these relations in \eqref{g2gen} and splitting up the resulting equations accordingly, we find the following constraints on the $SU(3)$ structure forms,
\bs\label{dwform}
\begin{align}\label{dO-H}
  d\Omega_-&=2d\hat\phi\wedge\Omega_- \\\label{dJ-H}
  dJ&=e^{-\Delta}\Omega_-'-2e^{-\Delta}{\hat\phi}'\,\Omega_-+2d\hat\phi\wedge J -J\wedge\varUpsilon-*\hat H \\\label{JdJ-H}
  J\wedge dJ&=J\wedge J\wedge d\hat\phi \\
  d\Omega_+&=e^{-\Delta}J\wedge J'-e^{-\Delta}{\hat\phi}' J\wedge J+2d\hat\phi\wedge\Omega_+ +\Omega_+\wedge\varUpsilon\label{dO} \\\label{HJ}
  *d\hat\phi&=-\frac{1}{2}\hat H\wedge J \\\label{beforelast}
  e^{-\Delta}{\hat\phi}'\,*1&=-\frac{1}{2}\hat H\wedge\Omega_- \\\label{last}
  0&=\hat H\wedge \Omega_+ \;,
\end{align}
\es
where the Hodge star and the exterior derivative now refer to six dimensions. We also recall that $\varUpsilon=d\Delta$ is the exterior derivative of the warp factor $\Delta$. A quick calculation allows to find its value. Taking the Hodge star dual of Eq.~\eqref{HJ}, we find
\be
  J\neg*\hat H=2d\phi \;,
\ee
where we used the relation~\eqref{starwedge}. Knowing that $2\cW_4=J\neg \,dJ$ by definition~\eqref{su3torsion}, we can conclude that the contribution from $*\hat H$ to the torsion class $\cW_4$ is given by $d\phi$ and, comparing~\eqref{dJ-H} and~\eqref{su3torsion}, that $\varUpsilon=0$. Therefore, matching up the first four of these equations with the general expressions for $dJ$ and $d\Omega$ in Eq.~\eqref{su3torsion}, we find that the torsion classes are constrained by
\begin{equation}\label{tc}
  \cW_{1-}=\cW_{2-}=0\;,\quad 2\cW_4=\cW_5=2d\hat{\phi} \;,
\end{equation} 
and arbitrary otherwise. We can compare this result with the constraints~\eqref{stroclass} characterizing Strominger's class of solutions. The only difference is that $\cW_{1+}$ and $\cW_{2+}$ can be non-zero and, as a consequence, the six-dimensional space $X$, while still having an almost complex structure, does no longer need to be complex. Furthermore, since $\cW_4$ and $\cW_5$ are non-vanishing and proportional to the dilaton, the $SU(3)$ structure is mildly more general than that for half-flat manifolds. We will refer to this structure as generalized half-flat.

Having established that $\varUpsilon=0$, we can set the warp factor $\Delta$ to zero without loss of generality and, as before, the $10$-dimensional metric for our solution becomes
\begin{equation}
  ds^2_{10}=\eta_{\alpha\beta}dx^\alpha dx^\beta+dy^2+g_{ab}\left(y,x^d\right)dx^adx^b \;.
\end{equation}
Here, for every value of the $y$ coordinate, $g_{ab}$ is the metric associated to the $SU(3)$ structure with torsion classes satisfying~\eqref{tc} and with $y$-dependence governed by
\bs\label{10dHflow}
\begin{align}\label{10dHflow1}
  d\Omega_+&=J\wedge J'-{\hat\phi}'J\wedge J+2d\hat\phi\wedge\Omega_+ \\\label{10dHflow2}
  dJ&=\Omega_-'-2{\hat\phi}'\Omega_-+2d\hat\phi\wedge J-*\hat H \;,
\end{align}
\es
where we recall that a prime means derivative with respect to $y$. These are the generalizations of Hitchin's flow equations~\eqref{hfe} in the presence of non-zero NS-NS flux and, again, we should note that these flow equations do not guarantee a well behaved domain wall everywhere in $y$. As a consistency check, we can assume that all fields are $y$-independent. In this case, we indeed recover the standard equations of the Strominger system for maximally symmetric four-dimensional backgrounds preserving four supercharges~\eqref{strominger}, as we should.

\subsection{The four-dimensional solution}

We would now like to discuss the above solutions from the viewpoint of the effective four-dimensional supergravity. In section~\ref{halflatcomp}, we have reviewed the structure of the four-dimensional theory resulting from compactification on half-flat mirror manifolds $X$ with vanishing $\hat H$ flux. Here, we want to perform the same dimensional reduction ansatz including non-trivial NS-NS flux and find the domain wall equations. This will enable us to show in the next subsection that the resulting vacuum geometry still lifts up to the ten-dimensional theory.

For this purpose, we will continue to assume that the internal manifold is described by the mirror half-flat properties~\eqref{hfmdef}. Hence, the superfields of the four-dimensional supergravity theory are still given by $\left(S,T^i,Z^a\right)$. The only difference comes from the NS-NS zero-mode expansion which now reads,
\begin{equation}\label{Hfluxexpansion}
  \hat H=H+db^i\wedge\omega_i+b^id\omega_i+\epsilon_a\beta^a+\mu^a\alpha_a \;,
\end{equation}
where we introduce the electric flux $\epsilon_a$ and the magnetic flux $\mu^a$. The Kahler potential remains the standard one as given in Eq.~\eqref{hfk}. However, the superpotential is now modified since it contains the additional contribution due to the flux. It can still be obtained from the heterotic Gukov-Vafa formula~\eqref{GukovW} which gives,
\begin{equation}
  W=e_iT^i+\epsilon_aZ^a+\mu^a\mathcal{G}_a\left(Z\right) \;,
\end{equation}
where $\mathcal{G}_a(Z)$ are the derivatives of the pre-potential (see appendix~\ref{appendixCY}). We have also set ${\cal Z}^0=1$ for simplicity since the four-dimensional supergravity is independent of this field.

To find the domain wall equations, we can follow the same general set-up as in the previous section. We can start with the domain wall Killing spinor equations~\eqref{DW} and, as before, look at the real and imaginary parts of the fields,
\begin{equation}
  S=a+ie^{-2\phi} \;, \quad T^i=b^i+iv^i \;, \quad Z^a=c^a+iw^a \;.
\end{equation}
Again, we can conclude by comparing the first equations~\eqref{DW1} for the dilaton $S$ with the second equation~\eqref{DW2} that the warp factor $B$ in the metric ansatz~\eqref{4dmetric} is determined by the dilaton
\be
  B=\phi \;.
\ee
Furthermore, the superpotential must be purely imaginary and the real parts of the superfields $S$ and $T^i$ must be constant as before. However, the real parts of the superfields $Z^a$ are not constant as $D_{a^*}W^*$ is not real (zero) anymore. Thus, we find for the real parts,
\be
  a\sim b^i\sim \mbox{constant} \;, \quad \partial_yc^a=-\sqrt{\frac{\tilde{\mathcal{K}}}{\mathcal{K}}}\mu^a \;,
\ee
where the condition for $\partial_yc^a$ follows straightforwardly from the real part of the flow equation~\eqref{DW1} for $Z^a$. It implies that Eq.~\eqref{DW3} is not trivial as opposed to the fluxless case. Combining the resulting equation with the condition of a purely imaginary superpotential, we obtain
\be
  b^ie_i+\epsilon_ac^a=\frac{1}{2}\tilde{\mathcal{K}}_{abc}c^ac^b\mu^c \;, \quad \tilde{\mathcal{K}}_a\mu^a=0 \;.
\ee
They represent non-trivial constraints for the flux parameters that must be satisfied in order for our scenario to take place. The remaining equations for the imaginary parts of the superfields are given by,
\bs\label{DWstepflux}
\begin{align}\label{DWstep1flux}
  &\phi'=-\frac{3}{2}\frac{1}{\sqrt{\cK\tilde\cK}}\,\im W \;,\\\label{DWstep2flux}
  &\cK_i\left(\frac{\cK'}{\cK}+2\phi'\right)-\cK'_i=-2\sqrt{\frac{\calk}{\tilde\calk}}e_i \;,\\\label{DWstep3flux}
  &\tilde\cK_a\left(\frac{\tilde\cK'}{\tilde\cK}+2\phi'\right)-\tilde\cK'_a=-2\sqrt{\frac{\tilde{\mathcal{K}}}{\mathcal{K}}}\left(\epsilon_a-\tilde{\mathcal{K}}_{abc}c^b\mu^c\right) \;.
\end{align}
\es
The equation for the dilaton can be re-written in a slightly more fashionable way. By contracting the two equations~\eqref{DWstep2flux} and~\eqref{DWstep3flux} with $v^i$ and $w^a$ respectively and using the result back into the dilatino equations, we obtain,
\be
  2\phi'=-\frac{1}{4}\left(\frac{\mathcal{K}'}{\mathcal{K}}+\frac{\tilde{\mathcal{K}}'}{\tilde{\mathcal{K}}}\right) \;.
\ee
It is easy to see that, for vanishing $\epsilon_a$ and $\mu^a$ fluxes, these equations reduce to the previous ones~\eqref{dweq}.

\subsection{Comparison between ten and four dimensions}

As before, we would like to show that this four-dimensional domain wall indeed matches our $10$-dimensional solution. For clarity, let us rewrite the relevant Killing spinor equations~\eqref{dwform} in terms of the mirror half-flat manifolds definition~\eqref{hfmdef}. First, we should note that the relations for the basis forms $\omega_i$ and $\alpha_0$ together with Eqs.~\eqref{dO-H}, \eqref{JdJ-H} and~\eqref{HJ} imply that
\begin{equation}
  db^i=d\hat\phi=0 \;.
\end{equation}
This comes from the fact that we assume our internal space to be genuine half-flat manifolds and not have the slightly generalized half-flat torsion classes~\eqref{tc}. This restriction is necessary for simplicity as the structure of effective four-dimensional fields for non vanishing $\cW_4$ and $\cW_5$ torsion classes is more complicated (we refer the reader to table~\ref{truncationtable}). Therefore, we are left with the Killing spinor equations
\bs\label{10dHeq}
\begin{align}
  \Omega_-'&=2\hat\phi'\Omega_-+dJ+*\hat H \;,\label{10dHeq3}\\
  J\wedge J'&=\hat\phi'J\wedge J+d\Omega_+ \;,\label{10dHeq4}\\
  2\hat\phi'*1&=\Omega_-\wedge\hat H \;,\label{10dHeq2}\\
  0&=\Omega_+\wedge\hat H \;.\label{10dHeq1}
\end{align}
\es
Let us also remind that the warp factor has been set to zero $\Delta=0$.

We can now expand these equations on the basis $\{\omega_i\}$ and $\{\alpha_A,\beta^A\}$ to obtain explicit equations for the moduli fields. For this, we insert the respective expansions~\eqref{Jexpansion} and~\eqref{Hfluxexpansion} of the Kahler form $J$, the complex structure $\Omega$ and the NS-NS field $\hat H$ into the above Killing spinor equations~\eqref{10dHeq}. The calculation is a bit tedious due to the Hodge $*$ operator (how to compute this term is explained in appendix~\ref{appendixCY}). First, we can easily deduce the two constraints,
\begin{equation}
  b^ie_i+\epsilon_ac^a=\frac{1}{2}\tilde{\mathcal{K}}_{abc}c^ac^b\mu^c \;, \quad \tilde{\mathcal{K}}_a\mu^a=0 \;,
\end{equation}
from the $\alpha_0$ component of~\eqref{10dHeq3} together with~\eqref{10dHeq1}. Then, the real parts of the moduli obey the following flow equations,
\begin{equation}
  a\sim b^i\sim{\rm constant} \;, \quad \partial_yc^a=-\frac{1}{{\cal Z}^0}\mu^a \;.
\end{equation}
It comes from the NS-NS flux ansatz~\eqref{fluxansatz} and~\eqref{Hy0}, and, for the last equation, we contract  the $\alpha_a$ component of~\eqref{10dHeq3} with $\tilde\cK_{abc}c^c$ and compare it to the $\beta^a$ component of the same equation~\eqref{10dHeq3}. Finally, we can write the flow equations for the dilaton and the imaginary parts of the moduli,
\begin{align}
  \frac{1}{2}\mathcal{K}_i'-\hat\phi'\mathcal{K}_i&={\cal Z}^0e_i \;,\\
  \frac{1}{2}\left(\frac{\partial_y{\cal Z}^0}{{\cal Z}^0}-\frac{\tilde{\mathcal{K}}'}{\tilde{\mathcal{K}}}\right)\tilde{\mathcal{K}}_a+\tilde{\mathcal{K}}_a'&=\frac{2}{{\cal Z}^0}\left(\epsilon_a-\tilde{\mathcal{K}}_{abc}c^b\mu^c\right) \;,\\
  \frac{3\partial_y{\cal Z}^0}{4{\cal Z}^0}+\frac{\tilde{\mathcal{K}}'}{4\tilde{\mathcal{K}}}&=\hat\phi' \;.
\end{align}
The first equation comes from Eq.~\eqref{10dHeq2} compared to the contraction of the $\alpha_a$ component of~\eqref{10dHeq3} with $w^a$. The second equation is simply~\eqref{10dHeq4}. Finally, the last equation is the $\alpha_a$ component of~\eqref{10dHeq3} with a bit of rearranging. These are the exhaustive set of constraints found to be equivalent to the Killing spinor equations~\eqref{10dHflow} for mirror half-flat manifolds with NS-NS flux.

Again, it is now easy to see the correspondence between the four-dimensional domain wall equations and the $10$-dimensional ones. For this, we need to insert the $10$-dimensional relations
\begin{equation}
  \calk=({\cal Z}^0)^2\tilde\calk \;, \quad \hat{\phi}'=\phi'+\frac{1\calk'}{2\calk} \;,
\end{equation}
into one or the other set of equations. It is straightforward to see that it will lead to equivalent equations. This means that the four-dimensional supergravity theory obtained from compactification on mirror half-flat manifolds with flux do indeed lift up to the correct ten-dimensional solutions.

\section{Calabi-Yau with flux and domain wall solutions}

It is interesting to realize that our flow equations~\eqref{10dHflow} imply the existence of an exact solution which involves Calabi-Yau manifolds and non-vanishing $\hat H$-flux. In this solution, the flux stress-energy in the Einstein equations~\eqref{Einstein}, instead of deforming away from a Calabi-Yau space, leads to a non-trivial variation of the moduli as one moves in the direction transverse to the domain wall. The full seven-dimensional manifold has $G_2$ structure with a non-vanishing Ricci tensor, while, at the same time, the six-dimensional fibers remain Ricci-flat at each fixed point in the coordinate $y$. The goal of this section is to present this solution in details.

\subsection{The ten-dimensional solution}

Let us first derive the ten-dimensional Killing spinor equations. Calabi-Yau manifolds are characterized by the property $dJ=0$ and $d\Omega=0$. Inserting this back into the generalized flow equations~\eqref{10dHflow}, we find
\bs\label{CYflow}
\begin{align}
  &J\wedge J'={\hat\phi}' J\wedge J \;, \\
  &\Omega_-'=2{\hat\phi}'\,\Omega_-+*\hat H \;, \\
  &2{\hat\phi}'\,*1=\Omega_-\wedge\hat H \;.
\end{align}
\es
A Calabi-Yau manifold $X$ with moduli varying along $y$ as dictated by the above flow equations will then be a solution of the Einstein equations~\eqref{Einstein}. To satisfy the full system of equations of motion, we also have the constraints on the flux $\hat H$ and the dilaton $\hat\phi$. From Eqs.~\eqref{dwform}, they can be written as
\begin{equation}\label{CYH}
  d\hat\phi=0\;,\quad\hat H\wedge J=0 \;, \quad \hat H\wedge \Omega_+=0 \;.
\end{equation}
We observe that these are the same characteristics as for the case of mirror half-flat manifold with flux.

We can now deploy the full range of Calabi-Yau moduli space technology to solve these differential equations for the various moduli fields. In principle, this amounts to taking the limit $e_i=0$ in our previous general discussion of section~\ref{halflatHcomp}. However, for the sake of clarity, we will repeat the required steps here. We recall the standard expansion of the Kahler form and the complex structure,
\begin{equation}
  J=v^i\omega_i \;, \quad \Omega={\cal Z}^A\alpha_A-\mathcal{G}_A\beta^A \;,
\end{equation}
in terms of harmonic two-forms $\{\omega_i\}$ and harmonic three-forms $\{\alpha_A,\beta^B\}$ on the Calabi-Yau manifold $X$. We have as well the expansion of the NS-NS flux,
\begin{equation}
  \hat H=\epsilon_a\beta^a+\mu^a\alpha_a \;.
\end{equation}
This will satisfy the second constraint of~\eqref{CYH} from the property of the basis forms. We also recall that we have set all components of $\hat H$ breaking four-dimensional Lorenz-invariance to zero, that is, $\hat H_{\mu MN}=0$. This implies the axions $a$ and $b^i$ have to be constant.
The $y$-dependence of the remaining moduli $v^i$ and ${\cal Z}^A$ is determined by the flow equations~\eqref{CYflow} and can be explicitly obtained by inserting the above expansions for $J$, $\Omega$ and $\hat H$ and looking at the coefficients of each basis forms separately. Working in the large complex structure limit, we find for the complex structure moduli
\begin{equation}
  \partial_yc^a=-\frac{1}{{\cal Z}^0}\mu^a \;, \quad \tilde{\mathcal{K}}_a'=\frac{2}{{\cal Z}^0}\left(\epsilon_a-\tilde{\mathcal{K}}_{abc}c^b\mu^c\right) \;,
\end{equation}
and for the dilaton and the Kahler moduli
\begin{equation}
  \hat\phi'=\frac{\tilde{\mathcal{K}}'}{\tilde{\mathcal{K}}} \;, \quad \hat\phi'=\frac{\partial_y{\cal Z}^0}{{\cal Z}^0}\;,\quad \hat\phi'=\frac{\partial_yv^i}{v^i} \;.
\end{equation}
We should point out that the $y$-dependence of $J$ and $\Omega$ implied by these solutions is consistent with the $SU(3)$ structure compatibility relations~\eqref{su3def}. Finally, we also have the conditions coming from the third constraint of~\eqref{CYH} together with the $\alpha_0$ component of the $\Omega_-'$ equation,
\begin{equation}\label{fluxconstraint}
  \tilde{\mathcal{K}}_a\mu^a=0 \;, \quad \epsilon_ac^a=\frac{1}{2}\tilde{\mathcal{K}}_{abc}c^ac^b\mu^c \;.
\end{equation}
They are constraints on the flux parameters and the different integration constants of the previous flow equations.

We can integrate the above differential equations in term of the new rescaled variable 
\be
  d y= {\cal Z}^0 d\tilde y \;.
\ee
We find for the complex structure moduli,
\begin{align}
  c^a&=-\mu^a\tilde y + \mathcal{C}^a \;,\\
  \tilde{\mathcal{K}}_a&=\tilde{\mathcal{K}}_{abc}\mu^b\mu^c\tilde y^2+2\left(\epsilon_a-\tilde{\mathcal{K}}_{abc}\mu^c\mathcal{C}^b\right)\tilde y+\tilde{\mathcal{K}}_{0a} \;,
\end{align}
where $\mathcal{C}^a$ and $\tilde{\mathcal{K}}_{0a}$ are integration constants. This then determines the dilaton and, therefore, the Kahler moduli $v^i$ and the ${\cal Z}^0$ field,
\begin{equation}
  \tilde{\mathcal{K}}\sim e^{\hat\phi} \;, \quad  {\cal Z}^0\sim e^{\hat\phi}   \;, \quad v^i\sim e^{\hat\phi} \;.
\end{equation}
Finally, we have the constraints~\eqref{fluxconstraint} on the flux parameters. When plugging the solution back into it, they turn out to be equivalent to the following. First, the flux parameters $\mu^a$ are constrained by
\begin{equation}
  \tilde{\mathcal{K}}_{abc}\mu^a\mu^b\mu^c=0 \;.
\end{equation}
Then, the integration constants $\mathcal{C}^a$ must be chosen such that
\begin{align}
  \epsilon_a\mu^a&=\tilde{\mathcal{K}}_{abc}\mu^a\mu^b\mathcal{C}^c \;, \\
  \epsilon_a\mathcal{C}^a&=\frac{1}{2}\tilde{\mathcal{K}}_{abc}\mu^a\mathcal{C}^b\mathcal{C}^c \;.
\end{align}
This could turn out to be non-trivial conditions on $\epsilon_a$ as it might not be possible to choose the constants $\mathcal{C}^a$ satisfying the above conditions for any flux parameters. The analysis should be carried on on a case by case basis with explicit intersection numbers.
Finally, we also have the condition
\begin{equation}
  \tilde{\mathcal{K}}_{0a}\mu^a=0 \;,
\end{equation}
on the $\tilde{\mathcal{K}}_{0a}$ integration constants which can always be satisfied by choosing the constants to vanish. Hence, provided that the fluxes and the integration constants satisfy the above non-trivial constraints, we find indeed a Calabi-Yau domain wall solution. 

\subsection{The four-dimensional solution}

As before, we now relate this $10$-dimensional Calabi-Yau domain wall solution to the four-dimensional supergravity obtained by compactification on the corresponding Calabi-Yau manifold with flux. The module fields in this four-dimensional supergravity are as usual,
\begin{equation}
  S=a+ie^{-2\phi} \;, \quad T^i=b^i+iv^i \;, \quad Z^a=c^a+iw^a \;,
\end{equation}
and the superpotential is given by the Gukov-Vafa formula~\eqref{GukovW},
\begin{equation}
  W=\epsilon_aZ^a+\mu^a\mathcal{G}_a\left(Z\right) \;.
\end{equation}
In the same way as in section~\ref{halflatHcomp}, the domain wall Killing spinor equations~\eqref{DW} tell us that the real parts of the superfields satisfy
\begin{equation}
  a\sim b^i\sim\mbox{constant} \;, \quad \partial_yc^a=-\sqrt{\frac{\tilde{\mathcal{K}}}{\mathcal{K}}}\mu^a \;.
\end{equation}
Again, the warp factor of the metric ansatz~\eqref{4dmetric} can be set to $B=\phi$ and we have the constraints from the superpotential being purely imaginary,
\be
  \tilde{\mathcal{K}}_a\mu^a=0 \;,\quad \epsilon_ac^a=\frac{1}{2}\tilde{\mathcal{K}}_{abc}c^ac^b\mu^c \;.
\ee
For the imaginary parts, we have from~\eqref{DW1} evaluated for the $\left(S,T^i,Z^a\right)$ fields respectively,
\begin{align}
  \cK_i\left(\frac{\cK'}{\cK}+2\phi'\right)-\cK'_i&=0 \;,\\
  \tilde\cK_a\left(\frac{\tilde\cK'}{\tilde\cK}+2\phi'\right)-\tilde\cK'_a&=-2\sqrt{\frac{\tilde{\mathcal{K}}}{\mathcal{K}}}\left(\epsilon_a-\tilde{\mathcal{K}}_{abc}c^b\mu^c\right) \;,\\
  -\frac{1}{4}\left(\frac{\mathcal{K}'}{\mathcal{K}}+\frac{\tilde{\mathcal{K}}'}{\tilde{\mathcal{K}}}\right)&=2\phi' \;.
\end{align}
However, this can be simplified further. Contracting the third equation with $v^i$, we realize that $\cK'=-6\phi'\cK$. This can then be plugged back into every equations to reduce the system in the following form,
\begin{equation}
  \phi'=-\frac{1}{2}\frac{\tilde{\mathcal{K}}'}{\tilde{\mathcal{K}}} \;, \quad \tilde{\mathcal{K}}_a\partial_yw^a=\sqrt{\frac{\tilde{\mathcal{K}}}{\mathcal{K}}}\left(\epsilon_a-\tilde{\mathcal{K}}_{abc}c^b\mu^c\right) \;, \quad \partial_yv^i=-2\phi'v^i \;.
\end{equation}
We note that this corresponds to Eqs.~\eqref{DWstepflux} in the limit where the half-flat flux parameters vanish, that is, $e_i=0$. 

The matching of these four-dimensional flow equations with the ten-dimensional ones~\eqref{CYflow} can be worked out in the same way as before, namely by inserting the definitions
\begin{equation}
  \calk=\left({\cal Z}^0\right)^2\tilde\calk \;, \quad \hat{\phi}'=\phi'+\frac{1\calk'}{2\calk} \;.
\end{equation}
Hence, we conclude that the four-dimensional domain wall solution is identical, upon up-lifting, to the $10$-dimensional Calabi-Yau domain wall solution.

We should note that for the case of vanishing magnetic flux $\mu^a=0$, the above equations reduce to
\begin{equation}
  a\sim b^i\sim c^a\sim \mbox{const} \;, \quad \epsilon_ac^a=0 \;,
\end{equation}
for the real parts, and
\begin{equation} 
  \phi'=-\frac{1}{2}\frac{\tilde{\mathcal{K}}'}{\tilde{\mathcal{K}}} \;, \quad \tilde{\mathcal{K}}_{ab}\partial_yw^a={\sqrt{\frac{\tilde{\mathcal{K}}}{\mathcal{K}}}}\epsilon_a \;, \quad \partial_yv^i=-2\phi'v^i \;,
\end{equation}
for the imaginary parts. These equations are ``mirror-symmetric'' to~\eqref{dweq} under the following correspondence
\be
  \left\{v^i,\;\mathcal{K}_{ijk},\;e_i\right\}\longleftrightarrow\left\{w^a,\;\tilde{\mathcal{K}}_{abc},\;\epsilon_a\right\}
\ee
and can, therefore, be integrated in the same way. This fact is not surprising and reflects the original construction of half-flat mirror manifolds~\cite{Gurrieri:2002wz} as type II mirror duals of Calabi-Yau manifolds with electric NS flux. In the present context, it suggests a symmetry between heterotic Calabi-Yau compactifications with electric NS flux and heterotic compactifications based on the associated half-flat mirror manifolds.\\

This closes the discussion of this chapter about the consistency of the dimensional reduction of heterotic string theory compactified on backgrounds preserving only two supercharges. In the next chapter, we will build explicit geometries that fulfills the aforementioned properties.
\clearemptydoublepage
\chapter{Nearly-Kahler homogeneous spaces compactification}\label{chaptercoset}

In this chapter, we would like to illustrate the previous result about mirror half-flat compactifications with some explicit examples. In particular, this will allow us to study the gauge sector which has been neglected so far. The presence of gauge bundles is one of the distinctive features of heterotic string compactification  and is responsible for many of the physically interesting properties as well as technical complications. For Calabi-Yau compactifications, the internal metric is not known explicitly and conclusions about gauge bundles are drawn using techniques from algebraic geometry. When working with fluxes and non-Calabi-Yau geometries, the internal manifolds are non-Kahler and, in the context of the previous chapter, non-complex as well. This renders the use of algebraic geometry not straightforwardly useful or, at worse, futile. One way to circumnavigate this difficulties is to work with manifolds whose metrics are known explicitly. This is the strategy adopted in this chapter.

For this purpose, we will first introduce the class of manifolds that will be used, namely the six-dimensional homogeneous spaces $SU(3)/U(1)^2$, $Sp(2)/SU(2)\times U(1)$ and $G_2/SU(3)$. They have this advantage that their metric can be calculated explicitly. We will derive the relevant properties and show that they satisfy the structure of half-flat mirror manifolds to justify them being solutions of heterotic string theory. Thereafter, we will describe vector bundle constructions over such spaces and give some explicit examples satisfying the heterotic supergravity constraints.

\section{Coset space geometry}

We start with a short review of the general formalism used in the construction of homogeneous manifolds. This formalism is well established~\cite{Opfermann:1998qd,Castellani:1999fz,Kapetanakis:1992hf,MuellerHoissen:1987cq,Camporesi:1990wm,Castellani:1983tb,Lust:1986ix,KashaniPoor:2007tr}, so we will be brief and let the interested reader refer to the appendix~\ref{appendixcoset} where we present a more detailed analysis on this subject. Here, we only outline the general strategy and collect the relevant formulas necessary to compute the required geometrical data.

Let $G$ be a Lie-group and $H$ a sub-Lie group of $G$. The coset space $G/H$ is defined as the set of left cosets which arise from the equivalence relation,
\begin{equation}
  \mathsf{g}\sim\mathsf{g}'\Leftrightarrow\mathsf{g}^{-1}\mathsf{g}'=\mathsf{h} \;,
\end{equation}
where $\mathsf{g}$ and $\mathsf{g}'$ are elements of $G$ and $\mathsf{h}$ is an element of $H$. We should think about the group $G$ as a principal bundle $G\left(G/H,H\right)$ with base space $G/H$ and fibers given by the orbits of $H$. This picture will be useful later when we consider the construction of vector bundles. In order to get an explicit description of the coset space, we can choose a representative for each cosets. It corresponds to a section of the principal bundle $G$ and, using the exponential map, it can be written for the coordinates $x$ as,
\begin{equation}
  L\left(x\right)=\exp\left(x^aK_a\right) \;,
\end{equation}
where $K_a$ are the generators of the Lie algebra of $G$ which are not generators of the Lie algebra of $H$. (More details about our conventions can be found in appendix~\ref{appendixcoset}.) Following a procedure similar to the one leading to left-invariant one forms on a Lie-group, we can define non-singular one-forms on $G/H$ as,
\begin{equation}\label{epsdef}
  L^{-1}dL=e^aK_a+\varepsilon ^i H_i \;,
\end{equation}
where $d$ is the exterior derivative on $G/H$ and $H_i$ are the generators of the Lie-algebra of $H$. From hereon, our convention is to have coset indices $a,b,c,\dots$ running over values $1,\ldots ,6$ and indices $i,j,k,\dots$, which label the generators of the sub-group $H$, range from $7,\ldots ,{\rm dim}(G)$. (Index conventions are summarized in appendix~\ref{appendixconventions}.) The right-hand side of this equation corresponds to the expansion of $L^{-1}dL$ on the Lie-algebra of $G$ and defines the one-form ``coefficients'' $e^a$ and $\varepsilon ^i$. The exterior derivative of these forms follows from the Maurer-Cartan structure equations,
\bs\label{Maurer}
\begin{align}
  de^a&=-\frac{1}{2}f_{bc}^{\phantom{bc}a}e^b\wedge e^c-f_{ib}^{\phantom{ib}a}\varepsilon^i\wedge e^b \;,\\
  d\varepsilon^i&=-\frac{1}{2}f_{ab}^{\phantom{ab}i}e^a\wedge e^b-\frac{1}{2}f_{jk}^{\phantom{jk}i}\varepsilon^j\wedge\varepsilon^k \;.
\end{align}
\es

The geometrical structures of coset spaces can then be written in terms of these one-forms. For six-dimensional manifolds and with suitable coefficients, it means for the $SU(3)$ structure,
\bs
\begin{align}
  g&=g_{ab}\,e^a\otimes e^b \;,\\
  J&=\frac{1}{2}\,J_{ab}\,e^a\wedge e^b \;,\\
  \Omega&=\frac{1}{3!}\,\Omega_{abc}\,e^a\wedge e^b\wedge e^c \;.
\end{align}
\es
However, the one-forms $e^a$ are not left-invariant and nothing ensures in general that the above structure is well-defined everywhere on the coset. To this end, we must impose for the above combinations of $e^a$ to lead to left-invariant objects. The relation
\begin{equation}\label{Lgaction}
  \mathsf{g}L\left(x\right)=L\left(x'\right)\mathsf{h} \;,
\end{equation}
allows to compute the transformations of $e^a$ under the left-action of $\mathsf{g}$. This equation simply follows from the definition of the coset. The ``gauge transformation'' with $\mathsf{h}$ on the right-hand side accounts for the fact that the group action, while leading to an element in the coset represented by $L(x')$, does not necessarily give the chosen representative $L(x')$. Looking at infinitesimal transformations, Eq.~\eqref{Lgaction} allows us to compute transformations of the $SU(3)$-structure. Imposing such infinitesimal transformations to vanish gives the conditions,
\begin{equation}\label{leftinv}
  f_{i(a}^{\phantom{i(a}c}g_{ b)c}=0 \;, \quad f_{i[a}^{\phantom{a]i}c}J_{b]c}=0 \;, \quad f_{i[a}^{\phantom{a]i}d}\Omega_{bc]d}=0 \;,
\end{equation}
on the $SU(3)$ structure coefficients in order for it to be left invariant and, therefore, well-defined everywhere.

Finally, we see that $e^a$ plays the role of vielbein and, so, the Levi-Civita connection can be calculated from the standard relations,
\be
  de^a+\omega^a_{\phantom{a}b}\wedge e^b=0 \;, \quad \omega_{ab}=-\omega_{ba} \;.
\ee
This allows to compute the Euler form $\gamma\left(TX\right)$ which is a function of the Riemann curvature two-form. From the generalized Gauss bonnet theorem~\cite{Kobayashi},
\begin{equation}\label{GaussBonnet}
  \chi\left(X\right)=\int_X\gamma\left(TX\right) \;,
\end{equation}
where $\chi$ is the Euler number, we can calculate the volume of $G/H$. Indeed the right-hand side of~\eqref{GaussBonnet} is a top form and, therefore, proportional to the volume. From this, we can write
\begin{equation}
  {\cal V}=\varrho^{-1}\,\chi \;,
\end{equation}
for the appropriate coefficient $\varrho$ which, knowing $\chi$, can be calculated from~\eqref{GaussBonnet}.

\section{Six-dimensional nearly-Kahler coset manifolds}

In this section, we would like introduce the particular six-dimensional manifolds on which we compactify heterotic string theory. We simply apply the formalism developed in the previous section to special cases, mainly following the results of Refs.~\cite{Chatzistavrakidis:2008ii,Chatzistavrakidis:2009mh}. Vector bundles and gauge connections on these manifolds will be discussed in the following section.

It is known~\cite{Butruille} that precisely four six-dimensional spaces within this class are half-flat manifolds, namely the cosets $SU(3)/U(1)^2$, $Sp(2)/SU(2)\times U(1)$, $G_2/SU(3)$ and $SU(2) \times SU(2)$. Since the last example $SU(2) \times SU(2)$ is less suited for bundle constructions, it will not be discussed explicitly and we will focus on the first three cases. We will see that the torsion classes of these manifolds satisfy the half-flat constraints~\eqref{halflat} and are, in fact, somewhat more special in a way that is referred to as ``nearly Kahler" in the literature. In practice, we will systematically construct an explicit set of forms on these spaces which satisfy the relations~\eqref{hfmdef} for half-flat mirror manifolds. This has been first exposed by House and Palti for the $SU(3)/U(1)^2$ case~\cite{House:2005yc, Palti:2006yz}. We will also present the corresponding families of $G$-invariant $SU(3)$ structures and their torsion classes. To avoid cluttering the main text, the relevant group-theoretical information, such as generators and structure constants, has been collected in appendix~\ref{appendixcoset}.

\subsection{$SU(3)/U(1)^2$}

We choose the usual Gell-Mann matrices for the $SU(3)$ generators $T_A$, however, relabeled in such a way that the coset generators, corresponding to the non-diagonal Gell-Mann matrices, carry indices from $1$ to $6$. The resulting generators and structure constants are given in appendix~\ref{appendixcoset}. Solving Eq.~\eqref{leftinv} shows that the most general $SU(3)$-invariant metric takes the form
\begin{equation}\label{su3metric}
  ds^2=R_1^2\,\left(e^1\otimes e^1+e^2\otimes e^2\right)+R_2^2\,\left(e^3\otimes e^3+e^4\otimes e^4\right)+R_3^2\,\left(e^5\otimes e^5+e^6\otimes e^6\right)
\end{equation}
where $R_1$, $R_2$ and $R_3$ are arbitrary real parameters representing the moduli. The corresponding $G$-invariant structure $(J,\Omega)$ is given by the forms,
\bs
\begin{align}\label{JSU3}
  J&=-R_1^2\,e^{12}+R_2^2\,e^{34}-R_3^2\,e^{56} \\\label{OmegaSU3}
  \Omega&=R_1R_2R_3\left(\left(e^{136}-e^{145}+e^{235}+e^{246}\right)+i\left(e^{135}+e^{146}-e^{236}+e^{245}\right)\right) \;.
\end{align}
\es
By computing the metrics associated to this family of $SU(3)$ structures, one can verify that the moduli $R_1$, $R_2$ and $R_3$ are indeed identical to the ones appearing in \eqref{su3metric}.

In general, $G$-invariant forms are spanned by the following set of basis elements for two- and three-forms:
\begin{equation}
  \left\{e^{12} \;, \quad e^{34} \;, \quad e^{56}\right\} \;, \quad \left\{e^{136}-e^{145}+e^{235}+e^{246} \;, \quad e^{135}+e^{146}-e^{236}+e^{245}\right\} \;.
\end{equation}
It is worth noting that there is no $G$-invariant one-forms nor five-forms. With a suitable re-definition, we can find a more convenient basis to unveil the structure of mirror half-flat manifold. We introduce the following linear combinations of $G$-invariant two-, three- and four-forms,
\bs
\begin{align}\label{eq_su3basisstart}
  \omega_1&=-\frac{1}{2\pi}\left(e^{12}+\frac{e^{34}}{2}-\frac{e^{56}}{2}\right) & \tilde\omega^1&=\frac{4\pi}{3{\cal V}_\odot}\left(2\,e^{1234}+e^{1256}-e^{3456}\right) \\
  \omega_2&=-\frac{1}{4\pi}\left(e^{12}+e^{34}\right) & \tilde\omega^2&=-\frac{4\pi}{{\cal V}_\odot}\left(e^{1234}+e^{1256}\right)\label{eq_su3end} \\
  \omega_3&=\frac{1}{2\pi}\left(e^{12}-e^{34}+e^{56}\right) & \tilde\omega^3&=\frac{2\pi}{3{\cal V}_\odot}\left(e^{1234}-e^{1256}+e^{3456}\right) \\\label{eq_su3basisend}
  \alpha_0&=\frac{3\pi}{4{\cal V}_\odot}\left(e^{136}-e^{145}+e^{235}+e^{246}\right) & \beta^0 &=\frac{1}{3\pi}\left(e^{135}+e^{146}-e^{236}+e^{245}\right)
\end{align}
\es
where ${\cal V}_\odot$ is the unitary volume of the coset space. It corresponds to the volume evaluated for the moduli $R_1=R_2=R_3=1$ and is given by,
\begin{equation}\label{vol}
  {\cal V}_\odot=\int_Xe^{123456}=4\left(2\pi\right)^3 \;.
\end{equation}
This ensures the proper normalization to satisfy the intersections~\eqref{intersections}. Moreover, it has been chosen such that it obeys the condition,
\begin{equation}\label{isec}
  \omega_r\wedge\omega_s=\cK_{rst}\,\tilde{\omega}^t \;,
\end{equation}
required so that the set of forms have the same intersection structure as the basis of cohomology classes (see appendix~\ref{appendixconventions}). They are, however, not harmonic and some of the above forms are not closed. The corresponding intersection numbers are,
\be
\begin{aligned}
  \cK_{111}&=6 \;, \quad \cK_{112}=3 \;, \quad \cK_{113}=6 \;,\\
  \cK_{122}&=1 \;, \quad \cK_{123}=3 \;, \quad \cK_{133}=0 \;,\\
  \cK_{222}=0 \;&, \quad \cK_{223}=2 \;, \quad \cK_{233}=0 \;, \quad \cK_{333}=-24 \;.
\end{aligned}
\ee
The exterior derivative can easily be computed in terms of the structure constants using the Maurer-Cartan structure equations~\eqref{Maurer}. We find the following intrinsic torsion parameters,
\begin{equation}\label{SU3e}
  e_1=0 \;, \quad e_2=0 \;, \quad e_3=1 \;,
\end{equation}
from the definition of the differential relation for mirror half-flat manifolds~\eqref{hfmdef}. In particular, from the general discussion of section~\ref{hfcomp}, this means that the $SU(3)$ structure defined above is mirror half-flat.

Writing $J$ in \eqref{JSU3} in terms of the above two-forms $\omega_i$ and comparing with Eq.~\eqref{Jexpansion}, we can
read off expressions for ``Kahler" moduli $v^i$, defined in the context of half-flat mirror manifolds. In terms of the radii $R_i$, they are given by
\begin{equation}
  v^1=\frac{4\pi}{3}\left(R_1^2+R_2^2-2R_3^2\right) \;, \quad v^2=-4\pi\left(R_2^2-R_3^2\right) \;, \quad v^3=-\frac{2\pi}{3} \left(R_1^2+R_2^2+R_3^2\right) \;.
\end{equation}
Note that the forms $\omega_1$ and $\omega_2$ are closed and, hence, define cohomology classes, whereas $\omega_3$ is not closed. Therefore, we expect two massless modes, $v^1$ and $v^2$, and a massive one, $v^3$. This anticipation is confirmed by looking at the superpotential for half-flat mirror compactifications which has been reviewed in chapter~\ref{chapterreview}. It is given by the Gukov-Vafa formula~\eqref{GukovW} and, for mirror half-flat manifolds, gives
\begin{equation}\label{W}
  W=e_iT^i \;, \quad \mbox{where } \quad {\rm Im}\,(T^i)=v^i \;.
\end{equation}
For the present case, in view of the torsion parameters~\eqref{SU3e}, this means,
\begin{equation}
  W=T^3 \;,
\end{equation}
so that $T^1$ and $T^2$ are indeed massless. Also note that the existence of only two $G$-invariant three-forms $\alpha_0$ and $\beta^0$ means that the analogues of complex structure moduli are not present in this particular model. Consequently, this would lead to difficulties if one would want to find the corresponding mirror Calabi-Yau.

Last, for the sake of completeness, let us present the corresponding torsion classes of this manifold. Explicitly, it we can be computed knowing the derivatives of $J$ and $\Omega$ and using Eqs.~\eqref{su3torsion}. They are given by~\cite{Chatzistavrakidis:2009mh},
\bs
\begin{align}
  \cW_1^+=&-\frac{R_1^2+R_2^2+R_3^2}{3R_1R_2R_3} \;,\\\nonumber
  \cW_2^+=&-\frac{2}{3R_1R_2R_3}\left[R_1^2\left(2R_1^2-R_2^2-R_3^2\right)e^{12}-R_2^2\left(2R_2^2-R_1^2-R_3^2\right)e^{34}\right. \\
  &\ph{\frac{2}{3R_1R_2R_3}[}\left.+R_3^2\left(2R_3^2-R_1^2-R_2^2\right)e^{56}\right] \;.
\end{align}
\es
It will be relevant to realize that on the locus in moduli space where the three radii are equal, $R_1=R_2=R_3\equiv R$, the torsion classes reduce to
\begin{equation}
  \cW_1^+ =-\frac{1}{R} \;, \quad \cW_2^+=0 \;.
\end{equation}
This shows that the $SU(3)$ structure is nearly-Kahler at this particular locus (the table~\ref{tablesu3torsion} summarizes properties of manifolds according to their torsion classes). This locus will play a special role and it is straightforward to see that for,
\be
  \partial_yR=-\frac{2}{9} \;,
\ee
Hitchin's flow equations~\eqref{hfe} are satisfied. Thus, we can build a seven-dimensional $G_2$ holonomy manifold when using the above fibration.

\subsection{$Sp(2)/SU(2)\times U(1)$}

We now turn to the second case. In order to obtain a mirror half-flat space, this coset is defined by taking the non-maximal embedding of $SU(2)$ into $Sp(2)$. Group-theoretical details, in particular generators and structure constants, are again given in appendix~\ref{appendixcoset}. We proceed in the same way as in the previous case. Solving Eq.~\eqref{leftinv}, the most general $Sp(2)$-invariant metric turns out to be
\begin{equation}\label{Spmetric}
  ds^2=R_1^2\,\left(e^1\otimes e^1+e^2\otimes e^2\right)+R_2^2\,\left(e^3\otimes e^3+e^4\otimes e^4\right)+R_1^2\,\left(e^5\otimes e^5+e^6\otimes e^6\right)
\end{equation}
with, this time, only two moduli $R_1$ and $R_2$. The corresponding $SU(3)$-structure forms are given by,
\bs
\begin{align}\label{JSp2}
  J&=-R_1^2\,e^{12}+R_2^2\,e^{34}-R_1^2\,e^{56} \;,\\\label{OmegaSp2}
  \Omega&=R_1^2R_2\left(\left(e^{136}-e^{145}+e^{235}+e^{246}\right)+i\left(e^{135}+e^{146}-e^{236}+e^{245}\right)\right) \;.
\end{align}
\es
These forms satisfy~\eqref{leftinv} and are indeed $G$-invariant.

In general, $G$-invariant forms are spanned by the following set of basis forms,
\begin{equation}
  \left\{e^{12}+e^{56} \;, \quad e^{34}\right\} \;, \quad\left\{e^{135}+e^{146}-e^{236}+e^{245} \;, \quad e^{136}-e^{145}+e^{235}+e^{246}\right\} \;.
\end{equation}
Again, there is no left-invariant one- nor five-forms. We can write a more suitable combination with the following,
\bs
\begin{align}\label{eq_sp2basisstart}
  \omega_1&=\frac{1}{2\pi}\left(e^{12}+2\,e^{34}+ e^{56}\right) & \tilde\omega^1&=\frac{\pi}{3{\cal V}_\odot}\left(e^{1234}+2\,e^{1256}+ e^{3456}\right) \\
  \omega_2&=\frac{1}{\pi}\left(e^{12}-e^{34}+e^{56}\right) & \tilde\omega^2&=\frac{\pi}{3{\cal V}_\odot}\left(e^{1234}-e^{1256}+e^{3456}\right) \\\label{eq_sp2basisend}
  \alpha_0&=\frac{3\pi}{{\cal V}_\odot}\left(e^{136}-e^{145}+e^{235}+e^{246}\right) & \beta^0&=\frac{1}{12\pi}\left(e^{135}+e^{146}-e^{236}+e^{245}\right)
\end{align}
\es
where the volume ${\cal V}_\odot$ is the unitary volume of the coset space defined as before~\eqref{vol} and assures the correct normalization to satisfy~\eqref{intersections}. For the present case, we have
\be
  {\cal V}_\odot=\frac{\left(2\pi\right)^3}{12} \;.
\ee
This basis has again been chosen such that it has the same structure as a basis of cohomology classes and obey~\eqref{isec}. The intersections are found to be,
\begin{equation}
  \cK_{111}=1 \;, \quad  \cK_{112}=1 \;, \quad \cK_{122}=0 \;, \quad \cK_{222}=-4 \;.
\end{equation}
It also satisfies the relevant relations~\eqref{hfmdef} for half-flat mirror manifolds provided that the torsion parameters are set to
\begin{equation}\label{eSp2}
  e_1=0 \;, \quad e_2=1 \;.
\end{equation}

We can expand $J$ from Eq.~\eqref{JSp2} in terms of the basis forms $\omega_i$ to obtain the Kahler moduli fields,
\begin{equation}
  v^1=-\frac{2\pi}{3}\left(R_1^2-R_2^2\right) \;, \quad v^2=-\frac{\pi}{3}\left(2R_1^2+R_2^2\right) \;.
\end{equation}
The form $\omega_1$ is closed while $\omega_2$ is not, so we expect $v^1$ to be massless and $v^2$ to be heavy. From the torsion parameters~\eqref{eSp2} the superpotential~\eqref{W} is now given by
\be
  W=T^2 \;,
\ee
which confirms this expectation. As before, there are no ``complex structure moduli" for this coset space.

Finally, the half-flat torsion classes are given by~\cite{Chatzistavrakidis:2009mh},
\bs
\begin{align}
  \cW_1^+=&-\frac{4R_1^2+2R_2^2}{3R_1^2R_2} \;,\\\nn
  \cW_2^+=&-\frac{4}{3R_1^2R_2}\left[R_1^2\left(R_1^2-R_2^2\right)e^{12}+2R_2^2\left(R_1^2-R_2^2\right)e^{34}\right. \\
  &\ph{\frac{4}{3R_1^2R_2}(}\left.+R_1^2\left(R_1^2-R_2^2\right)e^{56}\right] \;.
\end{align}
\es
When the two radii are equal, $R_1=R_2\equiv R$, they simplify to
\begin{equation}
  \cW_1^+=-\frac{2}{R} \;, \quad \cW_2^+=0 \;,
\end{equation}
and corresponds to a nearly-Kahler $SU(3)$ structure, as before. Again, for completeness, let us state that Hitchin's flow equations~\eqref{hfe} are satisfied when
\be
  \partial_yR=-\frac{1}{36} \;.
\ee
Thus, we can easily build a seven-dimensional $G_2$ holonomy manifold.

\subsection{$G_2/SU(3)$}

Details of the group theory are explicitly given in appendix~\ref{appendixcoset}. Following the same procedure as in the previous two cases, the most general $G_2$ invariant metric turns out to be
\begin{equation}
  ds^2=R^2\,\left(e^1\otimes e^1+e^2\otimes e^2+e^3\otimes e^3+e^4\otimes e^4+e^5\otimes e^5+e^6\otimes e^6\right)
\end{equation}
where $R$ is the only modulus. The corresponding $G_2$ invariant $SU(3)$-structure forms are given by
\bs
\begin{align}\label{G2J}
  J&=R^2\,\left(e^{12}-e^{34}-e^{56}\right) \;,\\
  \Omega&=R^3\left(\left(e^{136}+e^{145}-e^{235}+e^{246}\right)+ i\left(e^{135}-e^{146}+e^{236}+e^{245}\right)\right) \;.
\end{align}
\es
Since the sets of left invariant forms are one-dimensional, we only need to choose the appropriate normalization to satisfy the required intersection pattern,
\bs
\begin{align}
  \omega_1&=\frac{5}{3\pi}\left(-e^{12}+e^{34}+e^{56}\right)& \tilde\omega^1&=\frac{\pi}{5{\cal V}_\odot}\left(e^{1234}+e^{1256}-e^{3456}\right) \\
  \alpha_0&=\frac{10\pi}{\sqrt{3}\,{\cal V}_\odot}\left(e^{136}+e^{145}-e^{235}+e^{246}\right)& \beta^0&=\frac{\sqrt{3}}{40\pi}\left(e^{135}-e^{146}+e^{236}+e^{245}\right) \;.
\end{align}
\es
Here, the volume is $\cV_\odot=9(2\pi)^3/20$ and the corresponding single intersection number from~\eqref{isec} is found to be,
\be
  \cK_{111}=-100 \;.
\ee
Moreover, the above basis satisfies the half-flat mirror conditions~\eqref{hfmdef} with the intrinsic torsion parameter given by
\begin{equation}
  e_1=1 \;.
\end{equation}
This implies only one single Kahler modulus,
\begin{equation}
  v^1=-\frac{3\pi}{5}\,R^2 \;,
\end{equation}
which is a heavy mode since $\omega^1$ is not closed or, equivalently, since the superpotential is given by $W=T^1$. Once more, there are no complex structure moduli. The only non-vanishing torsion class is~\cite{Chatzistavrakidis:2009mh},
\begin{equation}
  \cW_1^+=-\frac{4}{\sqrt{3}\,R} \;,
\end{equation}
which is nearly-Kahler.

\section{Vector bundles}

So far, we have presented the gravitational sector of certain non-Calabi-Yau heterotic compactifications. We now come to the main point of this chapter which is the construction of gauge fields associated to these compactifications. To date, gauge fields in heterotic non-Calabi-Yau compactifications have been mainly addressed in a generic way, without providing explicit bundles and connections. Obviously, this restricts phenomenological applications of non-Calabi-Yau models considerably. One reason for this is the lack of suitable example manifolds on which to construct gauge bundles. For non-Calabi-Yau manifolds without an integral complex structure, the case considered in this chapter, an added complication is that powerful tools from algebraic geometry, which are essential in Calabi-Yau model building, cannot be directly applied. (An interesting new class of examples where one may be able to circumnavigate this problem has been recently found in Ref.~\cite{Larfors:2011zz, Larfors:2010wb}.) Here, we focus on a small class of half-flat coset manifolds suitable for heterotic compactifications which have the advantage of allowing for an explicit computation of most relevant gauge field quantities. Discussion about SU(3)-equivariant pseudo-holomorphic bundles over $SU(3)/U(1)^2$ can also be found in~\cite{Popov:2010rf}.

In the next sub-sections, we will explain the basic mathematical methods for constructing bundles with connections over coset spaces and for evaluating their properties. In particular, we will concentrate on how to construct line bundles. These can be used as building blocks to construct the higher rank bundles which are typically of interest in heterotic compactifications. We will also show how the index of bundles --- giving the number of chiral families in the low-energy theory --- can be computed from the Atiyah-Singer index theorem. These general constructions will then be applied to our particular coset examples in the following section. As we will see, gauge connections and their associated field strengths can be written down explicitly for these spaces. It is this feature, facilitated by the group structure of the manifolds, which allows us to check all relevant properties required for heterotic vacua.

\subsection{Associated vector bundles and line bundles}\label{assbundles}

We have mentioned before that the group $G$ can also be viewed as a principle bundle $G=G(G/H , H)$ over the coset space $X=G/H$. This observation is the starting point for constructing vector bundles $V$ over $G/H$. It is well-known that for each representation $\rho$ of $H$ on a vector space $F$ there is a vector bundle $V=V(G/H,F)$ over $G/F$ with typical fiber $F$ which is associated to the principle bundle $G$. More explicitly, this vector bundle can be constructed as follows. We start with the trivial vector bundle $G\times F$ over $G$ where the group $H$ is acting on the fiber $F$ via the representation $\rho$. On this vector bundle, we can introduce an equivariant map,
\begin{equation}
  E_\mathsf{h} : \quad \left(\mathsf{g},\xi\right)\in G\times F\;\rightarrow\; \left(\mathsf{g},\xi\right)\,\mathsf{h}=\left(\mathsf{g}\cdot\mathsf{h},\rho\left(\mathsf{h}^{-1}\right)\xi\right),
\end{equation}
which sends elements of $G\times F$ onto elements of a new bundle $V_\rho$ depending on the representation $\rho$.
The following diagram is then commutative,
\begin{equation}
  \begin{array}{ccc}
   G\times F & \stackrel{E_\mathsf{h}}{\longrightarrow} & V_\rho \\
   \phantom{\pi}\downarrow \pi && \phantom{\pi_{\hat E}}\downarrow \pi' \\
   G & \stackrel{\mathsf{h}}{\longrightarrow} & G/H \\
  \end{array}
\end{equation}
where $\pi$ and $\pi'$ are the bundle projections, which implies that $V_\rho$ is indeed a vector bundle. Hence, for every representation $\rho$ of $H$, we have a corresponding vector bundle $V_\rho$ over $G/H$ which is associated to the principle bundle $G$ and is defined as the set of equivalence classes under the relation $\left(\mathsf{g},\xi\right)\sim\left(\mathsf{g}\cdot\mathsf{h},\rho\left(\mathsf{h}^{-1}\right)\xi\right)$. A particularly useful fact for our purpose is that a connection on the principal bundle uniquely induces a connection on every associated vector bundles. This leaves us with finding a connection on $G(G/H,H)$. Fortunately, it is known~\cite{Nomizu} that the $G$-invariant connections of the principal bundle $G(G/H,H)$ are in one-to-one correspondence with reductive decompositions of the Lie algebra $\mathfrak{g}$ of $G$ and are explicitly given by
\begin{equation}
  A=\varepsilon^iH_i \;.
\end{equation} 
We recall, that the $H_i$ are a basis of the Lie-algebra of $H$ and the one-forms $\varepsilon^i$ on the coset have been defined in Eq.~\eqref{epsdef}. The induced connection $A_{(\rho)}$ on the associated vector bundle $V_\rho$ is then given by
\begin{equation}\label{eq_ginvconncection}
  A_{\left(\rho\right)}=\varepsilon^i\rho\left(H_i\right) \;.
\end{equation}
For simplicity of notation we will drop the index $\rho$ from now on. The curvature $F=dA+A\wedge A$ of this connection can be computed from the Maurer-Cartan structure equations~\eqref{Maurer}. This leads to
\begin{equation}\label{F}
  F=-\frac{1}{2}f_{ab}^{\phantom{ab}i}\rho\left(H_i\right)e^a\wedge e^b \;.
\end{equation}
Note that this curvature is independent of $\varepsilon^i$ and can be expressed solely in terms of the vielbein forms $e^a$ as a direct consequence of reductiveness, that is, of the structure constants satisfying~\eqref{eq_reductivity}. This fact is of considerable practical importance since it means that all subsequent calculations can be performed ``algebraically", merely based on the knowledge of structure constants. 

We would like to mention two specific types of associated vector bundles which will be relevant for the following discussion. The first is obtained by choosing the representation
\begin{equation}\label{torsionconnection}
  \rho\left(H_i\right)_{b}^{\phantom{cb}a}=f_{ib}^{\phantom{ib}a} \;,
\end{equation}
that is $\rho$ being induced by the adjoint representation of $G$. The corresponding bundle is the tangent bundle of $G/H$ and the gauge field defined by the above choice of representation provides a connection with torsion on this bundle. 

The second type arises for one-dimensional representations $\rho$ of $H$. Applying the above formalism to such representations leads to line bundles and connections on them. One choice, which is always possible, is of course the trivial representation of $H$. However, in this case, the associated line bundle is simply the trivial line bundle ${\cal O}_X$. Fortunately, for two of our examples, the corresponding sub-groups $H$ allow for non-trivial one-dimensional representations so that we can generate more interesting line bundles $L$. Since we know the curvature form of these bundles, it is possible to explicitly work out their first Chern class
\begin{equation}
  c_1\left(L\right)=\frac{i}{2\pi}\left[F\right]=p^r\omega_r \;.
\end{equation} 
Here, the square bracket denotes the cohomology class in $H^2(X)$. The last part of the equation is a linear combination of a suitable basis $\{\omega_r\}$ of $H^2(X)$ with integer coefficients $p^r$ to be determined explicitly for our examples. The line bundle $L$ is uniquely characterized by its first Chern class or, equivalently, by the integer vector ${\bf p}=(p^r)$, and will also be denoted as $L={\cal O}_X({\bf p})$. These line bundles will be used as building blocks for higher-rank bundles. In particular, we will consider sums of $n$ line bundles
\begin{equation}\label{V}
  V=\bigoplus_{i=1}^nL_i \quad {\rm where} \quad L_i={\cal O}_X\left({\bf p}_i\right) \;.
\end{equation}
For such sums, we require a vanishing total first Chern class $c_1(V)=0$ which means that, for all $r$, the integers $p_i^r$ must satisfy,
\begin{equation}
  \sum_{i=1}^np_i^r=0 \;.
\end{equation}
This guarantees that the structure group of $V$ is contained in $S(U(1)^n)$ which, for $1<n\leq 8$ allows for an embedding into one of the $E_8$ factor of the gauge group via the sub-group chain $S(U(1)^n)\subset SU(n)\subset E_8$. The low-energy gauge group in this $E_8$ sector is the commutant of the bundle structure group within $E_8$, as usual. For $S(U(1)^n)$ with $n=3,4,5$ this commutant is given by $S(U(1)^3)\times E_6$, $S(U(1)^4)\times SO(10)$ and $S(U(1)^5)\times SU(5)$ respectively and, therefore, contains phenomenologically interesting GUT groups as its non-abelian part.

For consistent heterotic vacua, the gauge bundle needs to satisfy further requirements. First of all, we need to satisfy the supersymmetry conditions~\eqref{HYM}, which remain the same even for the case of a non-integrable complex structure. Since we know the gauge field strengths $F$ on our bundles, as well as the $SU(3)$ structure forms $(J,\Omega)$, these conditions can be checked explicitly and this is what we will do for our examples. It will turn out that both the connection~\eqref{torsionconnection} as well as line bundle sums can satisfy the supersymmetry conditions. In addition, we need to satisfy the integrability condition for the Bianchi identity~\eqref{biancoho} and we now turn to a discussion of this task.

\subsection{Bianchi identity}

We recall from Eq.~\eqref{biancoho} that the integrability condition for the Bianchi identity is given by,
\begin{align}\label{biancoho1}
  \left[{\rm tr}R\wedge R\right]=\left[{\rm tr}F\wedge F\right]+\left[{\rm tr}\tilde F\wedge \tilde F\right] \;,
\end{align}
where the square bracket indicates cohomology classes in $H^4(X)$. Here, $R$ is the curvature two-form of the coset space $X$ and $F$, $\tilde{F}$ are the field strengths in the two $E_8$ sectors corresponding to the observable and hidden bundles $V$ and $\tilde{V}$ respectively. We also remind that $\hat H=0$ for the mirror half-flat solution at zeroth order in $\alpha'$. Moreover, solving the Bianchi identity in cohomology only will produce two-loop contributions to the field equations~\eqref{Einstein} and we expect our solution to get corrected to the next orders. Further analysis is required to ensure having a full consistent solution. Nonetheless, our study consists of an adequate starting point for such investigations. In terms of characteristic classes, Eq.~\eqref{biancoho1} can be written as
\begin{equation}\label{bc}
  p_1\left(TX\right)=2\left({\rm ch}_2(V)+{\rm ch}_2(\tilde{V})\right) \;.
\end{equation}
In practice, we will write those classes as a linear combinations of a basis $\{\tilde{\omega}^r\}$ of $H^4(X)$ dual to our earlier basis $\{\omega_r\}$ of the second cohomology. Of course we can anticipate the use of the basis presented in the previous section --- its subset that is harmonic --- and that is why we use the same notations. We will also need the intersection numbers~\eqref{isec}. This is valid as all the following relations hold in cohomology and, thus, the extra exact parts are irrelevant.

Which choice of gauge bundle should we make in order to satisfy the anomaly condition~\eqref{bc} for a given manifold, that is, for having a given first Pontryagin class on the left-hand side? One obvious attempt would be to set the ``observable" gauge field $F$ equal to the above curvature while choosing the hidden curvature to be trivial. This would obviously satisfy the Bianchi identity~\eqref{bianchi}, not just in cohomology, but point-wise on the coset space for a vanishing three-form $\hat H$. This choice is the analogue of the ``standard embedding" traditionally used in heterotic Calabi-Yau compactifications. In the present context, the problem is that the curvature tensor~\eqref{R} does not satisfy the supersymmetry conditions~\eqref{HYM} required for the gauge fields. Hence, we cannot choose a standard embedding in the conventional sense.

A related choice, somewhat reminiscent, is however possible. We can choose the observable gauge field on the tangent bundle specified by \eqref{torsionconnection} while the hidden gauge field is trivial. This will satisfy the anomaly condition~\eqref{bc} since both the Levi-Civita connection and~\eqref{torsionconnection} provide connections on the same bundle and, hence, will result in the same topological characteristics. Also, as we have mentioned earlier, the gauge field connection defined by~\eqref{torsionconnection} can indeed satisfy the supersymmetry conditions~\eqref{HYM} for our coset spaces, as we will show. Thus, this choice leads to a consistent and supersymmetric vacuum. Nevertheless, since the curvature forms~\eqref{F},~\eqref{torsionconnection} and~\eqref{R} are not the same --- in fact, the former is equal to the first term in the latter --- the right-hand side of the Bianchi identity does not vanish point-wise and a non-zero $\hat H$-field will be required at first order in $\alpha'$. For this reason it might not be appropriate to refer to this choice as ``standard embedding".

In addition, we would like to work with more general gauge fields rather than special choices resembling the standard embedding. Our focus will be on the simplest such class with abelian structure group. This means that the associated vector bundles are sums of line bundles as in Eq.~\eqref{V}. More precisely, we will allow for both an observable bundle $V$ and a hidden bundle $\tilde{V}$ of this kind, that is,
\begin{equation}
  V=\bigoplus_{i=1}^n{\cal O}_X\left({\bf p}_i\right) \;, \quad \tilde{V}=\bigoplus_{j=1}^m{\cal O}_X\left(\tilde{\bf p}_i\right) \;.
 \end{equation}
We demand vanishing first Chern classes $c_1(V)=c_1(\tilde{V})=0$ to allow for an embedding into the two $E_8$ factors. This translates into
\begin{equation}
  \sum_{i=1}^np_i^r=\sum_{j=1}^m\tilde{p}_j^r=0 \;,
\end{equation}
for all $r$. Using additivity of the Chern character and the fact that ${\rm ch}_2(L)=c_1(L)^2/2$ for a line bundle $L$, together with Eq.~\eqref{isec}, we find for the second Chern character ${\rm ch}_{2}(V)={\rm ch}_{2r}(V)\tilde{\omega}^r$ that
\begin{equation}\label{ch2V}
  {\rm ch}_{2r}\left(V\right)=\frac{1}{2}\,\cK_{rst}\sum_{i=1}^np_i^sp_i^t \;,
\end{equation}
and analogously for $\tilde{V}$. With this result, the anomaly condition~\eqref{bc} can be re-written as
\begin{equation}
  \cK_{rst}\left(\sum_{i=1}^np_i^sp_i^t+\sum_{j=1}^m\tilde{p}_j^s\tilde{p}_j^t\right)=p_{1r}\left(TX\right) \;.
\end{equation} 
Again, the use of cohomology classes implies that the Bianchi identity does not vanish point-wise and a non-vanishing $\hat H$-field is required at order $\alpha'$.

\subsection{Index formula}

One of the most basic topological invariant of bundles is the index. It gives the chiral asymmetry of zero modes of the Dirac operator and, hence, the net number of families in the four-dimensional theory. The index can be computed from the Atiyah-Singer index theorem~\cite{Nash:1991pb} which involves the A-roof genus
\begin{equation}
  \hat{A}\left(X\right)=1-\frac{1}{24}\,p_1\left(TX\right)+\dots
\end{equation}
of the manifold $X$. For a bundle $U$ on a six-dimensional manifold $X$, the index theorem takes the form
\begin{equation}\label{AS}
  {\rm ind}\left(U\right)=-\int_X\hat A\left(X\right)\wedge {\rm ch}\left(U\right)=-\int_X\left[{\rm ch}_3\left(U\right)-\frac{1}{24}\,p_1\left(TX\right){\rm ch}_1\left(U\right)\right] \;.
\end{equation}
For a line bundle $L$, we have ${\rm ch}_3(L)=c_1(L)^3/6$ where $c_1(L)=c_1^r(L)\omega_r$ is the first Chern class of $L$. Inserting this into the index formula~\eqref{AS} together with the definition~\eqref{isec} of the intersection numbers, it leads to
\begin{equation}
  {\rm ind}\left(L\right)=-\frac{1}{6}\,\cK_{rst}\,c_1^r\left(L\right)c_1^s\left(L\right)c_1^t\left(L\right)+\frac{1}{24}\,p_{1r}\left(TX\right)c_1^r\left(L\right) \;.
\end{equation}

In the following, we will consider sums of line bundles $V=\bigoplus_{i=1}^nL_i$, where $L_i={\cal O}_X({\bf p}_i)$, with vanishing first Chern class $c_1(V)=0$. For such vector bundles, the above formula simplifies to
\begin{equation}\label{indV}
  {\rm ind}\left(V\right)=-\frac{1}{6}\,\cK_{rst}\sum_{i=1}^np_i^rp_i^sp_i^t \;.
\end{equation} 
Hence, we only need to know the intersection numbers $\cK_{rst}$ of the manifold $X$ together with the integers $p_i^r$ characterizing the line bundles in order to work out the index. For the case when we consider some non-abelian bundles $V$ with vanishing first Chern class, we will use
\begin{equation}\label{index}
  {\rm ind}\left(V\right)=\frac{i}{6\left(2\pi\right)^3}\int_X{\rm tr}\left( F\wedge F\wedge F\right) \;.
\end{equation}  
It corresponds to the expression of the index~\eqref{AS} in terms of the curvature $F$ of $V$ explicitly. We will now use these formulas and the constraints derived previously from the Bianchi identity to some specific cases.

\section{Bundles over coset spaces}

In this section, we apply the above bundle constructions to the three coset spaces introduced earlier. Wherever possible, our focus will be on line bundle sums, although we will discuss some specific non-abelian bundles as well.

\subsection{$SU(3)/U(1)^2$}

Let us first specify some of the required coset properties for this case. The $SU(3)$ generators $\{T_A\}=\{K_a,H_i\}$ are split into the six coset generators $K_a$ given by the non-diagonal Gell-Mann matrices and the two generators $H_i$ of the sub-group $U(1)^2$ given by the two diagonal Gell-Mann matrices. The explicit matrices and the associated structure constants are presented in appendix~\ref{appendixcoset}. The second Betti number of this coset space is two, so we have two basis forms for $\{\omega_r\}$ and $\{\tilde{\omega}^r\}$, with $r=1,2$, for the second and fourth cohomology respectively. They can be choosen such that they are explicitly given by the forms in Eqs.~\eqref{eq_su3basisstart} and~\eqref{eq_su3end} introduced earlier. Let us recall that we have the following intersection numbers,
\begin{equation}\label{su3isec}
  \cK_{111}=6 \;, \quad \cK_{112}=3 \;, \quad \cK_{122}=1 \;, \quad \cK_{222}=0 \;,
\end{equation}
where we only wrote the one that are relevant for the cohomology classes. We also find, for the first Pontryagin class of the tangent bundle,
\begin{equation}
  p_1\left(TX\right)=0 \;.
\end{equation}  
This can be calculated by using explicitly the Riemann curvature two-from which can be computed from~\eqref{R}.

We start with discussing the possible non-abelian bundles. Using the explicit structure constants from appendix~\ref{appendixcoset}, we can verify that the Levi-Civita curvature~\eqref{omega} does not satisfy the supersymmetry equations~\eqref{HYM} and, hence, cannot be used as a gauge curvature. Let us consider the supersymmetry conditions for associated bundles specified by representations $\rho$ of the sub-group $H$ as introduced in sub-section~\ref{assbundles}. First, it can be checked that the  constraint $\Omega\,\neg\, F=0$ is always trivially satisfied. The constraint $J\,\neg\, F=0$ implies,
\begin{equation}
  J^{ab}f_{ab}^{\phantom{ab}i}\rho\left(H_i\right)=\left(\frac{2}{R_1^2}-\frac{1}{R_2^2}-\frac{1}{R_3^2}\right)\rho\left(H_7\right)+\left(\frac{\sqrt{3}}{R_3^2}-\frac{\sqrt{3}}{R_2^2}\right)\rho\left(H_8\right)=0 \;.
\end{equation}
In general, the two representation matrices are linearly independent and, so, we have two constraints on the moduli which are solved by
\begin{equation}\label{equalR}
  R_1^2=R_2^2=R_3^2\equiv R^2 \;.
\end{equation}
Consequently, all associated bundles are supersymmetric on the nearly-Kahler locus of the moduli space. In particular, this applies to the connection~\eqref{torsionconnection}. However, from Eq.~\eqref{index}, its index vanishes as one would expect for an associated vector bundle which corresponds to a real representation of the group $H$. Thus, it is not of particular interest from a physics point of view. 

Associated bundles which correspond to irreducible representations of the sub-group $H$ can be viewed as ``building blocks" for general associated bundles. In the present case, the sub-group $H=U(1)^2$ is abelian so that all irreducible representations are one-dimensional and, as a consequence, lead to line bundles. We characterize an irreducible representation $\rho$ by a pair of integer charges $(p,q)$ and, more specifically, define the representation by
\begin{equation}
  \rho\left(H_7\right)=-i\left(p+\frac{q}{2}\right) \;, \quad \rho\left(H_8\right)=-i\frac{q}{2\sqrt{3}} \;.
\end{equation}
We see from Eq.~\eqref{F} that the associated curvature form is given by
\begin{equation}
  \frac{F}{2\pi}=-ip\,\omega_1-iq\,\omega_2 \;,
\end{equation}
and the first Chern class of the associated line bundle $L$ is $c_1(L)=p\omega_1+q\omega_2$. From our earlier discussion it means that $L$ should be identified with ${\cal O}_X(p,q)$. Taking the observable and hidden bundles $V$ and $\tilde{V}$ as sums of line bundles with vanishing first Chern class, we can write
\begin{equation}\label{Vsu3}
  V=\bigoplus_{i}{\cal O}_X\left(p_i,q_i\right) \;, \quad \tilde{V}=\bigoplus_{j}{\cal O}_X\left(\tilde{p}_j,\tilde{q}_j\right) \;,
\end{equation}
where
\begin{equation}\label{su3c10}
  \sum_{i}^np_i=\sum_{j}^nq_j=0 \;,
\end{equation} 
and similarly for $\tilde{p}_i$ and $\tilde{q}_i$. As a result, each such sum of $n$ line bundles is determined by the $2n$ integers $p_i$ and $q_i$ subject to the constraints~\eqref{su3c10}. For the second Chern character, we find from Eqs.~\eqref{ch2V} and~\eqref{su3isec} relative to the basis $\{\tilde{\omega}^1,\tilde{\omega}^2\}$,
\begin{equation}
  {\rm ch}_2\left(V\right)=\left(\sum_{i}\left(3p_i^2+\frac{q_i^2}{2}+3p_iq_i\right),\sum_{i}\left(p_iq_i+\frac{3p_i^2}{2}\right)\right) \;,
\end{equation}
and similarly for $\tilde{V}$. Analogously, Eqs.~\eqref{indV} and~\eqref{su3isec} lead to the expression
\begin{equation}\label{indVsu3}
  {\rm ind}\left(V\right)=-\sum_i\left(p_i^3+\frac{1}{2}\,p_iq_i\left(q_i+3p_i\right)\right) \;,
\end{equation}
for the index of $V$. Again, the supersymmetry equations~\eqref{HYM} can be solved by the constraints,
\begin{equation}
  R_1^2=R_2^2=R_3^2\equiv R^2 \;.
\end{equation}
From the above result for the second Chern character --- remembering that the first Pontryagin class for this coset vanishes --- the two components of the anomaly cancellation condition~\eqref{bc} can be written as
\begin{align}\label{secondchernzero1}
  \sum_i\left(3p_i^2+\frac{q_i^2}{2}+3p_iq_i\right)+\sum_j\left(3\tilde p_j^2+\frac{\tilde q_j^2}{2}+3\tilde p_j\tilde q_j\right)=0 \;,\\\label{secondchernzero2}
  \quad \sum_i\left(p_iq_i+\frac{3p_i^2}{2}\right)+\sum_j\left(\tilde p_j\tilde q_j+\frac{3\tilde p_j^2}{2}\right)=0 \;.
\end{align}
We have now collected all results required for basic model building on this coset. The problem is to choose observable bundles $V$ with ${\rm rk}(V)=3,4,5$ specified by integers $p_i$, $q_i$ and corresponding hidden bundles $\tilde{V}$ with ${\rm rk}(\tilde{V})=2,\ldots ,8$ specified by integers $\tilde{p}_j$, $\tilde{q}_j$ subject to the following constraints:
\begin{itemize}
\item The first Chern classes of $V$ and $\tilde{V}$ vanish, that is, Eqs.~\eqref{su3c10} are satisfied.
\item The anomaly conditions~\eqref{secondchernzero1} and~\eqref{secondchernzero2} are satisfied.
\item The index~\eqref{indVsu3} of the observable bundle $V$ equals three to obtain a GUT model with three net families.
\end{itemize}
It is clear that there are many possible solutions to these constraints and we present a sample of examples in table~\ref{tab1}.

Let us conclude this sub-section on bundles over $SU(3)/U(1)^2$ by a remark on the quasi standard embedding discussed previously. We have seen earlier that the torsion connection~\eqref{torsionconnection} on the tangent bundle, while supersymmetric, has a vanishing index since it is associated to a real representation. It was, therefore, not suitable as a ``standard embedding". A related complex representation can be defined by considering $H$ as a sub-group of $SU(3)$ and by choosing the representation $\rho$ induced by the fundamental representation of $SU(3)$. This means setting 
\begin{equation}\label{sesu3}
  \rho\left(H_7\right)=\sqrt{3}\,\lambda_8 \;, \quad \rho\left(H_8\right)=\sqrt{3}\,\lambda_3 \;.
\end{equation}
The associated bundle for this representation has rank three and is, in fact, a sum of three line bundles. It turns out that it corresponds to the  example in the first row of table~\ref{tab1}. For this choice, the anomaly condition is satisfied for a trivial hidden bundle and the chiral asymmetry equals half the Euler number, three in this case. Thus, this bundle has two of the main characteristics of the standard embedding. Note, however, that it does not lead to a vanishing right-hand side of the Bianchi identity~\eqref{bianchi} and, therefore, the model receives corrections at order $\alpha'$.
\clearemptydoublepage

\begin{table}[p]
\begin{center}
\begin{tabular}{|c|c|c|c|c|} 
  \hline
  $\;n\;$ &$p_i$ & $q_i$ & $\tilde p_i$ & $\tilde q_i$\\ \hline \hline
  3&$\left(-1,-1,2\right)$&$\left(0,3,-3\right)$&$\left(0\right)$&$\left(0\right)$\\
  3&$\left(-3,0,3\right)$&$\left(3,1,-4\right)$&$\left(2,1,0,-1,-2\right)$&$\left(-4,-1,-3,4,4\right)$\\
  3&$\left(-3,0,3\right)$&$\left(3,2,-5\right)$&$\left(3,0,0,-1,-2\right)$&$\left(-4,-1,-3,4,4\right)$\\
  3&$\left(-2,-1,3\right)$&$\left(1,2,-3\right)$&$\left(2,1,0,-1,-2\right)$&$\left(-4,-3,-1,4,4\right)$\\
  3&$\left(-2,-1,3\right)$&$\left(2,2,-4\right)$&$\left(2,0,-1,-1\right)$&$\left(-3,-3,3,3\right)$\\
  3&$\left(-2,-1,3\right)$&$\left(4,-1,-3\right)$&$\left(2,1,0,-1,-2\right)$&$\left(-3,-4,1,2,4\right)$\\
  3&$\left(-2,-1,3\right)$&$\left(4,1,-5\right)$&$\left(3,0,0,-1,-2\right)$&$\left(-4,3,-3,2,2\right)$\\
  3&$\left(-2,0,2\right)$&$\left(1,1,-2\right)$&$\left(2,1,-1,-1,-1\right)$&$\left(-4,-3,3,2,2\right)$\\
  3&$\left(-2,0,2\right)$&$\left(1,3,-4\right)$&$\left(3,1,-1,-1,-2\right)$&$\left(-4,-3,2,1,4\right)$\\
  3&$\left(-1,0,1\right)$&$\left(-1,2,-1\right)$&$\left(2,2,-1,-1,-2\right)$&$\left(-3,-4,2,2,3\right)$\\
  3&$\left(-1,0,1\right)$&$\left(-1,3,-2\right)$&$\left(3,2,-1,-2,-2\right)$&$\left(-4,-4,1,4,3\right)$\\
  3&$\left(-1,0,1\right)$&$\left(4,-2,-2\right)$&$\left(3,1,-2,-2\right)$&$\left(-3,-3,3,3\right)$\\ \hline
  4&$\left(-3,-1,2,2\right)$&$\left(3,3,-3,-3\right)$&$\left(2,0,-1,-1\right)$&$\left(-3,-3,3,3\right)$\\
  4&$\left(-2,-1,1,2\right)$&$\left(1,2,-1,-2\right)$&$\left(1,1,1,-1,-2\right)$&$\left(-1,-3,-3,3,4\right)$\\
  4&$\left(-2,-1,1,2\right)$&$\left(1,2,0,-3\right)$&$\left(2,1,0,-1,-2\right)$&$\left(-3,-4,1,2,4\right)$\\
  4&$\left(-2,0,1,1\right)$&$\left(1,1,-1,-1\right)$&$\left(2,2,-1,-1,-2\right)$&$\left(-4,-4,4,2,2\right)$\\
  4&$\left(-2,0,1,1\right)$&$\left(1,2,-2,-1\right)$&$\left(1,1,1,-1,-2\right)$&$\left(-2,-2,-2,2,4\right)$\\
  4&$\left(-1,-1,-1,3\right)$&$\left(-1,2,2,-3\right)$&$\left(2,1,-1,-1,-1\right)$&$\left(-4,-3,3,2,2\right)$\\
  4&$\left(-1,-1,0,2\right)$&$\left(-1,2,1,-2\right)$&$\left(3,1,-1,-1,-2\right)$&$\left(-4,-4,2,2,4\right)$\\
  4&$\left(-1,-1,0,2\right)$&$\left(0,0,3,-3\right)$&$\left(2,2,-1,-1,-2\right)$&$\left(-3,-4,2,2,3\right)$\\
  4&$\left(-1,-1,0,2\right)$&$\left(3,3,-3,-3\right)$&$\left(3,1,-2,-2\right)$&$\left(-3,-3,3,3\right)$\\
  4&$\left(-1,-1,1,1\right)$&$\left(-1,3,-2,0\right)$&$\left(3,1,-1,-1,-2\right)$&$\left(-4,-3,2,1,4\right)$\\
  4&$\left(-1,0,0,1\right)$&$\left(-1,1,1,-1\right)$&$\left(2,1,1,-2,-2\right)$&$\left(-3,-1,-3,4,3\right)$\\
  4&$\left(-1,0,0,1\right)$&$\left(-1,1,2,-2\right)$&$\left(3,1,-2,-2\right)$&$\left(-4,-3,4,3\right)$\\ \hline
  5&$\left(-2,-1,-1,1,3\right)$&$\left(1,2,3,-3,-3\right)$&$\left(2,1,0,-1,-2\right)$&$\left(-3,-4,1,2,4\right)$\\
  5&$\left(-2,-1,-1,2,2\right)$&$\left(2,1,3,-3,-3\right)$&$\left(3,1,0,-2,-2\right)$&$\left(-4,0,-4,4,4\right)$\\
  5&$\left(-2,-1,1,1,1\right)$&$\left(1,2,-1,-1,-1\right)$&$\left(2,1,0,-1,-2\right)$&$\left(-3,-4,1,2,4\right)$\\
  5&$\left(-1,-1,0,0,2\right)$&$\left(0,3,-2,2,-3\right)$&$\left(3,2,-1,-1,-3\right)$&$\left(-4,-4,4,0,4\right)$\\
  5&$\left(-1,-1,0,0,2\right)$&$\left(0,3,-1,1,-3\right)$&$\left(3,1,-1,-1,-2\right)$&$\left(-3,-4,2,1,4\right)$\\
  5&$\left(-1,-1,0,0,2\right)$&$\left(3,3,-2,-1,-3\right)$&$\left(4,1,-1,-2,-2\right)$&$\left(-4,-4,0,4,4\right)$\\
  5&$\left(-1,-1,0,1,1\right)$&$\left(-1,3,2,-3,-1\right)$&$\left(3,2,-1,-2,-2\right)$&$\left(-4,-4,3,3,2\right)$\\
  5&$\left(-1,-1,0,1,1\right)$&$\left(-1,3,3,-3,-2\right)$&$\left(3,2,0,-2,-3\right)$&$\left(-4,-3,-1,4,4\right)$\\
  5&$\left(-1,0,0,0,1\right)$&$\left(-1,-1,2,2,-2\right)$&$\left(3,2,-1,-2,-2\right)$&$\left(-4,-4,1,4,3\right)$\\
  5&$\left(-1,0,0,0,1\right)$&$\left(-1,1,1,1,-2\right)$&$\left(-3,-1,1,1,2\right)$&$\left(4,3,-2,-1,-4\right)$\\
  5&$\left(-2,0,0,0,2\right)$&$\left(1,-2,1,2,-2\right)$&$\left(2,2,0,-2,-2\right)$&$\left(-3,-4,-1,4,4\right)$\\
  5&$\left(-1,-1,-1,1,2\right)$&$\left(-1,2,2,-1,-2\right)$&$\left(1,1,1,-1,-2\right)$&$\left(-1,-2,-3,2,4\right)$\\ \hline
\end{tabular}
\parbox{6in}{\caption{\it\small Sample of bundles over the base space $SU(3)/U(1)^2$ leading to models with four-dimensional GUT group and three generations. Observable and hidden bundles are specified by the integers $(p_i,q_i)$ and $(\tilde p_i, \tilde q_i)$ respectively. The ranks of the visible bundles are $n$ and the ranks of the hidden bundles are arbitrarily either four or five.}\label{tab1}}
\end{center}
\end{table}
\clearemptydoublepage

\subsection{$Sp(2)/SU(2)\times U(1)$}

The generators $\{T_A\}=\{K_a,H_i\}$ of $Sp(2)$ consist of six coset generators $K_a$ and the four generators $H_i$ of the sub-group $SU(2)\times U(1)$. The explicit matrices and associated structure constants are listed in appendix~\ref{appendixcoset}. The second Betti number of this manifold is one and the second and fourth cohomology groups are spanned by the forms $\omega_1$ and $\tilde{\omega}^1$ given in Eq.~\eqref{eq_sp2basisstart}. We also recall the intersection number
\begin{equation}
  \cK_{111}=1 \;,
\end{equation}
which is the only one relevant for the cohomology classes. The first Pontryagin class is found to be
\be
  p_1\left(TX\right)=4\,\tilde{\omega}^1 \;,
\ee
and is again calculated directly from the Riemann tensor formula~\eqref{R}.

As for the $SU(3)$ case, we can verify from the structure constants given in appendix~\ref{appendixcoset} that the Levi-Civita curvature does not satisfy the supersymmetry conditions. For associated bundles with representation $\rho$, the constraint $\Omega\,\neg\, F=0$ is trivially satisfied while the constraint $J\,\neg\, F=0$ implies
\begin{equation}
  J^{ab}f_{ab}^{\phantom{ab}i}\rho\left(H_i\right)=\left(\frac{4}{R_1^2}-\frac{4}{R_2^2}\right)\rho\left(H_{10}\right)=0 \;.
\end{equation}
This is solved in the region of moduli space where
\begin{equation}
  R_1^2=R_2^2\equiv R^2 \;.
\end{equation}
Therefore, all associated bundles satisfy the supersymmetry conditions in the nearly-Kahler locus of the moduli space. In particular, this applies to the connection~\eqref{torsionconnection}. However, as before, it has a vanishing index and turns out to be of limited interest.

Line bundles $L={\cal O}_X(p)$ are characterized by a single integer $p$. They can be constructed as associated bundles by choosing representations $\rho$ of the sub-group $SU(2)\times U(1)$ which are trivial on the $SU(2)$ part and have $U(1)$ charge $p$. Explicitly, this means
\begin{equation}
  \rho\left(H_7\right)=0 \;, \quad \rho\left(H_8\right)=0 \;, \quad \rho\left(H_9\right)=0 \;, \quad \rho\left(H_{10}\right)=ip \;.
\end{equation}
The associated field strength is $F/(2\pi)=-ip\omega_1$ which shows that the associated bundle has first Chern class $c_1(L)=p\omega_1$ and should indeed be identified with ${\cal O}_X(p)$. As before, the observable and hidden bundles $V$ and $\tilde{V}$ are taken as line bundle sums with vanishing first Chern class, thus
\begin{equation}\label{Vsp2}
  V=\sum_i{\cal O}_X\left(p_i\right) \;, \quad \tilde{V}=\sum_j{\cal O}_X\left(\tilde{p}_j\right) \;, \quad \sum_ip_i=\sum_j\tilde{p}_j=0 \;.
\end{equation} 
We find for the second Chern character and the index
\begin{equation}\label{indsp2}
  {\rm ch}_2\left(V\right)=\frac{1}{2}\sum_ip_i^2\tilde{\omega}^1 \;, \quad {\rm ind}\left(V\right)=-\frac{1}{6}\sum_ip_i^3 \;,
\end{equation}
and similarly for $\tilde{V}$. The anomaly condition then reads
\begin{equation}\label{anomsp2}
  \sum_ip_i^2+\sum_j\tilde p_j^2=4 \;.
\end{equation}
We should now study the model building options in analogy to what we did for $SU(3)/U(1)^2$. We need to choose bundles $V$ and $\tilde{V}$ specified by integers $p_i$ and $\tilde{p}_j$ as in~\eqref{Vsp2} which satisfy the anomaly condition~\eqref{anomsp2} and lead to an index~\eqref{indsp2} of three in order to obtain three chiral GUT families. However, unlike for the previous case, the combination of these conditions is quite restrictive. A quick look over all integers $p_i$ for ${\rm rk}(V)=3,4,5$ and all integers $\tilde{p}_j$ shows that there is only one solution which satisfies the anomaly condition. It is given by the rank four observable bundle,
\begin{equation}
  \left(p_i\right)=\left(1,1,-1,-1\right) \;,
\end{equation} 
and a trivial hidden bundle. Unfortunately, this model has vanishing index and so is not of physical interest. 

Finally, let us work out the quasi standard-embedding analogue to~\eqref{sesu3} for the present case. We recall that this is done by choosing the representation $\rho$ which is induced by the fundamental of $SU(3)$ via the embedding $SU(2)\times U(1)\subset SU(3)$. This means explicitly
\begin{equation}\label{seg2}
  \rho\left(H_7\right)=-2\lambda_1 \;, \quad \rho\left(H_8\right)=-2\lambda_2 \;, \quad \rho\left(H_9\right)=-2\lambda_3 \;, \quad \rho\left(H_{10}\right)=-2\sqrt{3}\,\lambda_8 \;,
\end{equation}
where $\lambda_i$ are the Gell-Mann matrices as given in appendix~\ref{gellmann}. This choice satisfies the anomaly condition for a trivial hidden bundle. From Eq.~\eqref{index}, we can calculate the index explicitly and we find two chiral families which equals half the Euler number as expected.

\subsection{$G_2/SU(3)$}

Last, let us look at the $G_2/SU(3)$ case. The generators $\{T_A\}=\{K_a,H_i\}$ of $G_2$ consist of the six coset generators $K_a$ and the eight generators $H_i$ of the sub-group $SU(3)$. The explicit matrices and structure constants are given in appendix~\ref{appendixcoset} as usual. The second Betti number of this coset vanishes so, unfortunately, there are no non-trivial line bundles. Hence, we have to consider non-abelian gauge fields in this case.

The Levi-Civita connection and the torsion connection~\eqref{torsionconnection} on the tangent bundle have the same properties as for the two previous cases. The former does not satisfy the supersymmetry conditions while the latter does but has a vanishing index.

The quasi standard embedding similar to~\eqref{sesu3} and~\eqref{seg2} is obtained here by choosing $\rho$ to be the fundamental representation of the $SU(3)$ sub-group. In practice, this means setting
\begin{equation}
  \rho\left(H_i\right)=-2\lambda_{i-6} \;.
\end{equation}
Note that, by our conventions, the index $i$ numbering the sub-group generators $H_i$ runs over the range $7,\ldots ,14$. The anomaly condition is satisfied with a trivial hidden bundle  and the number of generations is one and corresponds to half the Euler number as is expected from the standard embedding properties.
\clearemptydoublepage
\chapter{Conclusion and outlook}\label{chapterconclusion}

We have studied $10$-dimensional vacuum solutions of the heterotic string. In the prospect of discovering new classes of solutions, we concentrated on backgrounds preserving only two supercharges out of the original sixteen from the ten-dimensional heterotic supergravity. We also restricted ourselves to six internal compact dimensions. Thus, we studied geometries involving a warped product of a four-dimensional domain wall with a six-dimensional internal space as a general setting for flux compactifications on manifolds with $SU(3)$ structure. This allows more general classes of compactifications than for the standardly studied $N=1$ cases. In particular, the internal manifolds do not need to be complex anymore and their torsion classes $\cW_1^+$ and $\cW_2^+$ can be non-vanishing.\\

For the special case with vanishing flux and constant dilaton, the solution is a direct product of a $2+1$-dimensional domain wall world volume and a seven-dimensional manifold with $G_2$ holonomy. In turn, this $G_2$ manifold consists of a six-dimensional half-flat manifold varying along some direction $y$, transverse to the domain wall, as specified by Hitchin's flow equations. We have shown that these $10$-dimensional solutions form the basis for compactifications on half-flat mirror manifolds without flux as carried out in Ref.~\cite{Gurrieri:2004dt}. Specifically, we have verified that the BPS domain walls of the four-dimensional $N=1$ supergravity theories associated to these compactifications precisely lift up to our $10$-dimensional solutions. 

We have further generalized this picture to include non-vanishing flux and a non-constant dilaton. In this case, the $10$-dimensional space is still a direct product between the $2+1$-dimensional domain wall world volume and a seven-dimensional space. However, this seven-dimensional manifold now has $G_2$ structure rather than $G_2$ holonomy. As before, it can be thought of as the dependence of a six-dimensional manifold along the direction $y$ where the variation is described by a generalized version of Hitchin's flow equations. The torsion classes of the allowed spaces are constrained by the relations given in Eq.~\eqref{tc}. In particular, they imply that the six-dimensional manifolds are generalized half-flat and almost complex. Compared to Strominger's original class of complex non-Kahler manifolds, this opens up many more possibilities. In particular, flux compactifications on half-flat mirror manifolds are based on these solutions.

Moreover, we have also obtained a class of solutions consisting of an exact Calabi-Yau three-fold with NS-NS flux which varies in its moduli space as one moves along the direction $y$. For the case of purely electric NS-NS flux, they are the natural candidates ``mirrors'' of the solutions based on $G_2$ holonomy manifolds. This is analogous to the original type II mirror symmetry correspondence with NS-NS flux~\cite{Gurrieri:2002wz}.\\

In order to gain a better understanding of the gauge field sector in heterotic half-flat compactifications, we then turned on to finding explicit examples of the aforementioned scenario, focusing on the case of mirror-half flat manifolds with vanishing flux. For this purpose, we have studied the compactification of heterotic string on six-dimensional coset spaces $G/H$ with center of attention on the three manifolds $SU(3)/U(1)^2$, $Sp(2)/SU(2)\times U(1)$ and $G_2/SU(3)$. These spaces are half-flat and they solve the gravitational sector of the theory from the general results obtained in the context of heterotic domain wall vacua.

The group origin of the coset spaces facilitates the construction of gauge bundles and the computation of explicit connections on them. The supergravity equations can, therefore, be checked directly. Specifically, for each representation of the sub-group $H$, one has a vector bundle associated to the principal bundle $G=G(G/H,H)$. For the case $SU(3)/U(1)^2$, the irreducible representations of the sub-group $H=U(1)^2$ lead to line bundles and, in fact, all line bundles on this coset can be obtained in this way. Since the second Betti number of this space is two, these line bundles are characterized by two integers that correspond to the two charges of $U(1)^2$. The situation is analogous for $Sp(2)/SU(2)\times U(1)$. Taking the $SU(2)$ representation to be trivial, the $U(1)$ representations, specified by a single charge, lead to a one-integer family of line bundles in accordance with $b^2=1$ for this space. The second Betti number of $G_2/SU(3)$ vanishes so there are no non-trivial line bundles on this manifold as, indeed, there are no non-trivial one-dimensional representation of $H=SU(3)$. Of course, we can also consider higher-dimensional representations and we have presented some examples. One possible choice is the ``fundamental" representation of $H$, that is the representation induced by the fundamental of $SU(3)\supset H$. It turns out that this choice, for all three coset spaces, leads to a quasi standard embedding where the anomaly condition is satisfied for a trivial hidden bundle and the chiral asymmetry is given by half the Euler number. 

For the first two coset spaces, we have also shown that consistent vacua can be obtained by suitable sums of line bundles in the observable and hidden sector. The $Sp(2)/SU(2)\times U(1)$ case, where line bundles are labeled by only one integer, is quite restrictive and we have been able to find only one consistent model, unfortunately with a vanishing chiral asymmetry. The $SU(3)/U(1)^2$ case, however, allows for many consistent solutions with line bundles and we have presented a number of explicit examples with chiral asymmetry three and four-dimensional GUT group.\\

Our results open up new possibilities for heterotic string model building and they put heterotic half-flat compactifications on a more solid theoretical basis. It would be interesting to generalize the result by breaking more supersymmetry and, thus, keeping only one supercharge. This would lead to a $Spin(7)$ manifold times a string soliton and could then be compared to the corresponding quarter-BPS string state of the four-dimensional effective supergravity. If the solutions do lift up correctly, it would increase the class of $SU(3)$ structure manifolds suitable for heterotic compactification. It would also be interesting to study the lift of our solutions to heterotic M-theory~\cite{Horava:1996ma, Witten:1996mz, Lukas:1998yy}.

Another promising direction of research could be to look for constructions of more solutions. For example, mirror half-flat manifolds constitute a very large set: one such manifold is obtained for each Calabi-Yau three-fold --- with a mirror --- and a choice of electric NS flux. It would be very interesting to find an explicit mathematical construction for them, maybe inspired by mirror symmetry, as it would provide a large category of solutions.

Furthermore, we could also study more general non-abelian bundle constructions over $SU(3)/U(1)^2$, and possibly over $Sp(2)/SU(2)\times U(1)$ as well, based, for instance, on quotients or extensions of line bundle sums. Recently, a new class of $SU(3)$ structure manifolds which might be suitable for heterotic compactifications has been found using methods in toric geometry~\cite{Larfors:2011zz, Larfors:2010wb}. It might also be interesting to study bundles on this new classes.

Finally, model building within our set of bundles over $SU(3)/U(1)^2$ could be studied more systematically. An exhaustive list could be produced with a deeper scan of the Chern class parameters and, maybe, leading to the correct group theoretical data for the effective four-dimensional supergravity. The inclusion of non-perturbative effects and next orders in $\alpha'$ would then be required in order to lift the domain wall back to a maximally symmetric space.

\newpage\thispagestyle{empty}\be\ph{page blanche}\nn\ee
\clearemptydoublepage

\appendix
\chapter{Conventions and notations}\label{appendixconventions}
\section{Indices and differential geometry}

We will make use of several indices throughout the thesis and we would like to summarize in this section the different conventions adopted. The ten-dimensional space-time background geometry $M_{10}$ decomposes into a three-dimensional space $M_3$, a special direction $y$ (represented by an interval $I$) and an internal compact $6$-dimensional manifold $X$,
\be\label{decomposition}
  M_{10}=M_3\times I\times X \;.
\ee
We will alternatively see this geometry as a $3+7$ decomposition where we consider the $7$-dimensional space $Y=I\times X$ or as a $4+6$ decomposition where we consider the four-dimensional space $M_3\times I$. For this reason, we make use of the following sets of indices with their corresponding range:
\be\label{indices}
\begin{aligned}
  10d: & \quad M,N,P,...=0,1,...\,,9 \\
  7d: & \quad m,n,p,...=3,4,...\,,9 \\
  6d: & \quad a,b,c,...=4,5,...\,,9 \\
  4d: & \quad \mu,\nu,...=0,1,2,3 \\
  3d: & \quad \alpha,\beta,...=0,1,2 \\
  1d: & \quad M=\mu=m=3 \;.
\end{aligned}
\ee
When we will consider flat indices (from the vielbein basis), we will use the same letters and underline them: $\underline{m},\underline{a},$ etc.

Furthermore, for the case of the internal $6$-dimensional geometry being coset space, we use labels according to the Lie group it is referring to. For a Lie group $G$ and a subgroup $H \subset G$, we make use of the indices:
\be\label{cosetindices}
\begin{aligned}
  G: & \quad A,...=1,2,...\,,{\rm dim}\left(G\right) \\
  G/H: & \quad a,b,...=1,2,...\,,6 \\
  H: & \quad i,j,...=7,8,...\,,{\rm dim}\left(G\right) \;.
\end{aligned}
\ee
Evidently, the indices $a,b,...$ label the internal six-dimensional geometry and are associated to the corresponding coordinates in~\eqref{indices}. However, for the sake of clarity, the numbering ranges from $1,...,6$ in the context of Lie groups rather than $4,...,9$ as above. Which one is meant should be clear from the context. (Moreover, let us point out that the same letters $\{A,a,i,..\}$ will be used as well for the Kahler and complex structure modulii. However, we believe that the context in which they are used is clear enough to avoid any confusion.)\\

Let us now turn to our differential geometry notation. We mainly use the conventions of Ref.~\cite{Nakahara:2003nw}. First, it is useful to introduce the square bracket to denote full anti-symmetrization of indices. For any set of $r$ indices $\{i_1,...,i_r\}$, we have
\be
  [i_1...\,i_r]\equiv \frac{1}{r!}\sum_{\sigma\in S_p}{\rm sgn}\left(\sigma\right)\sigma\left(i_1\right)...\,\sigma\left(i_r\right) \;,
\ee
where $S_p$ is the group of permutations. Respectively, normal brackets are used for symmetrization of indices.
A general differential $r$-form is written as
\be\label{form}
  \omega=\frac{1}{r!}\,\omega_{i_1...\,i_r}dx^{i_1}\wedge...\wedge dx^{i_r} \;,
\ee
where $dx^{i}$ is used for the cotangent basis and $\omega_{i_1...i_r}=\omega_{[i_1...i_r]}$ is the anti-symmetric tensor corresponding to the form components. When the wedge product of a list of $r$ one-forms $e^a$ is to be taken, we simply write it as a list of indices,
\be
  e^{a_1...\,a_r}\equiv e^{a_1}\wedge...\wedge e^{a_r} \;,
\ee
to avoid lengthy notations. The Hodge star dual is defined by,
\be
  *\omega=\frac{\sqrt{|g|}}{r!\left(d-r\right)!}\,\omega_{i_1...\,i_r}\epsilon^{i_1...\,i_r}_{\ph{i_1...\,i_r}i_{r+1}...\,i_d}dx^{i_{r+1}}\wedge...\wedge dx^{i_d} \;,
\ee
where $d$ is the dimension of the space in which the $*$ is operating, $|g|$ is the determinant of the metric $g$ and $\epsilon$ is the Levi-Civita tensor whose indices are raised (lowered) with the metric. Sometimes, to avoid confusion, we write $*_d$ with the dimensionality $d$ explicitly as an index. The star operator allows us to define the scalar product for two $r$-forms,
\be\label{formproduct}
  \int_M\alpha\wedge *\bar\beta=\frac{1}{r!}\int_M d\cV\, \alpha_{i_1...\,i_r}\bar\beta^{i_1...\,i_r} \;,
\ee
where the bar $\bar\beta$ is for the complex conjugation (when both forms are real, it is redundant and can be ignored). We have also defined the infinitesimal volume element,
\be
  d\cV\equiv\sqrt{|g|}\,dx^{1}\wedge...\wedge dx^{d}=*1 \;.
\ee
Finally, we are making use of the interior product,
\be
  \left(\alpha\neg\beta\right)_{i_1...\,i_q}\equiv\frac{1}{r!}\,\alpha^{j_1...\,j_r}\beta_{j_1...\,j_ri_1...\,i_q} \;,
\ee
between an $r$-form $\alpha$ and a $q+r$-form $\beta$. It is the adjoint of the wedge product with respect to the scalar product~\eqref{formproduct} and it leads to the relation
\be\label{starwedge}
  *\left(\alpha\wedge\beta\right)=\left(-1\right)^{q+r}\alpha\neg *\beta \;.
\ee

Let us also introduce a small reminder about cohomology groups (based on~\cite{Candelas:1987is}). First, we have the result of Poincar\'e duality stating that for a $p$-cycle $a$ on a manifold $M$ of dimension $d$, there exist a $(d-p)$-form $\alpha$ (the Poincar\'e dual of $a$) such that
\be
  \int_a\omega=\int_M\alpha\wedge\omega \;,
\ee
for any closed form $\omega$. Then, let us consider a basis $\{z^j\}$ for the simplicial homology group $H_p$ of $p$-cycles and a basis $\{\omega_i\}$ for the de Rahm cohomology group $H^p$ of $p$-forms. De Rahm's theorem states that, with the correct choice of basis, we can have the normalization
\be
  \int_{z^j}\omega_i=\delta^{\ph{i}j}_i \;.
\ee
These two results together imply that we can find a basis $\{\omega_i\}$ for $H^p$ and $\{\omega^j\}$ for $H^{d-p}$ having the property,
\be\label{cohobasis}
  \int_M\omega_i\wedge\tilde\omega^j=\delta^{\ph{i}j}_i \;.
\ee
For the case of $6$-dimensional manifolds, we are considering basis $\{\omega_j\}$ of $H^2$ and $\{\tilde\omega^i\}$ of $H^4$. We define the triple intersection numbers as
\be
  \cK_{ijk}\equiv\int_M \omega_i\wedge \omega_j\wedge \omega_k \;.
\ee
Consequently, it follows from~\eqref{cohobasis} that the two basis are related by
\be\label{intercoho}
  \omega_i\wedge\omega_j=\cK_{ijk}\,\tilde\omega^k \;.
\ee
We will also consider extended basis including not only elements of cohomology groups but with more elements which are not closed. When this is the case, we will still choose them satisfying properties~\eqref{cohobasis} and~\eqref{intercoho}.

To conclude this section on differential geometry, we would like to list a couple of formulas about Chern classes~\cite{Nash:1991pb, Eguchi:1980jx}. The Chern character ${\rm ch}(V)$ of a vector bundle $V$ can be written in term of the bundle curvature $F$,
\be
  {\rm ch}\left(V\right)={\rm tr}\,\exp\frac{iF}{2\pi} \;.
\ee
For $6$-dimensional manifolds, this leads to
\be
  {\rm ch}\left(V\right)=k+c_1+\frac{1}{2}\left(c_1^2-2c_2\right)+\frac{1}{6}\left(c_1^3-3c_1c_2+3c_3\right) \;,
\ee
where $k$ is the fiber dimension of the bundle and the Chern classes are given by,
\be
\begin{aligned}
  c_0&=1 \\
  c_1&=\frac{i}{2\pi}{\rm tr}F \\
  c_2&=\frac{1}{2}\frac{1}{(2\pi)^2}({\rm tr}F\wedge F-{\rm tr}F\wedge {\rm tr}F) \\
  c_3&=\frac{i}{6}\frac{1}{(2\pi)^3}(-2{\rm tr}F\wedge F\wedge F+3{\rm tr}F\wedge F\wedge {\rm tr}F-{\rm tr}F\wedge {\rm tr}F\wedge {\rm tr}F) \;.
\end{aligned}
\ee

\section{Gamma matrices and spinor decomposition}

The $10$-dimensional gamma matrices are $32\times32$ matrices which we choose to be purely imaginary. They satisfy the usual Clifford algebra,
\begin{equation}
  \left\{\Gamma^{M},\Gamma^{N}\right\}=2g^{MN}\cdot\mathds{1}_{32} \;.
\end{equation}
The chirality operator is given by,
\begin{equation}
  \Gamma^{11}=\Gamma^{0}\Gamma^{1}...\,\Gamma^{9} \;.
\end{equation}
Furthermore, we need to choose a basis which splits in concordance with the geometry decomposition~\eqref{decomposition}. It is well known that we can build the gamma matrices as tensor products of Pauli matrices. Let us write the latter as,
\be
  \sigma^1=\left(\begin{matrix} 0 & 1 \\ 1 & 0 \end{matrix}\right) \;,\quad
  \sigma^2=\left(\begin{matrix} 0 & -i \\ i & 0 \end{matrix}\right) \;,\quad
  \sigma^3=\left(\begin{matrix} 1 & 0 \\ 0 & -1 \end{matrix}\right) \;.
\ee
We can now explicitly write our choice of gamma matrices satisfying the $3+7$ decomposition,
\begin{equation}
  \Gamma^{\alpha}=\tilde\gamma^{\alpha}\otimes\mathds{1}_8\otimes \sigma^{2} \;,\quad \Gamma^{m}=\mathds{1}_2\otimes\gamma^{m}\otimes \sigma^{1} \;,
\end{equation}
where each of the internal matrices $\gamma^{\alpha}$ and $\gamma^{m}$ satisfy their own Clifford algebras,
\begin{equation}
  \left\{\tilde{\gamma}^{\alpha},\tilde{\gamma}^{\beta}\right\}=2g^{\alpha\beta}\cdot\mathds{1}_{2} \;, \quad \left\{\gamma^{m},\gamma^{n}\right\}=2g^{mn}\cdot\mathds{1}_{8} \;.
\end{equation}
For the sake of clarity, let us write all the matrices explicitly. They can be written in terms of Pauli matrices when going into the vielbein frame. In three dimensions, we have the $2\times2$ purely real matrices
\be
  \tilde\gamma^{\underline{0}}=i\sigma^{2} \;, \quad \tilde\gamma^{\underline{1}}=\sigma^{1} \;, \quad \tilde\gamma^{\underline{2}}=\sigma^{3} \;,
\ee
and in $7$ dimensions, we have the $8\times8$ purely imaginary matrices:
\be
\begin{aligned}
  &\gamma^{\underline{3}}=\sigma^2\otimes\sigma^2\otimes\sigma^2 \;,\quad
  \gamma^{\underline{4}}=\sigma^2\otimes\mathds{1}_2\otimes\sigma^1 \;,\\
  &\gamma^{\underline{5}}=\sigma^2\otimes\mathds{1}_2\otimes\sigma^3 \;,\quad
  \gamma^{\underline{6}}=\sigma^1\otimes\sigma^2\otimes\mathds{1}_2 \;,\\
  &\gamma^{\underline{7}}=\sigma^3\otimes\sigma^2\otimes\mathds{1}_2 \;,\quad
  \gamma^{\underline{8}}=\mathds{1}_2\otimes\sigma^1\otimes\sigma^2 \;,\\
  &\gamma^{\underline{9}}=\mathds{1}_2\otimes\sigma^3\otimes\sigma^2 \;.
\end{aligned}
\ee
We can see that
\be
  \gamma^{\underline{3}}=-i\gamma^{\underline{4}}\gamma^{\underline{5}}\gamma^{\underline{6}}\gamma^{\underline{7}}\gamma^{\underline{8}}\gamma^{\underline{9}} \;,
\ee
which corresponds to the chirality operator in $6$ dimensions.

The irreducible spinor representations corresponding to our different geometries is summarized in table~\ref{tablespinordim}.
\begin{table}[ht]
\begin{center}
\begin{tabular}{|cc|cccc|}
  \hline
  Signature & d & Maj. & Weyl & M.-W. & Dirac \\\hline
  (1,2)&3&2&-&-&4 \\
  (1,3)&4&4&4&-&8 \\
  (0,6)&6&8&8&-&16 \\
  (0,7)&7&8&-&-&16 \\
  (1,9)&10&32&32&16&64 \\
  \hline
\end{tabular}
\parbox{6in}{\caption{\it\small Spinor representations in various dimensions $d$. The signature of the metric is given as well as the corresponding dimensions of the four types of representations: Majorana, Weyl, Majorana-Weyl and Dirac.}\label{tablespinordim}}
\end{center}
\end{table}
The Weyl representation is subject to the condition,
\be
  \Psi_{L,R}=P_{L,R}\Psi_{L,R} \;, \quad P_{L,R}\equiv1/2\left(\mathds{1}\pm\Gamma^{11}\right) \;,
\ee
where $\Psi$ is a generic spinor, $L,R$ stands for left- or right-handed and the projection operator $P$ is built with the ten-dimensional chirality operator $\Gamma^{11}$. When considering the $4+6$ decomposition and restricting ourselves on the internal space, the chirality operator in 6 dimensions $\gamma^3$ should be used. The Majorana representation is given by spinors being their own charge conjugate, $\Psi=\Psi^c$. This condition can be written with an appropriate matrix $X$,
\be
  \Psi=X\Psi^* \;, \quad XX^*=\mathds{1} \;.
\ee
In our above choice of basis, it turns out that $X=\mathds{1}_{32}$ and, therefore, Majorana spinors correspond to real spinors.

The ten-dimensional supergravity spinor $\epsilon$ is Majorana-Weyl and decomposes according to our choice of gamma matrices,
\begin{equation}
  \epsilon\left(x^m\right)=\rho\otimes\eta\left(x^m\right)\otimes\theta \;.
\end{equation}
The Majorana-Weyl condition implies $\rho=\rho^*$ for the three-dimensional spinor and $\eta=\eta^*$ for the $7$-dimensional spinor. The last component is appearing to make the dimension of the various spinors matches. It is constant and constrained by $\theta=\sigma^3\theta$ for positive chirality. The internal geometry can also be seen as a $7=1+6$ split. In that case, the seven-dimensional spinor decomposes as
\begin{equation}\label{eq_appendix_spinoransatz}
  \eta\left(x^m\right)=\frac{1}{\sqrt{2}}\left(\eta_+\left(x^m\right)+\eta_-\left(x^m\right)\right) \;,
\end{equation}
where $\eta_\pm$ are six-dimensional chiral spinors $\eta_\pm=\pm\gamma^3\eta$. Moreover they are subject to the ten-dimensional Majorana-Weyl condition which implies $\eta_\pm^*=\eta_\mp$.

Finally, the presence of the above covariantly constant spinor implies the existence of well-defined tensors on the geometry. This has drastic consequences for the possible choices of space-time solutions. These tensors are found from the following contractions,
\be
  T^{m_1...\,m_k}=\eta^\dagger\gamma^{m_1...\,m_k}\eta \;,
\ee
where the repetition of indices means the anti-symmetrization:
\begin{equation}
  \gamma^{m_1...\,m_k}=\gamma^{[m_1}\gamma^{m_2}...\,\gamma^{m_k]} \;.
\end{equation}
A useful fact is that several of the contractions vanish due to the nature of the spinor $\eta$. It can be summarized by the following properties
\be
\begin{aligned}
  \eta^T_+\gamma^{a_1...\,a_k}\eta_+&=0 \;, \quad\quad {\rm for}\;k\;{\rm even} \;,\\
  \eta^\dagger_+\gamma^{a_1...\,a_k}\eta_+&=0 \;, \quad\quad {\rm for}\;k\;{\rm odd} \;,
\end{aligned}
\ee
in term of the six-dimensional spinor $\eta_+$. The next appendix is dedicated to a systematic study of such tensors in dimensions six and seven.
\clearemptydoublepage
\chapter{Torsion classes}\label{appendixtorsion}
\section{$G$-structures and torsion classes}

In this appendix, we would like to review a couple of facts about $G$-structures. In particular, we will concentrate on the case of $SU(3)$ and $G_2$ structures in $6$ and $7$ dimensions. More details can be found in the literature, e.g. in Refs.~\cite{Koerber:2010bx, Agricola:2006tx, Grana:2005jc, Kaste:2003zd}.

Let us consider a $d$-dimensional manifold and its frame bundle. In general, the structure group of the frame bundle is contained in $GL(d,\mathbb{R})$. However, it can happen that it is actually smaller and takes value in some subgroup $G\subset GL(d,\mathbb{R})$. When this is the case, it is said for the manifold to admit a $G$-structure. Such a reduction implies, or is forced upon, the existence of globally defined objects on the manifold. Indeed, the transition functions are restricted to preserve global existence. Alternatively, if the transition functions are such that the structure group is reduced, then it is possible to find the said objects. Now what are they? They are tensors or, simply, some spinors (of which tensors can be built). For instance, the existence of a metric for a Riemannian manifold implies a reduction to $O(d)$. If it is further orientable, the structure group is $SO(d)$. For the dimensions we are interested in, the tensors are summarized in table~\ref{tablegstruc} and are explicitly given in the next sections.
\begin{table}[ht]
\begin{center}
\begin{tabular}{|c|c|l|}
  \hline
  $d$ & Group $G$ & Tensors \\
  \hline
  $7$ & $G_2$ & $\varphi$, $\varPhi$ \\
  $7$ & $SU(3)$ & $J$, $\Omega$, $v$ \\
  $6$ & $SU(3)$ & $J$, $\Omega$ \\
  \hline
\end{tabular}
\parbox{6in}{\caption{\it\small Frame bundle structure groups with respect to their corresponding globally defined tensors, given by differential forms, according to the dimension $d$ of the manifold.}\label{tablegstruc}}
\end{center}
\end{table}

An alternative description of $G$-structure is given in terms of connections on the tangent bundle. A manifold admits a $G$-structure if there exists a connection $\nabla^{\left(T\right)}$ whose holonomy takes value in $G$,
\be
  {\rm hol}\left(\nabla^{\left(T\right)}\right)\subset G\;.
\ee
What is then the relation with the previous definition? The answer is that the tensors are covariantly constant with respect to this connection and, reciprocally, they can be defined this way. In general, the connection will not be the Levi-Civita connection but a more general one with torsion,
\be
  \nabla^{\left(T\right)}=\nabla+\tau \;,
\ee
where $\nabla$ is the Levi-Civita connection and $\tau$ the contorsion. The possible existence of such torsion plays an important role and, in the physics literature, the difference is emphasized by the terminology. The term $G$-structure is meant for the connection with torsion, whereas the term $G$-holonomy implicitly implies the Levi-Civita connection even thought, strictly speaking, both cases are holonomy.

The contorsion tensor is antisymmetric in its last tow indices $\tau_{mnp}=\tau_{m[np]}$ and, therefore, can be viewed as a one-form taking value in $\mathfrak{so}\left(n\right)$, the Lie algebra of $SO(n)$. It can be decomposed into two parts,
\begin{equation}
  \tau_m=\tau_m^0+\tau_m^G \;,
\end{equation}
where $\tau_m^0$ takes value in $\mathfrak{g}$, the Lie algebra of $G$, and $\tau_m^G$ takes value in the orthogonal complement $\mathfrak{g}^\bot$ in $\mathfrak{so}\left(n\right)$. The reason for this decomposition is that the action of $\tau_m^G$ on the $G$-invariant tensors vanishes. Hence, the fact that the invariant tensors are covariantly constant under $\nabla^{(T)}$ and that  the holonomy of $\nabla^{(T)}$ is contained in $G$ only depends on $\tau_m^0$. For this reason, $\tau_m^0$ is also called the intrinsic (con)-torsion. It can be decomposed into its irreducible representation content under the group $G$. These irreducible parts of $\tau_m^0$ are called torsion classes and they can be used to characterize the $G$-structure.

\section{$G_2$ structures}

Let us look more closely at the 7-dimensional manifolds with structure group $G_2$. The torsion is a one form with its one-form index transforming as the fundamental of $SO(7)$ and otherwise taking values in the adjoint of $SO(7)$. Hence, the two relevant decompositions under $G_2$ are
\begin{equation}
  {\bf 7}_{SO(7)}\rightarrow{\bf 7}_{G_2} \;, \quad {\bf 21}_{SO(7)}\rightarrow\left({\bf 7}+{\bf 14}\right)_{G_2} \;.
\end{equation}
The intrinsic torsion only takes values in ${\mathfrak g}_2^\bot={\bf 7}_{G_2}$ and its $G_2$ representation content is thus given by
\begin{equation}
  {\bf 7}\otimes{\bf 7}={\bf 1}+{\bf 14}+{\bf 27}+{\bf 7} \;.
\end{equation} 
The representations on the right-hand side correspond to the four torsion classes ${\cal X}_1,\ldots,{\cal X}_4$ associated to a $G_2$ structure and, consequently, the con-torsion takes value
\begin{equation}
  \tau^0\in \cX_1\oplus{\cal X}_2\oplus{\cal X}_3\oplus{\cal X}_4 \;.
\end{equation}

The corresponding covariantly constant tensors can be written in terms of a spinor contracted with gamma matrices as in the previous appendix. They are given by
\begin{equation}\label{formg2spindef}
  \varphi_{mnp}=-i\eta^\dagger\gamma_{mnp}\eta \;, \quad \varPhi_{mnpq}=\eta^\dagger\gamma_{mnpq}\eta \;,
\end{equation}
and can be contracted with the basis one-forms of the cotangent space as in~\eqref{form} to obtain differential forms $\varphi$ and $\Phi$. The two resulting forms are not independent but are related via Hodge duality,
\begin{equation}
  \varphi=*_7\,\varPhi \;,
\end{equation}
where $*_7$ is the seven-dimensional Hodge star operator. Useful contractions formulas for these tensors can also be found in the appendix of Ref.~\cite{House:2004hv}. The four torsion classes are characterized by the exterior derivative of such forms. We have,
\begin{equation}\label{g2torsion}
  d_7\varphi=4{\cal X}_1\varPhi+3{\cal X}_4\wedge\varphi-*_7{\cal X}_3 \;, \quad d_7\varPhi=4{\cal X}_4\wedge\varPhi-2*_7{\cal X}_2 \;,
\end{equation}
where $d_7$ means the exterior derivative in 7 dimensions. These equations often offer the most straightforward way to determine the torsion classes by computing the exterior derivatives of $\varphi$ and $\varPhi$. Some properties of $7$-dimensional $G_2$ structures can be characterized by these classes and an illustration sample is given in table~\ref{tableg2torsion}.
\begin{table}[ht]
\begin{center}
\begin{tabular}{|c|c|}
  \hline
  Torsion class & Properties (name) \\
  \hline
  ${\cal X}_1$ & nearly parallel \\
  ${\cal X}_2$ & almost parallel \\
  ${\cal X}_3$ & balanced \\
  ${\cal X}_4$ & locally conformally parallel \\
  \hline
\end{tabular}
\parbox{6in}{\caption{\it\small Sample of $7$-dimensional $G_2$ structure properties determined in terms of its non-vanishing torsion class.}\label{tableg2torsion}}
\end{center}
\end{table}

\section{$SU(3)$ structures}

\subsection{Six dimensions}

We now move on to $SU(3)$ structures on six-dimensional manifolds. The torsion takes value in the lie algebra $\mathfrak{so}(6)$ while its one-form index transforms under the fundamental representation of $SO(6)$. Hence, the relevant decomposition reads,
\begin{equation}
  {\bf 6}_{SO(6)}\rightarrow\left({\bf 3}+\bar{\bf 3}\right)_{SU(3)} \;, \quad {\bf 15}_{SO(6)}\rightarrow\left({\bf 1}+{\bf 3}+\bar{\bf 3}+{\bf 8}\right)_{SU(3)} \;.
\end{equation} 
Since $\mathfrak{so}(6)^\bot={\bf 1}+{\bf 3}+\bar{\bf 3}$, the intrinsic torsion contains the irreducible $SU(3)$ representations,
\begin{equation}
  \left({\bf 3}+\bar{\bf 3}\right)\otimes\left({\bf 1}+{\bf 3}+\bar{\bf 3}\right)=\left({\bf 1}+{\bf 1}\right)+\left({\bf 8}+{\bf 8}\right)+\left({\bf 6}+\bar{\bf 6}\right)+\left({\bf 3}+\bar{\bf 3}\right)+\left({\bf 3}+\bar{\bf 3}\right) \;,
\end{equation}
which give rise to five torsion classes defined, respectively, by
\begin{equation}
  \tau^0\in\cW_1\oplus\cW_2\oplus\cW_3\oplus\cW_4\oplus\cW_5 \;.
\end{equation}

The characterizing tensors are written again as contractions of a covariatly constant spinor and gamma matrices. They are given by,
\begin{equation}\label{jomdefspin}
  J_{ab}=-i\eta_+^\dagger\gamma_{ab}\eta_+ \;, \quad \Omega_{abc}=\eta^\dagger_+\gamma_{abc}\eta_- \;,
\end{equation}
and satisfy the compatibility relations,
\begin{equation}\label{su3def}
  J\wedge J\wedge J=i\frac{3}{4}\,\Omega\wedge\bar\Omega \;, \quad \Omega\wedge J=0 \;,
\end{equation}
where, again, the notation without any indices implies differential forms as in~\eqref{form}. One first comment about these tensors is that it implies for the manifolds to be almost complex. When the first index of $J_{ab}$ is raised with the metric, we obtain an almost complex structure (which is in general not integrable). Indeed, the relation
\be
  J^{a}_{\ph{a}b}J^{b}_{\ph{b}c}=-\delta^{a}_{\ph{a}c} \;,
\ee
can be verified from the definition~\eqref{jomdefspin} and using properties of gamma matrices such as Fierz identities~\cite{Becker:2007zj}. A couple of other useful properties are the Hodge star,
\be
  *J=\frac{1}{2}\,J\wedge J \;, \quad *\Omega_{\pm}=\pm\Omega_{\mp} \;,
\ee
where we write the real and imaginary part of $\Omega$ as $\Omega=\Omega_++i\Omega_-$ which are related via contraction with the almost complex structure:
\be
  J^{d}_{\ph{d}a}\,\Omega_{\pm dbc}=\mp\,\Omega_{\mp abc} \;.
\ee
Finally, the torsion classes can be obtained from the exterior derivative of the differential forms,
\begin{equation}\label{su3torsion}
  dJ=-\frac{3}{2}\,{\rm Im}\left(\cW_1\bar\Omega\right)+\cW_4\wedge J+\cW_3 \;, \quad d\Omega=\cW_1J\wedge J+\cW_2\wedge J+\bar\cW_5\wedge\Omega \;,
\end{equation}
where the torsion classes are subject to the relations,
\begin{equation}
  \cW_3\wedge\Omega=\cW_3\wedge J=\cW_2\wedge J\wedge J=0 \;.
\end{equation}
The last two equations mean that the forms $\cW_3$ and $\cW_2\wedge J$ are primitives. Properties of six-dimensional manifolds can be specified by these classes and an illustration sample is given in table~\ref{tablesu3torsion}. In particular, half-flat manifolds are characterized by the following conditions,
\bs
\begin{align}
  d\Omega_-&=0 \;,\\
  J\wedge\ dJ&=0 \;,
\end{align}
\es
which play a significant role in this thesis.
\begin{table}[ht]
\begin{center}
\begin{tabular}{|c|c|}
  \hline
  Torsion classes & Properties (name) \\
  \hline
  $\cW_1=\cW_2=0$ & Complex \\
  $\cW_1=\cW_3=\cW_4=0$ & Symplectic \\
  $\cW_1=\cW_2=\cW_3=\cW_4=0$ & Kahler \\
  $\cW_2=\cW_3=\cW_4=\cW_5=0$ & Nearly-Kahler \\
  $\cW_{1-}=\cW_{2-}=\cW_4=\cW_5=0$ & Half-flat \\
  $\cW_1=\cW_2=\cW_3=\cW_4=\cW_5=0$ & Calabi-Yau \\
  \hline
\end{tabular}
\parbox{6in}{\caption{\it\small Sample of $6$-dimensional $SU(3)$ structure properties determined in terms of  their vanishing torsion classes.}\label{tablesu3torsion}}
\end{center}
\end{table}

\subsection{Seven dimensions}

It is also possible for a $7$-dimensional manifold to have an $SU(3)$ structure group~\cite{Friedrich:1995dp}. In that case, it is defined by a triplet of forms $\{J,\Omega,v\}$, where $J$ and $\Omega$ are obtained in a similar way as before and $v$ is a one-form. Intuitively, $v$ singles out a special direction and a complementary six-dimensional space on which $J$ and $\Omega$ can be thought of as defining an $SU(3)$ structure in the six-dimensional sense. In addition to the usual relations~\eqref{su3def} for a six-dimensional $SU(3)$ structure, its $7$-dimensional counterpart must also satisfy:
\be
  v\neg J=0 \;,\quad v\neg\Omega \;,\quad v\neg v=1 \;,
\ee
and for the contractions,
\be
  J^{m}_{\ph{m}n}J^{n}_{\ph{n}p}=-\delta^{m}_{\ph{m}p}+v^mv_p \;, \quad J^{q}_{\ph{q}m}\,\Omega_{\pm qnp}=\mp\;\Omega_{\mp mnp} \;.
\ee
The $7$-dimensional Hodge star operator acts as follows,
\be
  *_7\left(J\wedge v\right)=\frac{1}{2}\,J\wedge J \;, \quad *_7\Omega_{\pm}=\pm\Omega_{\mp}\wedge v \;.
\ee
From the spinor expressions~\eqref{formg2spindef} and~\eqref{jomdefspin} together with equation~\eqref{eq_appendix_spinoransatz}, we can show~\cite{Friedrich:1995dp} that a $7$-dimensional $SU(3)$ structure gives rise to a $G_2$ structure via
\begin{equation}
  \varphi=v\wedge J+\Omega_- \;, \quad \varPhi=v\wedge\Omega_++\frac{1}{2}\,J\wedge J \;.
\end{equation}
The same game can be played for the torsion classes, however it is not relevant for our purpose and therefore neglect its presentation. We refer the reader to Ref.~\cite{Lukas:2004ip} if he is interested.
\clearemptydoublepage
\chapter{Calabi-Yau moduli space geometry}\label{appendixCY}

In this appendix, we would like to review the geometry of Calabi-Yau moduli space. This is motivated by the fact that the same properties hold for mirror half-flat manifolds, a fact implied from their mirror symmetry origin. All the material is well known and can be found in Ref.~\cite{Strominger:1985ks} and in the classic paper by Candelas and de la Ossa~\cite{Candelas:1990pi}. For this reason, we will skip some details and concentrate on the relevant information for our purpose, mostly collecting the relevant formulas.

By moduli we mean metric deformations preserving the Calabi-Yau property. It is in one-to-one correspondence with harmonic forms and there will be $h^{1,1}+h^{1,2}$ fields parameterizing the moduli space $\mathfrak{M}$. Moreover, it decomposes as a direct product of two non-interacting components,
\be
  \mathfrak{M}=\mathfrak{M}_{\rm K}\times \mathfrak{M}_{\rm CS} \;,
\ee
where ${\rm K}$ and ${\rm CS}$ stand for Kahler and complex structure. These two parts are related to deformations of the Kahler form and the complex structure respectively. The high symmetries of Calabi-Yau spaces imply constraints on the nature of the space of deformations. For instance, we can define a metric on the moduli space and show that it is Kahler. Let us recall that a Kahler metric is a hermitian metric which can locally be written in terms of a Kahler potential $K$ as in~\eqref{KIJ}. Furthermore, for the case of the moduli space, it turns out to be  more special in that the Kahler potential itself can be written in term of another function, called the prepotential, which turns out to be a symplectic invariant holomorphic function homogeneous of degree two. Such geometries have been dubbed special Kahler.

\section{Kahler moduli space}

We start with a description of the Kahler moduli space. Let us write the basis of $H^2$ by $\{\omega_i\}$ as in appendix~\ref{appendixconventions}, where $i,j,\ldots=1,\ldots,h^{1,1}(X)$. We can then expand the Kahler form $J$ and the $\hat B$ fields,
\begin{equation}
  \hat B=b^i\omega_i \;, \quad J=v^i\omega_i \;.
\end{equation}
It can be shown that the natural combination for the Kahler moduli space is to consider the complexified Kahler cone with the combination $\hat B+iJ$. Thus, the corresponding moduli space coordinates are defined by
\be
  T^i=b^i+iv^i \;.
\ee
We should also note that the Kahler form $J$ corresponds to the $SU(3)$ structure form presented in the previous appendix~\ref{appendixtorsion} and, therefore, the volume of the Calabi-Yau space $X$ is given by,
\be
  {\cal V}=\frac{1}{6}\int_XJ\wedge J\wedge J \;.
\ee
The metric appearing in the deformation of the Calabi-Yau metric is  given by
\begin{equation}\label{Jkahlermetric}
K^{\left(1\right)}_ {ij}=\frac{1}{4\mathcal{V}}\int_X\omega_i\wedge*\omega_j\;,
\end{equation}
and we can see that it is a Kahler metric from the calculation of $*\omega_i$~\cite{Strominger:1985ks},
\be
  *\omega_i=-J\wedge\omega_i+\frac{3}{2}\,\frac{\int_X J\wedge J\wedge \omega_i}{\int_X J\wedge J\wedge J}\,J\wedge J \;.
\ee
The Kahler potential is thus given by,
\be
  K^{\left(1\right)}_{ij}=\frac{\partial^2K^{\left(1\right)}}{\partial T^i\partial\bar{T}^j} \;, \quad K^{\left(1\right)}=-\ln\left(\frac{4}{3}\int_XJ\wedge J\wedge J\right) \;,
\ee
which corresponds to the logarithm of the volume of the Calabi-Yau.

We can simplify the above expressions by introducing more convenient notations. Let us write the intersection numbers,
\begin{equation}
  {\cal K}_{ijk}=\int_X\omega_i\wedge\omega_j\wedge\omega_k
\end{equation}
and the following contractions with the Kahler moduli $v^i$,
\begin{equation}\label{metric1}
  {\cal K}={\cal K}_{ijk}v^iv^jv^k \;, \quad {\cal K}_i={\cal K}_{ijk}v^jv^k \;, \quad {\cal K}_{ij}={\cal K}_{ijk}v^k \;.
\end{equation}
This notations imply ${\cal K}=6{\cal V}$. The Kahler metric~\eqref{Jkahlermetric} can thus easily be re-written in the following manner,
\begin{equation}
  K^{\left(1\right)}_i\equiv \frac{\partial K^{\left(1\right)}}{\partial T^i}=\frac{3i}{2}\frac{\mathcal{K}_i}{\mathcal{K}} \;, \quad K^{\left(1\right)}_ {ij}=\frac{9}{4}\frac{\mathcal{K}_i\mathcal{K}_j}{\mathcal{K}^2}-\frac{3}{2}\frac{\mathcal{K}_{ij}}{\mathcal{K}} \;,
\end{equation}
where we also gave the first order derivative $K^{(1)}_i$ which is used later. Another useful formula is the inverse metric,
\be
  {K^{\left(1\right)}}^{ij}=-\frac{2}{3}{\cal K}\left({\cal K}^{ij}-3\frac{v^iv^j}{{\cal K}}\right) \;,
\ee
where we define ${\cal K}^{ij}$ from the property ${\cal K}^{ij}{\cal K}_{jk}=\delta^i_{\ph{i}k}$. We also have the following contraction
\be
  {K^{\left(1\right)}}^{ij}K_j=-2iv^i \;.
\ee

Finally, as mentioned in this appendix introductory words, the geometry is in fact special Kahler. To see this, we need to introduce a new field $T^0$. It could be the inverse of the dilaton but its precise nature we are not interested in. We can write the prepotential
\be
  \mathcal{F}=-\frac{1}{6}\mathcal{K}_{ijk}\frac{T^iT^jT^k}{T^0} \;,
\ee
giving the Kahler potential,
\be
  K^{\left(1\right)}=-\ln\left(i\left(\bar T^I\mathcal{F}_I-T^I\bar{\mathcal{F}}_I\right)\right) \;,
\ee
where $\mathcal{F}_I\equiv\partial\mathcal{F}/\partial T^I$ and $I=0,1,\dots,h^{1,1}$. It is straightforward to verify that this is exactly equivalent to~\eqref{Jkahlermetric} when we set $T^0$ to one. The whole special Kahler geometry is independent from a rescaling by $T^0$ and the meaning of this new coordinate is to make $T^I$ correspond to homogeneous coordinates of a projective space. 

\section{Complex structure moduli space}

A special choice of coordinates must be made to unveil the nature of the complex structure moduli. This is done by introducing the real symplectic basis $\left\{\alpha_A,\beta^A\right\}$, where $A,B,\ldots = 0,1,\ldots ,h^{2,1}(X)$, having the intersections,
\be\label{sympintersec}
  \int_X\alpha_A\wedge\beta^B=\delta_A^{\ph{A}B}\;,\quad \int_X\alpha_A\wedge\alpha_B=\int_X\beta^A\wedge\beta^B=0 \;.
\ee
The holomorphic three-form thus becomes,
\begin{equation}\label{omegasymp}
  \Omega={\cal Z}^A\alpha_A-\mathcal{G}_A\beta^A \;,
\end{equation}
where ${\cal Z}^A$ and $\mathcal{G}_A$ correspond to the periods of $\Omega$ on the symplectic basis. As for the Kahler moduli, it turns out that the moduli space is a projective space and the complex structure moduli is given by definition 
\be
  Z^a= c^a+iw^a \;,
\ee
where
$Z^a={\cal Z}^a/{\cal Z}^0$. Now, by considering the Calabi-Yau metric deformations, we find the following metric on the space of deformations,
\begin{equation}
  K^{\left(2\right)}_{a b}=-\frac{\int_X\chi_a\wedge\bar\chi_b}{\int_X\Omega\wedge\bar\Omega} \;,
\end{equation}
where $\chi_a$ are a set of $\left(2,1\right)$-forms defined from the result by Kodaira stating,
\begin{equation}\label{Kodaira}
  \frac{\partial\Omega}{\partial z^a}=-\frac{\partial K^{\left(2\right)}}{\partial Z^a}\,\Omega+\chi_a \;.
\end{equation}
It implies that we can write the metric with a Kahler potential coming from a prepotential giving the same special Kahler geometry as before,
\begin{equation}
  K^{\left(2\right)}=-\ln\left(i\int_X \Omega\wedge\bar\Omega\right)=-\ln\left(i\left(\bar \cZ^A\mathcal{G}_A-\cZ^A\bar{\mathcal{G}}_A\right)\right) \;.
\end{equation}
Here ${\cal G}_A=\partial{\cal G}/\partial{\cal Z}^A$
and, again, we see that the Kahler potential comes from the logarithm of the Calabi-Yau space volume,
\begin{equation}
  \mathcal{V}=\frac{i}{\|\Omega\|^2}\int_X\Omega\wedge\bar\Omega \;,
\end{equation}
where we define $3!\|\Omega\|^2=\Omega_{uvw}\bar\Omega^{uvw}$.

The properties of Kahler and complex structure moduli has striking similarities which led to the conjecture of mirror symmetry. For each Calabi-Yau $X$, there exists a mirror Calabi-Yau $\tilde X$ whose Kahler and complex structure moduli are reversed. This conjecture allows us to introduce intersections of the mirror Calabi-Yau $\tilde X$
\begin{equation}
  \tilde{\mathcal{K}}=\tilde{\mathcal{K}}_{abc}w^aw^bw^c \;, \quad \tilde{\mathcal{K}}_a=\tilde{\mathcal{K}}_{abc}w^bw^c \;, \quad \tilde{\mathcal{K}}_{ab}=\tilde{\mathcal{K}}_{abc}w^c \;,
\end{equation}
where we also wrote the relevant contractions. In the large complex structure limit, the prepotential $\cG$ of $X$ is given by
\begin{equation}
  \mathcal{G}=-\frac{1}{6}\tilde{\mathcal{K}}_{abc}\frac{\cZ^a\cZ^b\cZ^c}{\cZ^0} \;,
\end{equation}
leading to the same expressions for the Kahler metric as in the Kahler moduli space case,
\begin{equation}
  K^{\left(2\right)}_{a}=\frac{\partial K^{\left(2\right)}}{\partial Z^a}=\frac{3i}{2}\frac{\tilde{\mathcal{K}}_a}{\tilde{\mathcal{K}}} \;, \quad
  K^{\left(2\right)}_{a\bar b}=\frac{9}{4}\frac{\tilde{\mathcal{K}}_a\tilde{\mathcal{K}}_b}{\tilde{\mathcal{K}}^2}-\frac{3}{2}\frac{\tilde{\mathcal{K}}_{ab}}{\tilde{\mathcal{K}}} \;.
\end{equation}
We also have the inverse metric
\be
  {K^{\left(2\right)}}^{ab}=-\frac{2}{3}\tilde{{\cal K}}\left(\tilde{{\cal K}}^{ab}-3\frac{w^aw^b}{\tilde{{\cal K}}}\right) \;,
\ee
where by definition $\tilde{{\cal K}}^{ab}\tilde{{\cal K}}_{bc}=\delta^a_{\ph{a}c}$. The relation
\be
  {K^{\left(2\right)}}^{ab}K_b=-2iw^a
\ee
can easily be verified.

\section{More on the symplectic basis}

In this section, we would like to present the Hodge star of the symplectic basis $\left\{\alpha_A,\beta^A\right\}$. First, we should note that for the holomorphic basis,
\be
  *\Omega=-i\Omega,\quad *\chi_a=i\chi_a \;.
\ee
Now, for the symplectic forms, the Hodge star dual forms can be expanded on the original basis and, thus, we can write in general
\bs
\begin{align}
  *\alpha_A&=A_{A}^{\ph{A}B}\alpha_B+B_{AB}\beta^B \;,\\
  *\beta^A&=C^{AB}\alpha_B+D^A_{\ph{A}B}\beta^B \;,
\end{align}
\es
by definition of the different matrices. From the intersections~\eqref{sympintersec}, these matrices must obey the following properties 
\bs
\begin{align}
  B_{AB}=\int\alpha_A\wedge *\alpha_B&=\int\alpha_B\wedge *\alpha_A=B_{BA} \\
  C^{AB}=-\int\beta^A\wedge *\beta^B&=-\int\beta^B\wedge *\beta^A=C^{BA} \\
  A_{A}^{\ph{A}B}=-\int\beta^B\wedge *\alpha_A&=-\int\alpha_A\wedge *\beta^B=-D^{B}_{\ph{B}A} \;.
\end{align}
\es
It is a little exercise to use the expression of $\Omega$ in terms of the symplectic basis~\eqref{omegasymp} together with Kodaira's formula~\eqref{Kodaira} to calculate the expressions for $*\alpha$ and $*\beta$. The results can conveniently be written in terms of the matrix,
\be
  \cM_{AB}=\bar\cG_{AB}+2i\,\frac{\im\cG_{AC}\,\cZ^C\,\im\cG_{BD}\,\cZ^D}{\cZ^C\,\im \cG_{CD}\,\cZ^D} \;,
\ee
and the inverse matrix
\be
  \left(\im\cM\right)^{-1}=-\frac{6}{\tilde{\cal K}}\left(\begin{matrix}1&c^b\\c^a&\frac{{K^{\left(2\right)}}^{ab}}{4}+c^ac^b\end{matrix}\right) \;.
\ee
We obtain,
\bs
\begin{align}
  A&=\left(\re\cM\right)\left(\im\cM\right)^{-1} \\
  B&=-\left(\im\cM\right)-\left(\re\cM\right)\left(\im\cM\right)^{-1}\left(\re\cM\right) \\
  C&=\left(\im\cM\right)^{-1} \;.
\end{align}
\es
\clearemptydoublepage
\chapter{Coset space formalism}\label{appendixcoset}

\section{Coset space geometry}

In this appendix, we would like to review the construction of homogeneous spaces~\cite{Castellani:1999fz,Kapetanakis:1992hf,MuellerHoissen:1987cq,Camporesi:1990wm,Castellani:1983tb,Lust:1986ix,KashaniPoor:2007tr}. Let us consider a Lie group $G$ with some subgroup $H$ being a Lie group itself. The coset $G/H$ is defined as the set of equivalence relations
\begin{equation}
  \mathsf{g}\sim\mathsf{g}'\Leftrightarrow\mathsf{g}^{-1}\mathsf{g}'=\mathsf{h}\in H \;.
\end{equation}
This means two elements $\mathsf{g}$ and $\mathsf{g}'$ of $G$ are considered to be equivalent if they can be related by right multiplication with some element $\mathsf{h}$ of the subgroup $H$. A useful way to think about the group $G$ in this context is as a principal bundle $G(G/H,H)$ with base space $G/H$ and fibers given by the orbits of $H$. The Lie algebra $\mathfrak{g}$ of $G$ can be written as a direct sum,
\begin{equation}\label{decomp}
  \mathfrak{g}=\mathfrak{h}\oplus\mathfrak{k} \;,
\end{equation}
where $\mathfrak{h}$ is the Lie algebra of the subgroup $H$ and $\mathfrak{k}$ is the remainder. In the following, we will adopt the conventions
\begin{equation}
  \quad T_A\in\mathfrak{g} \;, \quad H_i\in\mathfrak{h} \;, \quad K_a\in\mathfrak{k} \;,
\end{equation}
to denote the Lie algebra basis elements in those various parts. Here, indices run over the appropriate ranges summarized in~\eqref{cosetindices}. The structure constants split up accordingly and we also require a basis such that they satisfy,
\begin{equation}\label{eq_reductivity}
  f_{ia}^{\phantom{ia}j}=0 \;, \quad f_{ij}^{\phantom{ij}a}=0 \;,
\end{equation}
which means the Lie algebra $\mathfrak{g}$ decomposes reductively. Explicitly, the commutation relations take the form,
\be
\begin{aligned}
  \left[K_a,K_b\right]&=f_{ab}^{\phantom{ab}c}K_c+f_{ab}^{\phantom{ab}i}H_i \;,\\\label{structconst}
  \left[H_i,K_a\right]&=f_{ia}^{\phantom{ia}b}K_b \;,\\
  \left[H_i,H_j\right]&=f_{ij}^{\phantom{ij}k}H_k \;.
\end{aligned}
\ee
In practice, the relevant geometrical information about the coset is contained in the structure constants which are collected in the last section of this appendix.

In order to find an explicit description of the coset space, we can choose one representative for each coset. Such representative can be written using the exponential map,
\begin{equation}\label{expo}
  L\left(x\right)=\exp\left(x^aK_a\right) \;.
\end{equation}
In the following, we adopt the conventions that $\{x^a,z^i\}$ stand for the coordinates relative to the basis $\{K_a,H_i\}$, respectively. The above representative can be viewed as a section of the principal bundle $G(G/H,H)$. A non-singular set of one-forms on $G/H$ can be obtained following a procedure analogous to the one leading to left-invariant one-forms on $G$. First, define the Lie algebra valued one-form
\begin{equation}
  \varTheta=L^{-1}dL \;,
\end{equation}
where $d$ is the exterior derivative on $G/H$. Then expand $V$ in terms of the chosen Lie algebra basis as
\begin{equation}\label{Vdef}
  \varTheta=e^aK_a+\varepsilon ^i H_i
\end{equation}
with form ``coefficients'' $e^a$ and $\epsilon^i$. It can be shown that the one-forms $e^a$ are non-singular. Thus, they form a basis for the cotangent space on $G/H$ and can be used as vielbein. The algebra of their exterior derivatives follows from the Maurer-Cartan structure equations on $G$. Using the above commutation relations, we obtain
\bs\label{d}
\begin{align}
  de^a&=-\frac{1}{2}f_{bc}^{\phantom{bc}a}e^b\wedge e^c-f_{ib}^{\phantom{ib}a}\varepsilon^i\wedge e^b \;,\\
  d\varepsilon^i&=-\frac{1}{2}f_{ab}^{\phantom{ab}i}e^a\wedge e^b-\frac{1}{2}f_{jk}^{\phantom{jk}i}\varepsilon^j\wedge\varepsilon^k \;.
\end{align}
\es
While the forms $e^a$ are left-invariant when viewed as forms on the group $G$, this is no longer the case when they descend to the coset $G/H$.

Another useful geometrical quantity is the Riemann curvature tensor. The Levi-Civita connection one-form $\omega^a_{\phantom{a}b}$ associated to the vielbein $e^a$ on the coset space is determined by the standard relations $de^a+\omega^a_{\phantom{a}b}\wedge e^b=0$ and $\omega_{ab}=-\omega_{ba}$. For reductive homogeneous space, we can explicitly find,
\begin{equation}\label{omega}
  \omega_{cb}^{\phantom{cb}a}e^c=D_{cb}^{\phantom{cb}a}e^c+f_{ib}^{\phantom{ib}a}\varepsilon^i \;\;\mbox{where }\;\; D_{cb}^{\phantom{cb}a}=\frac{1}{2}f_{cb}^{\phantom{cb}a}-\frac{1}{2}\left(g^{am}f_{cm}^{\phantom{cm}n}g_{nb}+g^{am}f_{bm}^{\phantom{bm}n}g_{cn}\right)
\end{equation}
which leads to the curvature two-form $R^a_{\phantom{a}b}=\frac{1}{2}R^a_{\phantom{a}bcd}\,e^c\wedge e^d$ with
\begin{equation}\label{R}
  R^a_{\phantom{a}bcd}=-f_{cd}^{\phantom{cd}i}f_{ib}^{\phantom{ib}a}-f_{cd}^{\phantom{cd}m}D_{mb}^{\phantom{mb}a}+D_{cm}^{\phantom{cm}a}D_{db}^{\phantom{db}m}-D_{dm}^{\phantom{dm}a}D_{cb}^{\phantom{cb}m} \;.
\end{equation}

Finally, to be able to integrate a top form given by the vielbein $e^a$, we will also be interested in the volume of the coset. One practical way to achieve this is to look at the generalized Gauss-Bonnet theorem which relates the Euler number to the integration of the Euler form~\cite{Kobayashi}. The theorem states that
\begin{equation}
  \chi\left(X\right)=\int_X\gamma\left(TX\right) \;,
\end{equation}
where $\gamma$ is the Euler form on $X$. In six dimensions, it is given in term of the Riemann curvature tensor by
\begin{equation}
  \gamma\left(TM\right)=\frac{-1}{2^3\left(2\pi\right)^33!}\sum\epsilon^{a_1...a_6}R_{a_1a_2}\wedge...\wedge R_{a_5a_6}\equiv\varrho\,e^1\wedge...\wedge e^6 \;.
\end{equation}
Here, the last equality defines the constant $\varrho$. In practice, the idea is to calculate this constant by plugging the Riemann tensor~\eqref{R} in the above relation and write it as a constant times the top form $e^1\wedge...\wedge e^6$. We can thus calculate the volume using the generalized Gauss-Bonnet theorem,
\begin{equation}\label{vchi}
  {\cal V}=\varrho^{-1}\,\chi \;.
\end{equation}
The Euler number $\chi$ is given as usual from the alternated sum of Betti numbers and to calculate the Betti numbers it is sufficient to count the number of harmonic forms on the coset.

\section{Left-invariant structures}

In order to find the existing forms on the coset space, we need to work out transformations under the group action. This will allow us to find left-invariant forms under this action and ensure that they are well defined everywhere. First, we should realize that
\begin{equation}\label{magic}
  \mathsf{g}L\left(x\right)=L\left(x'\right)\mathsf{h} \;.
\end{equation}
Indeed, acting on the left by the group action will lead us to another section $L(x')$. However, this section does not necessarily corresponds to the formula~\eqref{expo} and must therefore be compensated by a gauge transformation $\mathsf{h}$. Any sections are related this way by the definition of the coset equivalence class. This implies for the one-form,
\begin{equation}\label{transformer}
  \varTheta\left(x'\right)=\mathsf{h}\,\varTheta\left(x\right)\mathsf{h}^{-1}+\mathsf{h}d\mathsf{h}^{-1} \;.
\end{equation}
The second term of this equation can be discarded when projecting down onto the coset space. We then deduce the transformation for the vielbein one-forms,
\begin{equation}\label{etrafo}
  e^a\left(x'\right)=D_b^{\phantom{b}a}\left(\mathsf{h}^{-1}\right)e^b\left(x\right) \;,
\end{equation}
where we note the adjoint representation of $H^{-1}$ by $D$.

For an infinitesimal $G$-action $\mathsf{g}={\bf 1}+\epsilon^AT_A$, the associated gauge transformation $\mathsf{h}$ in Eq.~\eqref{magic} can be written as $\mathsf{h}={\bf 1}-\epsilon^A{W_A}^iH_i$, with ``compensator" functions ${W_A}^i$. Expanding the exponentials in Eq.~\eqref{magic}, these functions can be calculated order by order. However, their explicit form will not be needed in the present context. Inserting them into Eq.~\eqref{etrafo}, the infinitesimal transformation of the vielbein becomes
\begin{equation}\label{einftrafo}
  e^a\left(x'\right)-e^a\left(x\right)=\epsilon^AW_A^{\phantom{A}i}f_{bi}^{\phantom{bi}a}e^b\left(x\right) \;.
\end{equation}
This transformation law will be crucial in a moment when we establish which structures are $G$-invariant on the coset $G/H$. In general, we can give a structure in terms of the vielbein one-forms as follows,
\bs
\begin{align}
  g&=g_{ab}\,e^a\otimes e^b \;,\\
  J&=\frac{1}{2}\,J_{ab}\,e^a\wedge e^b \;,\\
  \Omega&=\frac{1}{3!}\,\Omega_{abc}\,e^a\wedge e^b\wedge e^c \;.
\end{align}
\es
For it to be left-invariant, we want that evaluated at one point $(x)$, we have the same form than the one evaluated at $(x')$. Using the above transformations, we find for the metric
\begin{equation}
  g_{ab}\,e^a\left(x'\right)\otimes e^b\left(x'\right)=g_{ab}\, e^a\left(x\right)\otimes e^b\left(x\right)+g_{ab}\,\epsilon^AW_A^i\left(f_{ci}^{\phantom{ci}a}e^c\otimes e^b+f_{di}^{\phantom{di}b}e^d\otimes e^a\right) \;.
\end{equation}
This implies the condition for left-invariance,
\begin{equation}\label{ginv}
  f_{i(a}^{\phantom{i(a}c}g_{ b)c}=0 \;.
\end{equation}
Here the brackets denote symmetrization of indices. We can work out the corresponding relations for the differential forms $(J, \Omega)$ in the same manner. We obtain,
\begin{equation}\label{Jinv}
  f_{i[a}^{\phantom{a]i}c}J_{b]c}=0 \;, \quad f_{i[a}^{\phantom{a]i}d}\Omega_{bc]d}=0 \;,
\end{equation}
where we have, this time, anti-symmetrization of indices. Once the $SU(3)$ structure established, it is easy to work out the corresponding torsion classes from the Maurer-Cartan relations~\eqref{d}. It turns out that an exhaustive list of nearly-Kahler manifolds in six dimensions is given by the four cosets $SU(3)/U(1)^2$, $Sp(2)/SU(2)\times U(1)$, $G_2/SU(3)$ and $SU(2)\times SU(2)$. However, we will concentrate on the first three as $SU(2)\times SU(2)$ is not well suited for the construction of vector bundles. The respective data of these cosets are summarized in the next section.

\newpage
\section{Coset data}

In this section, we simply summarize all the relevant geometrical data of coset spaces which are being used in this thesis.

\subsection{$SU(3)/U(1)^2$}\label{gellmann}

\begin{itemize}
 \item Gell-Mann matrices:
\be
\begin{aligned}
&\lambda_1
=
-\frac{i}{2}\left(
\begin{matrix}
 0 & 1 & 0 \\
 1 & 0 & 0 \\
 0 & 0 & 0
\end{matrix}
\right)
,\;
\lambda_2
=
\frac{1}{2}\left(
\begin{matrix}
 0 & -1 & 0 \\
 1 & 0 & 0 \\
 0 & 0 & 0
\end{matrix}
\right)
,\;
\lambda_3
=
-\frac{i}{2}\left(
\begin{matrix}
 1 & 0 & 0 \\
 0 & -1 & 0 \\
 0 & 0 & 0
\end{matrix}
\right)
,\\
&\lambda_4
=
-\frac{i}{2}\left(
\begin{matrix}
 0 & 0 & 1 \\
 0 & 0 & 0 \\
 1 & 0 & 0
\end{matrix}
\right)
,\;
\lambda_5
=\frac{1}{2}\left(
\begin{matrix}
0 & 0 & -1 \\
 0 & 0 & 0 \\
 1 & 0 & 0
\end{matrix}
\right)
,\;
\lambda_6
=-\frac{i}{2}\left(
\begin{matrix}
 0 & 0 & 0 \\
 0 & 0 & 1 \\
 0 & 1 & 0
\end{matrix}
\right)
,\\
&
\lambda_7
=\frac{1}{2}\left(
\begin{matrix}
  0 & 0 & 0 \\
 0 & 0 & -1 \\
 0 & 1 & 0
\end{matrix}
\right)
,\;
\lambda_8
=-\frac{i}{2\sqrt{3}}
\left(
\begin{matrix}
 1 & 0 & 0 \\
 0 & 1 & 0 \\
 0 & 0 & -2
\end{matrix}
\right)
.
\end{aligned}
\ee
\end{itemize}

\begin{itemize}
 \item Generators:
\be
\begin{aligned}
  K_1&=\lambda_1 \;, & K_2&=\lambda_2 \;, & K_3&=\lambda_4 \;, & K_4&=\lambda_5 \;,\\
  K_5&=\lambda_6 \;, & K_6&=\lambda_7 \;, & H_7&=\lambda_3 \;, & H_8&=\lambda_8 \;.
\end{aligned}
\ee
\end{itemize}

\begin{itemize}
 \item Structure constants:
\be
\begin{aligned}
  f_{12}^{\phantom{12}7}&=1 \;,\\
  f_{13}^{\phantom{13}6}&=-f_{14}^{\phantom{14}5}=f_{23}^{\phantom{23}5}=f_{24}^{\phantom{24}6}=f_{73}^{\phantom{73}4}=-f_{75}^{\phantom{75}6}=1/2 \;,\\
  f_{34}^{\phantom{34}8}&=f_{56}^{\phantom{56}8}=\sqrt{3}/2 \;.
\end{aligned}
\ee
\end{itemize}

\begin{itemize}
 \item Basis of left-invariant forms:
\begin{equation}
  e^{12} , \quad e^{34} , \quad e^{56} , \quad  e^{136}-e^{145}+e^{235}+e^{246} , \quad e^{135}+e^{146}-e^{236}+e^{245} .
\end{equation}
\end{itemize}

\begin{itemize}
 \item Betti and Euler numbers:
\bea
  &b_0=1 \;, \quad\quad b_2=2 \;, \quad\quad b_4=2 \;, \quad\quad b_6=1 \;,\\
  &\chi=6 \;.
\eea
\end{itemize}

\newpage
\begin{itemize}
 \item SU(3) structure:
\be
\begin{aligned}
  ds^2=&R_1^2\left(e^1\otimes e^1+e^2\otimes e^2\right)+R_2^2\left(e^3\otimes e^3+e^4\otimes e^4\right) 
  +R_3^2\left(e^5\otimes e^5+e^6\otimes e^6\right) ,\\
  J=&-R_1^2\,e^{12}+R_2^2\,e^{34}-R_3^2\,e^{56} ,\\
  \Omega=&R_1R_2R_3\left(\left(e^{136}-e^{145}+e^{235}+e^{246}\right)+i\left(e^{135}+e^{146}-e^{236}+e^{245}\right)\right) .
\end{aligned}
\ee
\end{itemize}

\begin{itemize}
 \item Torsion classes:
\bs
\begin{align}
  \cW_1^+=&-\frac{R_1^2+R_2^2+R_3^2}{3R_1R_2R_3} \;,\\\nn
  \cW_2^+=&-\frac{2}{3R_1R_2R_3}\left[R_1^2\left(2R_1^2-R_2^2-R_3^2\right)e^{12}-R_2^2\left(2R_2^2-R_1^2-R_3^2\right)e^{34}\right. \\
  &\ph{\frac{2}{3R_1R_2R_3}[}\left.+R_3^2\left(2R_3^2-R_1^2-R_2^2\right)e^{56}\right] \;.
\end{align}
\es
\end{itemize}

\newpage
\subsection{$Sp(2)/SU(2)\times U(1)$}

\begin{itemize}
 \item Generators:
\be
\begin{aligned}
&
K_1
=\frac{1}{\sqrt{2}}
\left(
\begin{matrix}
 0 & 0 & 1 & 0 \\
 0 & 0 & 0 & 1 \\
 -1 & 0 & 0 & 0 \\
 0 & -1 & 0 & 0
\end{matrix}
\right)
,\;
K_2=\frac{i}{\sqrt{2}}\left(
\begin{matrix}
 0 & 0 & 0 & 1 \\
 0 & 0 & 1 & 0 \\
 0 & 1 & 0 & 0 \\
 1 & 0 & 0 & 0
\end{matrix}
\right) 
,\\
&
K_3= \left(
\begin{matrix}
 i & 0 & 0 & 0 \\
 0 & -i & 0 & 0 \\
 0 & 0 & 0 & 0 \\
 0 & 0 & 0 & 0
\end{matrix}
\right)
,\;
K_4
=
\left(
\begin{matrix}
 0 & 1 & 0 & 0 \\
 -1 & 0 & 0 & 0 \\
 0 & 0 & 0 & 0 \\
 0 & 0 & 0 & 0
\end{matrix}
\right)
,\;
K_5
=\frac{1}{\sqrt{2}}\left(
\begin{matrix}
 0 & 0 & 0 & 1 \\
 0 & 0 & -1 & 0 \\
 0 & 1 & 0 & 0 \\
 -1 & 0 & 0 & 0
\end{matrix}
\right) 
,\\
&
K_6=\frac{i}{\sqrt{2}}\left(
\begin{matrix}
 0 & 0 & -1 & 0 \\
 0 & 0 & 0 & 1 \\
 -1 & 0 & 0 & 0 \\
 0 & 1 & 0 & 0
\end{matrix}
\right)
,\;
H_7
=
\left(
\begin{matrix}
 0 & 0 & 0 & 0 \\
 0 & 0 & 0 & 0 \\
 0 & 0 & i & 0 \\
 0 & 0 & 0 & -i
\end{matrix}
\right)
,\\
&
H_8
=
\left(
\begin{matrix}
 0 & 0 & 0 & 0 \\
 0 & 0 & 0 & 0 \\
 0 & 0 & 0 & -1 \\
 0 & 0 & 1 & 0
\end{matrix}
\right)
,\;
H_9=\left(
\begin{matrix}
 0 & 0 & 0 & 0 \\
 0 & 0 & 0 & 0 \\
 0 & 0 & 0 & -i \\
 0 & 0 & -i & 0
\end{matrix}
\right)
,\;
H_{10}=\left(
\begin{matrix}
 0 & i & 0 & 0 \\
 i & 0 & 0 & 0 \\
 0 & 0 & 0 & 0 \\
 0 & 0 & 0 & 0
\end{matrix}
\right)
.
\end{aligned}
\ee
\end{itemize}

\begin{itemize}
 \item Structure constants:
\be
\begin{aligned}
  f_{13}^{\phantom{13}6}&=-f_{14}^{\phantom{14}5}=f_{23}^{\phantom{23}5}=f_{24}^{\phantom{24}6}=1 \;,\\
  f_{71}^{\phantom{71}6}&=-f_{72}^{\phantom{72}5}=f_{81}^{\phantom{81}5}=f_{82}^{\phantom{82}6}=f_{91}^{\phantom{91}2}=-f_{95}^{\phantom{95}6}=f_{10\;1}^{\phantom{10\;1}2}=f_{10\;5}^{\phantom{10\;5}6}=1 \;,\\
  f_{78}^{\phantom{78}9}&=f_{10\;3}^{\phantom{10\;3}4}=2 \;.
\end{aligned}
\ee
\end{itemize}

\begin{itemize}
 \item Basis of left-invariant forms:
\begin{equation}
  e^{12}+e^{56} , \quad e^{34} , \quad e^{135}+e^{146}-e^{236}+e^{245} , \quad e^{136}-e^{145}+e^{235}+e^{246} .
\end{equation}
\end{itemize}

\begin{itemize}
 \item Betti and Euler numbers:
\bea
  &b_0=1 \;, \quad\quad b_2=1 \;, \quad\quad b_4=1 \;, \quad\quad b_6=1 \;,\\
  &\chi=4 \;.
\eea
\end{itemize}

\newpage
\begin{itemize}
 \item SU(3) structure:
\be
\begin{aligned}
  ds^2&=R_1^2\left(e^1\otimes e^1+e^2\otimes e^2\right)+R_2^2\left(e^3\otimes e^3+e^4\otimes e^4\right)+R_1^2\left(e^5\otimes e^5+e^6\otimes e^6\right) ,\\
  J&=-R_1^2\,e^{12}+R_2^2\,e^{34}-R_1^2\,e^{56} ,\\
  \Omega&=R_1^2R_2\left(\left(e^{136}-e^{145}+e^{235}+e^{246}\right)+i\left(e^{135}+e^{146}-e^{236}+e^{245}\right)\right) .
\end{aligned}
\ee
\end{itemize}

\begin{itemize}
 \item Torsion classes:
\bs
\begin{align}
  \cW_1^+=&-\frac{4R_1^2+2R_2^2}{3R_1^2R_2} \;,\\\nn
  \cW_2^+=&-\frac{4}{3R_1^2R_2}\left[R_1^2\left(R_1^2-R_2^2\right)e^{12}+2R_2^2\left(R_1^2-R_2^2\right)e^{34}\right. \\
  &\ph{\frac{4}{3R_1^2R_2}(}\left.+R_1^2\left(R_1^2-R_2^2\right)e^{56}\right] \;.
\end{align}
\es
\end{itemize}

\newpage
\subsection{$G_2/SU(3)$}

\begin{itemize}
 \item Generators:
\begin{align}\nn
K_1=\frac{1}{\sqrt{3}}\left(
\begin{matrix}
 0 & 2 & 0 & 0 & 0 & 0 & 0 \\
 -2 & 0 & 0 & 0 & 0 & 0 & 0 \\
 0 & 0 & 0 & 0 & 0 & 0 & 0 \\
 0 & 0 & 0 & 0 & 0 & 0 & 1 \\
 0 & 0 & 0 & 0 & 0 & 1 & 0 \\
 0 & 0 & 0 & 0 & -1 & 0 & 0 \\
 0 & 0 & 0 & -1 & 0 & 0 & 0
\end{matrix}
\right)
&,\;
K_2=\frac{1}{\sqrt{3}}\left(
\begin{matrix}
 0 & 0 & 2 & 0 & 0 & 0 & 0 \\
 0 & 0 & 0 & 0 & 0 & 0 & 0 \\
 -2 & 0 & 0 & 0 & 0 & 0 & 0 \\
 0 & 0 & 0 & 0 & 0 & 1 & 0 \\
 0 & 0 & 0 & 0 & 0 & 0 & -1 \\
 0 & 0 & 0 & -1 & 0 & 0 & 0 \\
 0 & 0 & 0 & 0 & 1 & 0 & 0
\end{matrix}
\right),\\\nn
K_3=\frac{1}{\sqrt{3}}\left(
\begin{matrix}
 0 & 0 & 0 & 0 & -2 & 0 & 0 \\
 0 & 0 & 0 & 0 & 0 & 1 & 0 \\
 0 & 0 & 0 & 0 & 0 & 0 & -1 \\
 0 & 0 & 0 & 0 & 0 & 0 & 0 \\
 2 & 0 & 0 & 0 & 0 & 0 & 0 \\
 0 & -1 & 0 & 0 & 0 & 0 & 0 \\
 0 & 0 & 1 & 0 & 0 & 0 & 0
\end{matrix}
\right)
&,\;
K_4=\frac{1}{\sqrt{3}}\left(
\begin{matrix}
 0 & 0 & 0 & -2 & 0 & 0 & 0 \\
 0 & 0 & 0 & 0 & 0 & 0 & 1 \\
 0 & 0 & 0 & 0 & 0 & 1 & 0 \\
 2 & 0 & 0 & 0 & 0 & 0 & 0 \\
 0 & 0 & 0 & 0 & 0 & 0 & 0 \\
 0 & 0 & -1& 0 & 0 & 0 & 0 \\
 0 & -1 & 0 & 0 & 0 & 0 & 0
\end{matrix}
\right)
,\\\nn
K_5=\frac{1}{\sqrt{3}}\left(
\begin{matrix}
 0 & 0 & 0 & 0 & 0 & 0 & 2 \\
 0 & 0 & 0 & 1 & 0 & 0 & 0 \\
 0 & 0 & 0 & 0 & -1 & 0 & 0 \\
 0 & -1 & 0 & 0 & 0 & 0 & 0 \\
 0 & 0 & 1 & 0 & 0 & 0 & 0 \\
 0 & 0 & 0 & 0 & 0 & 0 & 0 \\
 -2 & 0 & 0 & 0 & 0 & 0 & 0
\end{matrix}
\right)
&,\;
K_6=\frac{1}{\sqrt{3}}\left(
\begin{matrix}
 0 & 0 & 0 & 0 & 0 & 2 & 0 \\
 0 & 0 & 0 & 0 & 1 & 0 & 0 \\
 0 & 0 & 0 & 1 & 0 & 0 & 0 \\
 0 & 0 & -1 & 0 & 0 & 0 & 0 \\
 0 & -1 & 0 & 0 & 0 & 0 & 0 \\
 -2 & 0 & 0 & 0 & 0 & 0 & 0 \\
 0 & 0 & 0 & 0 & 0 & 0 & 0
\end{matrix}
\right)
,\\\nn
H_7=\left(
\begin{matrix}
 0 & 0 & 0 & 0 & 0 & 0 & 0 \\
 0 & 0 & 0 & 0 & 0 & 0 & 0 \\
 0 & 0 & 0 & 0 & 0 & 0 & 0 \\
 0 & 0 & 0 & 0 & 0 & 0 & -1 \\
 0 & 0 & 0 & 0 & 0 & 1 & 0 \\
 0 & 0 & 0 & 0 & -1 & 0 & 0 \\
 0 & 0 & 0 & 1 & 0 & 0 & 0
\end{matrix}
\right)
&,\;
H_8=\left(
\begin{matrix}
 0 & 0 & 0 & 0 & 0 & 0 & 0 \\
 0 & 0 & 0 & 0 & 0 & 0 & 0 \\
 0 & 0 & 0 & 0 & 0 & 0 & 0 \\
 0 & 0 & 0 & 0 & 0 & -1 & 0 \\
 0 & 0 & 0 & 0 & 0 & 0 & -1 \\
 0 & 0 & 0 & 1 & 0 & 0 & 0 \\
 0 & 0 & 0 & 0 & 1 & 0 & 0
\end{matrix}
\right)
,\\\nn
H_9=\left(
\begin{matrix}
 0 & 0 & 0 & 0 & 0 & 0 & 0 \\
 0 & 0 & 0 & 0 & 0 & 0 & 0 \\
 0 & 0 & 0 & 0 & 0 & 0 & 0 \\
 0 & 0 & 0 & 0 & -1 & 0 & 0 \\
 0 & 0 & 0 & 1 & 0 & 0 & 0 \\
 0 & 0 & 0 & 0 & 0 & 0 & 1 \\
 0 & 0 & 0 & 0 & 0 & -1 & 0
\end{matrix}
\right)
&,\;
H_{10}=\left(
\begin{matrix}
 0 & 0 & 0 & 0 & 0 & 0 & 0 \\
 0 & 0 & 0 & 0 & 0 & -1 & 0 \\
 0 & 0 & 0 & 0 & 0 & 0 & -1 \\
 0 & 0 & 0 & 0 & 0 & 0 & 0 \\
 0 & 0 & 0 & 0 & 0 & 0 & 0 \\
 0 & 1 & 0 & 0 & 0 & 0 & 0 \\
 0 & 0 & 1 & 0 & 0 & 0 & 0
\end{matrix}
\right)
,
\end{align}
\begin{equation}
\begin{aligned}
&
H_{11}=\left(
\begin{matrix}
 0 & 0 & 0 & 0 & 0 & 0 & 0 \\
 0 & 0 & 0 & 0 & 0 & 0 & 1 \\
 0 & 0 & 0 & 0 & 0 & -1 & 0 \\
 0 & 0 & 0 & 0 & 0 & 0 & 0 \\
 0 & 0 & 0 & 0 & 0 & 0 & 0 \\
 0 & 0 & 1 & 0 & 0 & 0 & 0 \\
 0 & -1 & 0 & 0 & 0 & 0 & 0
\end{matrix}
\right)
,\;
H_{12}=\left(
\begin{matrix}
 0 & 0 & 0 & 0 & 0 & 0 & 0 \\
 0 & 0 & 0 & 1 & 0 & 0 & 0 \\
 0 & 0 & 0 & 0 & 1 & 0 & 0 \\
 0 & -1 & 0 & 0 & 0 & 0 & 0 \\
 0 & 0 & -1 & 0 & 0 & 0 & 0 \\
 0 & 0 & 0 & 0 & 0 & 0 & 0 \\
 0 & 0 & 0 & 0 & 0 & 0 & 0
\end{matrix}
\right)
,\\
&
H_{13}=\left(
\begin{matrix}
 0 & 0 & 0 & 0 & 0 & 0 & 0 \\
 0 & 0 & 0 & 0 & -1 & 0 & 0 \\
 0 & 0 & 0 & 1 & 0 & 0 & 0 \\
 0 & 0 & -1 & 0 & 0 & 0 & 0 \\
 0 & 1 & 0 & 0 & 0 & 0 & 0 \\
 0 & 0 & 0 & 0 & 0 & 0 & 0 \\
 0 & 0 & 0 & 0 & 0 & 0 & 0
\end{matrix}
\right)
,\;
H_{14}=\frac{1}{\sqrt{3}}\left(
\begin{matrix}
 0 & 0 & 0 & 0 & 0 & 0 & 0 \\
 0 & 0 & -2 & 0 & 0 & 0 & 0 \\
 0 & 2 & 0 & 0 & 0 & 0 & 0 \\
 0 & 0 & 0 & 0 & 1 & 0 & 0 \\
 0 & 0 & 0 & -1 & 0 & 0 & 0 \\
 0 & 0 & 0 & 0 & 0 & 0 & 1 \\
 0 & 0 & 0 & 0 & 0 & -1 & 0
\end{matrix}
\right)
.
\end{aligned}
\end{equation}
\end{itemize}

\begin{itemize}
 \item Structure constants:
\be
\begin{aligned}
&f_{7\; 10}^{\phantom{7\; 10}13}=
-f_{7\; 11}^{\phantom{7\; 11}12}=
f_{73}^{\phantom{73}6}=
-f_{74}^{\phantom{74}5}=1 ,\\
&f_{8\; 10}^{\phantom{8\; 10}12}=
f_{8\; 11}^{\phantom{8\; 11}13}=
-f_{83}^{\phantom{83}5}=
-f_{84}^{\phantom{84}6}=
f_{9\; 10}^{\phantom{9\; 10}11}=
-f_{9\; 12}^{\phantom{9\; 12}13}=
-f_{93}^{\phantom{93}4}=
f_{95}^{\phantom{95}6}=1 ,\\
&f_{10\; 1}^{\phantom{10\; 1}6}=
f_{10\; 2}^{\phantom{10\; 2}5}=
-f_{11\; 1}^{\phantom{11\; 1}5}=
f_{11\; 2}^{\phantom{11\; 2}6}=
f_{12\; 1}^{\phantom{12\; 1}4}=
f_{12\; 2}^{\phantom{12\; 2}3}=
-f_{13\; 1}^{\phantom{13\; 1}3}=
f_{13\; 2}^{\phantom{13\; 2}4}=1 ,\\
&f_{10\; 11}^{\phantom{10\; 11}14}=
f_{12\; 13}^{\phantom{12\; 13}14}=\sqrt{3},\quad\quad
f_{78}^{\phantom{78}9}=2 ,\\
&f_{14\; 1}^{\phantom{14\; 1}2}=
f_{13}^{\phantom{13}6}=
f_{14}^{\phantom{14}5}=
-f_{23}^{\phantom{23}5}=
f_{24}^{\phantom{24}6}=2/\sqrt{3} ,\\
&f_{14\; 3}^{\phantom{14\; 3}4}=
f_{14\; 5}^{\phantom{14\; 5}6}=1/\sqrt{3} .
\end{aligned}
\ee
\end{itemize}

\begin{itemize}
 \item Basis of left-invariant forms:
\begin{equation}
  e^{12}-e^{34}-e^{56} , \quad e^{136}+e^{145}-e^{235}+e^{246} , \quad e^{135}-e^{146}+e^{236}+e^{245} .
\end{equation}
\end{itemize}

\begin{itemize}
 \item Betti and Euler numbers:
\bea
  &b_0=1 \;, \quad\quad  b_6=1 \;,\\
  &\chi=2 \;.
\eea
\end{itemize}

\newpage
\begin{itemize}
 \item SU(3) structure:
\be
\begin{aligned}
  ds^2&=R^2\left(e^1\otimes e^1+e^2\otimes e^2+e^3\otimes e^3+e^4\otimes e^4+e^5\otimes e^5+e^6\otimes e^6\right) ,\\
  J&=R^2\,\left(e^{12}-e^{34}-e^{56}\right) ,\\
  \Omega&=R^3\left(\left(e^{136}+e^{145}-e^{235}+e^{246}\right)+ i\left(e^{135}-e^{146}+e^{236}+e^{245}\right)\right) .
\end{aligned}
\ee
\end{itemize}

\begin{itemize}
 \item Torsion class:
\begin{equation}
  \cW_1^+=-\frac{4}{\sqrt{3}\,R} \;.
\end{equation}
\end{itemize}
\clearemptydoublepage

\addcontentsline{toc}{chapter}{Bibliography}
\bibliographystyle{plain}

\clearemptydoublepage

\newpage\thispagestyle{empty}\be\ph{page blanche}\nn\ee

\end{document}